\PassOptionsToPackage{unicode}{hyperref}
\PassOptionsToPackage{hyphens}{url}
\PassOptionsToPackage{dvipsnames,svgnames,x11names}{xcolor}
\documentclass[
  american,
  a4paper,
  11pt,ukrainian]{article}
\usepackage{xcolor}
\usepackage[margin=25mm]{geometry}
\usepackage{amsmath,amssymb}
\usepackage[T2A,T1]{fontenc}
\usepackage[utf8]{inputenc}
\usepackage{textcomp} 
\DeclareUnicodeCharacter{03C7}{\ensuremath{\chi}}
\DeclareUnicodeCharacter{03C8}{\ensuremath{\psi}}
\DeclareUnicodeCharacter{2194}{\ensuremath{\leftrightarrow}}
\usepackage{lmodern}
\IfFileExists{upquote.sty}{\usepackage{upquote}}{}
\IfFileExists{microtype.sty}{
  \usepackage[]{microtype}
  \UseMicrotypeSet[protrusion]{basicmath} 
}{}
\makeatletter
\@ifundefined{KOMAClassName}{
  \IfFileExists{parskip.sty}{%
    \usepackage{parskip}
  }{
    \setlength{\parindent}{0pt}
    \setlength{\parskip}{6pt plus 2pt minus 1pt}}
}{
  \KOMAoptions{parskip=half}}
\makeatother
\usepackage[shorthands=off]{babel}
\newcommand{\forcecyrillic}{\fontencoding{T2A}\selectfont}
\usepackage{bookmark}
\IfFileExists{xurl.sty}{\usepackage{xurl}}{} 
\hypersetup{
  pdftitle={From Volterra Series to Kunchenko Stochastic Polynomials: Half a Century of Non-Gaussian Estimation Methodology},
  pdfauthor={S. V. Zabolotnii},
  pdflang={en-US},
  colorlinks=true,
  linkcolor={Maroon},
  filecolor={Maroon},
  citecolor={Blue},
  urlcolor={Blue},
  pdfcreator={LaTeX via pandoc}}

\begin{document}
\selectlanguage{american}

\phantomsection\label{from-volterra-series-to-kunchenko-stochastic-polynomials-half-a-century-of-non-gaussian-estimation-methodology}
\addcontentsline{toc}{section}{From Volterra Series to Kunchenko Stochastic Polynomials}
\begin{center}
{\LARGE\bfseries From Volterra Series to Kunchenko Stochastic
Polynomials: Half a Century of Non-Gaussian Estimation Methodology\par}

\vspace{0.5em}
\emph{On the 87th anniversary of the birth of the founder of the
Cherkasy scientific school}

\vspace{0.4em}
\textbf{S. V. Zabolotnii}

\emph{Cherkasy State Business College}\\
\emph{ORCID: 0000-0003-0242-2234}
\end{center}

\subsection{Abstract}\label{abstract}

This paper reconstructs the half-century evolution of the scientific
school founded by Yuriy P. Kunchenko (1939--2006) as the development of
a semiparametric methodology for non-Gaussian estimation. Its starting
point is Kunchenko's 1972/1973 candidate dissertation, where Volterra
series were applied to estimating unknown parameters of random
processes; the trajectory is followed through the work of his students
and successors in 2006--2026. Kunchenko stochastic polynomials are
presented not merely as a local research tradition, but as a coherent
family of moment-cumulant procedures: the polynomial maximization method
for parameter estimation, polynomial criteria for statistical hypothesis
testing, and decomposition in the space with a generating element for
recognition problems. The paper also accounts for the institutional
structure of the school: a verified genealogy of 15 defended
dissertations, international collaborations in Poland, Slovakia, and
Germany, and the recent R package EstemPMM. A 2026 paper on
Volterra-based processing of stochastic signals is used as a diagnostic
case showing that Kunchenko's original problem has reappeared in applied
radio engineering at the level of problem structure: a nonlinear
functional of a random process is used to construct an estimation or
adaptation procedure. A formal bridge is built between finite Volterra
models and generalized Kunchenko stochastic polynomials on vector
arguments; at the same time, the MMSE/L2 criterion in the 2026 paper is
separated from PMM, since the former is a covariance projection for
kernel adaptation whereas the latter is a parameter-dependent
moment-based procedure for estimating parameters of a process or
distribution. Claims about PMM efficiency are stated conditionally:
gains are expected for non-Gaussian classes where the required moments
exist, the centered correlant matrix is nondegenerate, and the variance
reduction coefficient is below one. The concluding research program
turns the historical reconstruction into a set of testable statistical
and signal-processing tasks.

\textbf{Keywords:} Kunchenko stochastic polynomials; polynomial
maximization method; Volterra series; non-Gaussian random processes;
moment-cumulant description; space with generating element;
semiparametric methods; nonlinear signal processing; Cherkasy scientific
school; adaptive parameter estimation; variance reduction coefficient;
cumulant description perforation; second-order least squares (SLS).

\subsection{1. Introduction: history as the trajectory of the
method}\label{introduction-history-as-the-trajectory-of-the-method}

May 26, 2026 marks the eighty-seventh anniversary of the birth of Yuriy
Petrovych Kunchenko (1939, Rostov-on-Don --- 2006, Cherkasy) ---
Ukrainian radio physicist and mathematician, Doctor of Physical and
Mathematical Sciences, Honored Worker of Science and Technology of
Ukraine, founder of the Cherkasy Scientific School of Nonlinear
Statistical Analysis of Non-Gaussian Signals. His early work with the
Volterra series is important not only as a historical fact. It already
laid a way of thinking, which will later become recognizable for the
whole school: the parameter of a random process should be estimated not
only through linear statistics or a completely given distribution
function, but through polynomial functionals that are able to use
information of moments of higher orders.

That is why this article should not be read as a biographical essay or
as a local chronicle of the departmental tradition. Its subject is the
path of the statistical method. From Ph.D.~thesis 1972/1973 to
monographs 1987--2006, from parameter estimation of random variables to
statistical hypothesis testing and pattern recognition, from handwritten
normal equations to the reproducible R package EstemPMM --- the same
idea unfolds before us: non-Gaussianity is not noise to be eliminated,
but a source of additional statistical information.

This review article has three objectives. The first is to reconstruct
the conceptual evolution of the school on the basis of verified
bibliographic and institutional data. The second is to show that the
three applied branches of the school --- parameter estimation,
hypothesis testing and pattern recognition --- are derivatives of the
same formal apparatus of stochastic polynomials. The third is to explain
why the recent Volterra problem of stochastic signal processing,
published in 2026, in fact takes us back to the original type of
Kunchenko's formulation and at the same time opens up a modern research
agenda.

Structurally, the text moves in three layers. Chapters 2--4 reproduce
the evolution from Volterra's formulation to the mature apparatus of
stochastic polynomials. Chapters 5-6 show the school as an institutional
and research network. Chapters 7--10 translate history into methodology:
they compare Kunchenko's apparatus with related statistical traditions,
analyze the case of 2026, and build a formal bridge between the Volterra
model and the stochastic polynomial. Concluding chapters identify open
issues and formulate next steps.

\subsection{2. Point zero: Volterra's series in the 1972/1973
dissertation.}\label{point-zero-volterras-series-in-the-19721973-dissertation.}

Kunchenko's PhD thesis has the following initial data: manuscript dated
1972, place of defense --- Tomsk State University, volume --- 164 pages,
specialty --- 04.01.03 ``Radiophysics, including quantum radiophysics''.
According to the memorial article of ChDTU, the defense took place in
1973 in the specialized council at Tomsk University {[}7{]}. This
ambiguity of date --- typical for Soviet dissertations of a discrepancy
between the year of completion of the manuscript and the year of the
procedural defense --- has no conceptual consequence, and hereafter we
will refer to the dissertation as the work of 1972/1973.

What constitutes the intellectual context of this work? The beginning of
the 1970s in the Soviet radio physics school --- the period of maturity
of the theory of random processes, formulated in the monographs of
Pugachev, Tikhonov, Sosulin. At the same time, the Western school
(Wiener, Volterra, later Schetzen, Rugh) actively developed functional
series as a tool for describing nonlinear systems with memory. The point
of intersection of these two lines --- namely, the use of Volterra's
series to solve statistical problems, not system-identification, but
parametric-evaluation ones --- constitutes the subject field of
Kunchenko's dissertation.

The problem of the dissertation is posed as follows. Let the random process
\(\xi(t)\) be observed, the statistical characteristics of which depend
on the unknown parameter \(\theta\) (or the vector of parameters
\(\overrightarrow{\theta}\)). The Volterra series allows representing
the functional from \(\xi(t)\) in the form of a finite-dimensional
nonlinear expansion of the products of the process values
\hspace{0pt}\hspace{0pt}at different moments in time, with kernels
\(h_{n}\left( \tau_{1},\ldots,\tau_{n} \right)\) carrying information
about the nonlinear structure of the dependence of the observed quantity
on \(\theta\). The problem of estimating \(\theta\) is, therefore,
reduced to finding such kernels \(h_{n}\) that minimize the mean square
error of approximation of the statistic calculated from the sample to
its theoretical value as a function of the unknown parameter.
Technically, this is a system of normal equations of the type
\(C_{x} \cdot h = r_{yx}\), where \(C_{x}\) is the process moment matrix
and \(r_{yx}\) is the cross-correlation vector.

In this formulation, all four elements are already present that will
become the conceptual basis of the school in the following decades. First,
\textbf{focusing on evaluation, not on system identification}: the
Volterra kernel in Kunchenko does not describe a physical channel, but
is a tool for obtaining parameter estimates. Secondly,
\textbf{polynomial approximation structure}: the estimate is not based
on a single linear statistic, but on the products of process values
\hspace{0pt}\hspace{0pt}of different orders, which opens up the
possibility of involving information of higher moments. Third,
\textbf{the moment-based nature of the description}: kernel coefficients are
expressed through the moments (cumulants) of the process, and not
through its full distribution function. Fourth, \textbf{focusing on
non-Gaussian processes}: for Gaussian processes, the higher cumulants
are identical to zero, and the polynomial expansion loses its advantages
over the linear one; all the effectiveness of the method is manifested
precisely in the presence of non-Gaussian characteristics.

None of these four elements in 1972 had yet been formalized as a
separate theory. Their synthesis into an independent apparatus --- a
matter of the next 15 years, culminating in a doctoral dissertation in
1988 and a monograph in 1987. But it is important to record: the
conceptual germ of the entire future school is already present in the
candidate's dissertation. The stochastic polynomial as a mathematical object
has not yet been named, the space with a generating element has not yet
been introduced, the moment-cumulant description has not yet been
systematized --- however, the methodological setting is already set:
\textbf{estimate the parameters of non-Gaussian random processes through
nonlinear polynomial functionals from the moments of observed
statistics, and not through linear statistics and a priori distribution
function}.

It is this setup that has evolved in the following decades into a family
of interconnected devices---and it is this that is repeatedly and
independently reconstructed in 2026 by the authors of the article
{[}6{]}, which is discussed in § 8. Between these two points lies half a
century of the school's inner workings, which the following chapters
successively unfold.

\subsection{3. Transition 1973--1987: from Volterra kernels to nonlinear
estimation of non-Gaussian
signals}\label{transition-19731987-from-volterra-kernels-to-nonlinear-estimation-of-non-gaussian-signals}

The fifteen years that separate the candidate's thesis of 1972/1973 from
the first monograph of 1987, ``Nonlinear estimation of parameters of
non-Gaussian radiophysical signals'' (Kyiv: Vyshcha shkola, 191 p.)
{[}8{]}, constitute the least visible, but conceptually central period
of the school's evolution. This period covers the completion of the
Tomsk stage of Kunchenko's biography (1962--1978, Department of
Radiophysics of TSU and the Siberian Physical and Technical Institute)
and his Kirovohrad stage (1979--1990, associate professor, later head of
the Department of Higher Mathematics of the Kirovohrad Institute of
Agricultural Engineering). The period ends with the defense of a
doctoral dissertation in 1988 in the specialized council at Kharkiv
State University in the same specialty ``radiophysics, including quantum
radiophysics'' as the candidate's {[}7{]}.

The documentary trace of this period in publicly available sources is
fragmentary. Open library catalogs give the exact name and code of the
candidate (§ 2), and allow locating the 1987 monograph, but do not
preserve a complete list of articles and reports between these two
dates. This is not an exceptional situation --- for the Soviet authors
of the radiophysical profile of the 1970s--1980s, the main publication
channel was through journals, as well as through classified in-house
R\&D reports and conference proceedings, digitized only
selectively.

Conceptually, however, the direction of the movement can be clearly
traced by comparing two reference points --- the candidate's dissertation
of 1972/1973 and the 1987 monograph.

In the monograph of 1987, all the basic elements that will make up the
mature apparatus of the school are already present. The concept of a
stochastic polynomial as a random variable, which is a linear
combination of functions (power, trigonometric, exponential) from an
observed random variable, with a matrix of centered correlations and the
volume of the body of the polynomial as key invariants {[}4{]}. The
formulation of the parameter estimation problem through the maximization
of a stochastic polynomial, in which the estimates are located as the
points of the global maximum of the mathematical expectation of the
polynomial considered as a function of the unknown parameter. Proof of
the asymptotic efficiency of such estimates with increasing degree of
the polynomial. The transfer of this apparatus from estimation of
parameters of random variables to a wider class of problems ---
estimation of parameters of signals on a non-Gaussian background.

The key conceptual shift from the 1972/1973 candidate to the 1987
monograph --- \textbf{transition from the functional description
``input-output'' to the moment-cumulant description of the
distribution}. In the original Volterra formulation, the \(h_{n}\)
kernel described the physical signal transmission channel, even though
the goal was parameter estimation rather than system identification. In
the mature apparatus of 1987, the polynomial structure is transferred
from the space of signals to the space of moments: the coefficients
\(h_{i}\) cease to be functions of time delays
\(\tau_{1},\ldots,\tau_{n}\) and become coefficients of the expansion by
the moment or cumulant characteristics of a random variable. This shift
frees the theory from being tied to the specific physics of the channel
and makes it universal. The same apparatus now works for estimating the
parameters of the sample, and for estimating the parameters of the
process, and for constructing decisive rules for testing hypotheses ---
because all these problems are reduced to operations with moments.

The second conceptual shift of this period --- \textbf{restriction of
the class of studied distributions due to ``closeness to Gaussian''}. In
the monograph of 1987 and works of the following years, the idea of
\hspace{0pt}\hspace{0pt}perforation of the cumulant description {[}4{]}
is consistently developed: for many practically important problems, the
cumulant coefficients above a certain order are so small that they can
be neglected by zeroing them in the model. By combining different types
of perforation, three classes of non-Gaussian random variables are
distinguished --- asymmetric, kurtosis, and asymmetric-kurtosis, which
are called ``close to Gaussian''. This classification will become the
basis for the canonical English-language monograph of 2002
\emph{Polynomial Parameter Estimates of Close to Gaussian Random
Variables} (Aachen: Shaker Verlag, 396 p.) {[}9{]}, the title of which
directly reflects this theoretical course.

The third shift --- \textbf{emergence of a moment criterion of the
quality of the formation of decisive rules}. Already in the monograph of
1987 and in the articles of the late 1980s --- early 1990s, the idea
that the polynomial apparatus is applicable not only to estimation, but
also to hypothesis testing is formulated: the decomposition of the
logarithm of the likelihood ratio into a stochastic series with
coefficients optimal according to moment criteria provides an
alternative to the classical Neyman-Pearson criterion, without requiring
knowledge of the full distribution function {[}4{]}. For the first time,
this branch receives systematic design in the report of Yu.P. Kunchenko
at the IEEE International Symposium on Information Theory in Ulm in
1997. {[}12a{]} --- this fact should be emphasized separately: the
apparatus of the school was present in international IEEE literature a
quarter of a century before the current period of its continuation. This
branch later crystallized in the works of Kunchenko with V.V. Palaginim
and S.S. Martynenko 1998--2006 {[}12, 13{]} and will form the foundation
of the signal-detector branch of the school (§ 6.2).

Thus, the period 1973--1987 is a period in which the idea, the germ of
which is present in the candidate's dissertation, is formalized into an
independent apparatus. Kunchenko's stochastic polynomial as a mathematical object,
PMM as an evaluation method, moment-cumulant description as a form of
information, perforation as a classification technique, random variables
close to Gauss as the target class. All this --- the result of a
relatively quiet internal course of 15 years of work, documented only at
two reference points, but conceptually continuous.

\subsection{4. Maturity of the school 1991--2006: design of the
apparatus of stochastic
polynomials}\label{maturity-of-the-school-19912006-design-of-the-apparatus-of-stochastic-polynomials}

The period between the monograph of 1987 and the death of Yuriy Petrovych
on August 6, 2006 --- these are seventeen years, during which the school
from the thematic direction of one author turned into a full-fledged
institutional formation with its own department (since 1990 --- the
department of radio engineering of the Cherkasy Institute of Engineering
and Technology, now ChDTU), its own terminology, its own canonical
corpus of monographs and its own first generations of students. This
period opens with Kunchenko's move from Kirovohrad to Cherkasy in early
1990 and closes with a posthumous monograph in 2006, which became the
final synthesis of the theory.

\subsubsection{4.1. Canonical corpus of
monographs}\label{canonical-corpus-of-monographs}

The book canon of the school, created during this period, consists of
six monographs that form an internally consistent sequence. The first
--- the already mentioned ``Nonlinear estimation of parameters of
non-Gaussian radiophysical signals'' of 1987 {[}8{]} --- serves as a
bridge between the candidate period and the departmental work in
Cherkasy. The second --- ``Estimation of the parameters of random
variables by the polynomial maximization method'' (Kyiv: Naukova dumka,
1991/1992; co-authored with Yu.G. Lega) {[}2{]} --- becomes the first
systematic presentation of PMM as a method. The third --- ``Polynomial
estimates of parameters close to Gaussian random variables. Part 1''
(Cherkasy, 2001) together with ``Part 2'', written in co-authorship with
S.V. Zabolotnii --- fixes the Cherkasy stage and derives the
classification of values \hspace{0pt}\hspace{0pt}close to Gaussian
(asymmetric, excess, asymmetric-excess) in a form directly oriented to
applied problems {[}10{]}.

The fourth monograph --- \emph{Polynomial Parameter Estimates of Close
to Gaussian Random Variables} (Aachen: Shaker Verlag, 2002, 396 p.)
{[}9{]} --- constitutes the main English-language canon of the school.
Its appearance is connected with Kunchenko's guest stay at the Institute
of Communication Technology of the University of Hanover in 2001--2002,
where he gave a course of lectures on the polynomial maximization
method. This monograph remains to this day the only complete exposition
of the PMM apparatus in English, and it is cited by all international
publications of the school from 2010--2026 without exception as the main
entry point.

The fifth and sixth monographs of 2003--2006 --- ``Polynomial
approximations in space with a generating element'' (K.: Naukova Dumka,
2003, 243 p.) {[}3a{]}, its Ukrainian-language version ``Polynomial
approximations in space with a generating element'' (K.: Naukova Dumka,
2005) {[}3{]} and the final ``Stochastic polynomials'' (K.: Naukova
dumka, 2006, 275 p.) {[}11{]} --- form the geometric interpretation of
the apparatus. Here, a space with a generating element (Kunchenko space)
is introduced as an abstract formal environment in which a stochastic
polynomial exists as an object regardless of a specific class of basis
functions. It is this geometrization that opens the third branch of the
school---a layout in space with a generative element for pattern
recognition, to which we will return in § 6.3.

\subsubsection{4.2. The formal core of the
apparatus}\label{the-formal-core-of-the-apparatus}

The mature form of the apparatus of stochastic polynomials, designed in
the monographs of 2002--2006, in a compact presentation by the students
of the {[}4{]} school looks like this. Let \(\xi\) --- an observed
random variable with an unknown parameter \(\overrightarrow{\theta}\).
We consider the set of functions \(\{\varphi_{i}(\xi)\}_{i = 1}^{S}\),
which are linearly independent on the range of admissible values
\hspace{0pt}\hspace{0pt}of the parameter. \textbf{Generalized stochastic
polynomial} of degree \(S\) is defined as a random variable

\[
\eta_{S} = h_{0} + \sum_{i = 1}^{S}h_{i}\varphi_{i}(\xi),\quad\left| h_{i} \right| < \infty,
\]

where coefficients \(h_{i}\) --- non-random constants. If
\(\varphi_{i}(\xi) = \xi^{i}\), the polynomial is called a power
polynomial and is completely described in terms of initial moments
\(\alpha_{i}\left( \overrightarrow{\theta} \right) = E\{\xi^{i}\}\). If
\(\varphi_{i}(\xi) = \cos(ik\xi)\) or \(\varphi_{i}(\xi) = \sin(ik\xi)\)
--- is trigonometric and described by the characteristic function.

The key characteristic of a polynomial is the \textbf{matrix of centered
correlations}

\[
F_{i,j}\left( \overrightarrow{\theta} \right) = \Psi_{i,j}\left( \overrightarrow{\theta} \right) - \Psi_{i}\left( \overrightarrow{\theta} \right)\Psi_{j}\left( \overrightarrow{\theta} \right),
\]

where
\(\Psi_{i,j}\left( \overrightarrow{\theta} \right) = E\{\varphi_{i}(\xi)\varphi_{j}(\xi)\}\),
\(\Psi_{i}\left( \overrightarrow{\theta} \right) = E\{\varphi_{i}(\xi)\}\).
The variance of the polynomial is expressed through the matrix
\(F_{S} = \parallel F_{i,j} \parallel_{i,j = 1}^{S}\) as a quadratic
form

\[
\sigma_{\eta}^{2} = \sum_{i = 1}^{S}\sum_{j = 1}^{S}h_{i}h_{j}F_{i,j}\left( \overrightarrow{\theta} \right).
\]

Kunchenko calls the matrix \(F_{S}\) \textbf{body of stochastic
polynomial}, its determinant
\(\Delta_{S}\left( \overrightarrow{\theta} \right) = \left| F_{S} \right|\)
--- \textbf{body volume}. The non-degeneracy condition
\(\Delta_{S}\left( \overrightarrow{\theta} \right) > 0\), which follows
from the properties of the Gram determinant under linear independence
\(\{\varphi_{i}\}\), ensures the existence and uniqueness of the
solution to the parameter estimation problem. For a power-law
polynomial, the correlations are expressed directly in terms of moments:
\(\Psi_{i,j} = \alpha_{i + j}\),
\(F_{i,j} = \alpha_{i + j} - \alpha_{i}\alpha_{j}\).

\subsubsection{4.3. Polynomial maximization
method}\label{polynomial-maximization-method}

PMM is built on the property proved by Kunchenko: under certain
conditions, the mathematical expectation of the stochastic polynomial as
a function of the parameter \(\theta\) on the coefficients \(h_{i}\) has
a global maximum around the real value of this parameter. Formally, the
estimate \({\widehat{\theta}}_{\text{PMM}}\) is located as the maximum
point of the function

\[
L_{S}\left( \theta;x_{1},\ldots,x_{N} \right) = \sum_{i = 1}^{S}h_{i}^{*}(\theta) \cdot \frac{1}{N}\sum_{n = 1}^{N}\varphi_{i}\left( x_{n} \right),
\]

where \(h_{i}^{*}(\theta)\) --- optimal coefficients that minimize the
variance of the estimate for the given \(\theta\). The solution of the
system of linear equations
\(F_{S}(\theta) \cdot \overrightarrow{h} = \overrightarrow{b}(\theta)\),
where \(\overrightarrow{b}(\theta)\) --- a vector depending on the
derivatives of \(\Psi_{i}\) by parameter, gives an explicit form of
optimal coefficients.

The central result is the asymptotic behavior of the estimate variance.
To estimate the mean \(\theta = \mu\) of an asymmetric non-Gaussian
distribution with \(c_{2},c_{3},c_{4},\ldots\) cumulants, the variance
of the estimate obtained PMM of power \(S = 2\) is expressed as

\[
\sigma_{\widehat{\mu},\text{PMM-2}}^{2} = \frac{c_{2}}{N} \cdot g_{2},\quad g_{2}\left( \overrightarrow{\theta} \right) = 1 - \frac{c_{3}^{2}}{c_{2}\left( 2c_{2} + c_{4}/c_{2} \right)}.
\]

In the standard dimensionless form due to the coefficients of asymmetry
\(\gamma_{3} = c_{3}/c_{2}^{3/2}\) and kurtosis
\(\gamma_{4} = c_{4}/c_{2}^{2}\):

\[
g_{2} = 1 - \frac{\gamma_{3}^{2}}{2 + \gamma_{4}}.
\]

This formula --- \textbf{coefficient of variation reduction} PMM of the
second degree --- occupies one of the central places in the entire
apparatus of the school. Its meaning is easy to interpret. For the
Gaussian distribution, \(\gamma_{3} = 0\), \(\gamma_{4} = 0\), so
\(g_{2} = 1\) --- PMM coincides with the classical sample mean, there is
no gain. For distributions that satisfy the required moment conditions
and \(g_{2} < 1\), the estimate PMM-2 has a smaller asymptotic variance
than the sample mean; the amount of reduction is determined by the
characteristics of the distribution, not just the sample size \(N\).

Generalizing to \(S \geq 3\) in regular classes can reduce \(g_{S}\),
bringing the estimate closer to the Cramér--Rao efficiency limit. In the
\(S \rightarrow \infty\) limit, the PMM estimate is asymptotically
equivalent to the maximum likelihood estimate, while \textbf{not
requiring knowledge of the full distribution function}: knowledge of the
moments up to the \(2S\)th order is sufficient. It is this property that
makes PMM a semiparametric method --- a compromise between the
parametric approach (which requires a priori full distribution) and the
nonparametric (which ignores the information of higher moments) {[}4{]}.

\subsubsection{4.4. Perforation of the cumulant description and
classification of close to Gaussian
quantities}\label{perforation-of-the-cumulant-description-and-classification-of-close-to-gaussian-quantities}

The practical application of formula (6) and its generalizations to
\(S \geq 3\) rests on a problem that Kunchenko formulates as the
\textbf{problem of non-closure}: to evaluate the \(S\)-th coefficient,
moments up to the \(2S\)-th order are required, which generates an
infinite chain of {[}4{]} moment equations. The standard technique of
zeroing moments above a certain order (the so-called statistical
linearization) often turns out to be incorrect and leads to significant
errors.

The alternative proposed by Kunchenko --- \textbf{perforation of the
cumulative description}. The idea is that in many practically important
problems, the cumulant coefficients above a certain order are so small
that they can be set to zero, while certain cumulants of higher orders
remain informative and are taken into account during the synthesis of
the algorithm, and the rest take arbitrary values. By combining
different types of perforation, three classes of close to Gaussian
random variables are distinguished:

\begin{itemize}
\item
  \textbf{asymmetric} (informative \(\gamma_{3}\), zeroed
  \(\gamma_{4} = \gamma_{5} = \ldots = 0\));
\item
  \textbf{excessive} (informative \(\gamma_{4}\), zeroed
  \(\gamma_{3} = \gamma_{5} = \ldots = 0\));
\item
  \textbf{asymmetric-excessive} (both informative, \(\gamma_{3}\) and
  \(\gamma_{4}\)).
\end{itemize}

This classification is not only a mathematical convenience, but a
specific tool for modeling radio engineering signals and interference,
for which it is empirically known that deviations from the Gaussian
distribution are concentrated mainly in the third and fourth moments.
For each of the three classes, the school's mature apparatus gives
closed formulas for grades, variances, and conditions of applicability.

\subsubsection{4.5. The end of the Cherkasy
period}\label{the-end-of-the-cherkasy-period}

The last years of Yuriy Petrovich's life (2003--2006) are spent at a
busy pace: international business trips, writing monographs, a number of
joint articles with students on moment quality criteria of the
Neyman-Pearson type for testing statistical hypotheses {[}12, 13{]}
(continuing the line started in IEEE ISIT 1997 {[}12a{]}), working in a
position vice-rector for scientific research and international relations
of ChSTU. In 2006, just before his death, Kunchenko defended two theses:
A.V. Goncharova --- ``Estimation of the constant signal parameter in the
case of close to Gaussian additive disturbances'' (spec. 05.01.02) and
T.V. Vorobakolo --- ``Methods and algorithms for estimating the angle of
arrival of a harmonic signal on an antenna array in case of non-Gaussian
interference'' (spec. 01.05.02) {[}7{]}. On January 20, 2006, Yuriy
Petrovych was awarded the honorary title ``Honored Worker of Science and
Technology of Ukraine'' by the decree of the President of Ukraine.

Until the moment of death, the school exists as a mature structure with
its own institutional basis. Sixteen years of work in Cherkasy. Six
candidate theses were defended under Kunchenko's supervision: V.V.
Palagin 1999, S.V. Zabolotnii 2000, O.S. Havrysh 2001, S.S. Martynenko
2003, A.V. Goncharov and T.V. Vorobakolo 2006 {[}7{]}. Six monographs and
more than 150 scientific works. Separate theoretical apparatus,
terminology, classification of problems. The first composition of the
radio engineering department of ChSTU, which will become the
institutional basis of the school for the next two decades. The further
fate of the school --- how it survived the loss of its founder, how it
developed into three modern branches, how it reached the international
level --- is considered in the next section.

\subsection{5. School as an institutional phenomenon: genealogy and
geography}\label{school-as-an-institutional-phenomenon-genealogy-and-geography}

Looking at the history of Kunchenko's school from the point of view of
2026, we see a structure that goes beyond the usual ``teacher ---
students''. This is a multigenerational network built on three levels:
Kunchenko's direct students who defended themselves under his leadership
in 1999--2006; the second generation, headed by V.V. Palagin and S.V.
Zabolotnii with the defense of doctoral theses in 2013 and 2015; and
independent applications of the apparatus, unfolding outside the direct
genealogical line, but supported by the canonical monographs of the
school. The official register of protected dissertations is maintained
on the department's website and, as of 2026, includes fifteen
dissertations on the subject of the scientific school of Professor Yu.P.
Kunchenko {[}39{]}.

A separate note. During Kunchenko's Cherkasy period, other dissertations
were also defended at the Department of Radio Engineering of ChSTU, in
which Yuriy Petrovych was not the supervisor: two doctoral theses (V.M.
Sharapov, 1998; Yu.G. Lega, 2001) and one candidate's (O.O. Sytnyk,
1997) {[}7{]}. In the list of students of the school in the following
subsections, these three dissertations are not given, since their
authors were not students of Kunchenko in the formal sense, although
they belong to the common institutional context of the department.

\subsubsection{5.1. The first generation: direct students of
Kunchenko}\label{the-first-generation-direct-students-of-kunchenko}

\textbf{Volodymyr Vasyliovych Palagin}. Candidate's degree, 1999.
``Algorithms for detection of signals against the background of
non-Gaussian interference according to the criterion of asymptotic
normality'' (special. 05.12.01 ``Theoretical radio engineering'';
scientific supervisor --- Yu.P. Kunchenko) {[}14, 39{]}. Palagin worked
at the radio engineering department of the ChDTU since 1992 (that is, he
was with Kunchenko practically from the moment the department was
created). In 2013, he defended his doctoral thesis ``Mathematical
models, methods and means of detecting and distinguishing signals
against the background of non-Gaussian interference'' with the specialty
01.05.02 ``Mathematical modeling and computational methods''; Doctor of
Technical Sciences, Professor A.F. Verlan acted as a scientific
consultant {[}5, 39{]}. Since 2014 --- head of the department of robotics
and telecommunication systems and cyber security of ChSTU. Palagin is
the institutional successor of the scientific school at ChDTU.

\textbf{Serhiy Vasyliovych Zabolotnii} came to Kunchenko's school as a
3rd-year student. The thesis of 1995 was devoted to the distribution of
random variables and processes in space with a generating element --- a
problem that later became the first statement of his candidate's thesis.
The topic was later reformulated by Kunchenko to be related to the
polynomial maximization method, which reflected one of the typical
moments of the school's work: the internal mobility of students between
the conceptual branches of the apparatus. Candidate's degree in 2000
``Nonlinear algorithms for determining the parameters of non-Gaussian
random sequences in the channels of information and measurement
systems'' was defended on November 5, 2016 with the specialty
``Information and measurement systems'' under the scientific supervision
of Yu.P. Kunchenko {[}15, 39{]}. Doctoral 2015 ``Information technology
of probabilistic diagnosis of disorder parameters of non-Gaussian
sequences'' defended at the Ukrainian Academy of Typography on 05.13.06
``Information technologies''; the scientific consultant was L.S. Sikora
{[}4, 39{]}. Although the formal specialty of the doctorate went beyond
the school's radio engineering profile, it retained the central role of
the apparatus of stochastic polynomials and moment-cumulant
description, transferring it to the field of diagnostic information
technologies.

\textbf{Yuriy Hryhorovych Lega} does not formally belong to Kunchenko's
students in the genealogical sense. He was a member of the first team of
the department founded by Kunchenko in 1990, and co-authored the book
``Estimation of parameters of random variables by the polynomial
maximization method'' in 1991/1992. {[}2{]}. In 2001, he defended his
doctoral dissertation on a topic similar to the Kunchenko school; in
2006--2014 --- rector of ChDTU. Lega --- an institutional ally and
co-creator of the Cherkasy stage of the school, who after Kunchenko's
death partially transferred the administrative weight and in 2013 acted
as supervisor of the candidate's thesis of one of the representatives of
the second generation of the school --- V.V. Koval (Ph.D.~2013, spec.
01.05.02, topic ``Models and methods of polynomial estimation of signal
parameters against the background of multiplicative and
additive-multiplicative disturbances'').

The rest of Kunchenko's direct students from the canonical list
{[}39{]}:

\begin{itemize}
\item
  \textbf{O.S. Havrysh} --- Ph.D.~2001, special ``Radio physics'', the
  topic ``Algorithms for measuring the parameters of harmonic and
  polyharmonic signals with non-Gaussian interference'';
\item
  \textbf{S.S. Martynenko} --- Ph.D.~2003, special ``Radio physics'',
  topic ``Signal detection against the background of non-Gaussian
  interference by polynomial algorithms'';
\item
  \textbf{A.V. Goncharov} --- Ph.D.~2006, special 01.05.02, topic
  ``Estimation of the parameter of a constant signal with close to
  Gaussian additive disturbances'';
\item
  \textbf{T.V. Vorobakolo} --- Ph.D.~2006, special 01.05.02, topic
  ``Methods and algorithms for estimating the angle of incidence of a
  harmonic signal on an antenna array in case of non-Gaussian
  interference''.
\end{itemize}

In the modern structure of the school A.V. Goncharov rather played the
role of a successor to Kunchenko's unfinished supervision: he in effect
brought in V.V. Filipov, who had started as Kunchenko's graduate student. In
addition, Goncharov was the formal manager of O.M. Tkachenko, while the
conceptual leadership of this work belongs to S.V. Zabolotnii.

\subsubsection{5.2. Second generation: two main
lines}\label{second-generation-two-main-lines}

Two decades after Kunchenko's death, the school has at least eight
second-generation defended dissertations. Its main genealogical
structure unfolds along two vertical lines --- Palagin and Zabolotnii;
next to them, there are separate related protections associated with
other representatives of the first generation and the institutional
environment of the department. This structure is reflected in the
official register of ChDTU {[}39{]}.

\textbf{Palagin's line} --- the most fully presented signal-detector
branch:

\begin{itemize}
\item
  \textbf{O.V. Ivchenko} --- Ph.D.~2015, special 01.05.02, topic
  ``Mathematical models, methods and means of estimating parameters of
  non-Gaussian correlated random processes''; scientific supervisor ---
  V.V. Palagin;
\item
  \textbf{S.A. Leleko} --- Ph.D.~2018, special 01.05.02, topic
  ``Mathematical models and methods of detecting signals against the
  background of non-Gaussian disturbances according to the moment
  quality criterion''; scientific supervisor --- V.V. Palagin;
\item
  \textbf{D.A. Vedernikov} --- PhD 2021, special. 122 ``Computer
  Sciences'', the topic ``Mathematical models, methods and means of
  estimating the constant signal parameter against the background of
  non-Gaussian correlated disturbances''; scientific supervisor --- V.V.
  Palagin {[}16{]};
\item
  \textbf{D.O. Smirnov} --- PhD 2025, special. 122 ``Computer
  Sciences'', the topic ``Mathematical models, methods and means of
  detecting a constant signal against the background of non-Gaussian
  correlated disturbances''; scientific supervisor --- V.V. Palagin
  {[}17{]};
\item
  \textbf{O.S. Zorin} --- PhD defense planned for 2026, spec. 152
  ``Metrology and information-measuring technology'', topic ``Models and
  methods of evaluation and detection of signals against the background
  of non-Gaussian disturbances in information-measuring systems'';
  scientific supervisor --- V.V. Palagin {[}39{]}.
\end{itemize}

\textbf{Zabolotnii Line} --- regression-evaluation branch:

\begin{itemize}
\item
  \textbf{V.O. Selin} --- Ph.D.~2013, special 01.05.02, topic ``Models
  and methods of nonlinear estimation of parameters of polynomial trends
  with a non-Gaussian stochastic component''; scientific supervisor ---
  S.V. Zabolotnii;
\item
  \textbf{A.V. Chepynoga} --- Ph.D.~2016, special 01.05.02, topic
  ``Methods of polynomial estimation of parameters of polygaussian
  models with moment-cumulant description''; scientific supervisor ---
  S.V. Zabolotnii;
\item
  \textbf{O.M. Tkachenko} --- PhD 2021, special. 122 ``Computer
  science'', topic ``Polynomial methods and means of estimating
  regression parameters using non-Gaussian error models'' {[}18{]}. The
  conceptual leadership belongs to S.V. Zabolotnii, with whom O.M.
  Tkachenko jointly published all the key works of his dissertation
  2017--2021 {[}40, 41{]}.
\end{itemize}

Separately within the second generation is the dissertation \textbf{V.V.
Filipova} --- Ph.D.~2016, special 01.05.02, topic ``Methods of joint
estimation of constant signal parameters and non-Gaussian disturbances
using truncated stochastic polynomials''. The formal scientific
supervisor was \textbf{A.V. Goncharov}, but genealogically this defense
should be read as bringing to completion the work of graduate student
Kunchenko after the death of the founder of the school. Therefore, it is
important not as a separate vertical line, but as an example of the
successive completion of the topic, which was started even in the
Kunchenkov scientific context.

\subsubsection{5.3. Geography of international
collaborations}\label{geography-of-international-collaborations}

The international footprint of the school is real, but not massive. It
is a network of targeted bilateral collaborations rather than a broad
Western takeover of the apparatus. Reference points on the map:

\textbf{Poland} --- the most productive direction. Tandem S.V.
Zabolotnii (Cherkasy) + Z.L. Warsza (Industrial Institute of Automation
and Measurements, Warsaw) has given about fifteen joint publications
since 2015, in particular, in the annual series Springer AISC/LNNS
\emph{Automation} (Warsaw) {[}21--25{]}. Topics --- polynomial
estimation of measurement parameters with asymmetric distributions,
regression, ARIMA models.

\textbf{Slovakia} --- Kosice Technical University. The collaboration of
V.V. Palagin (ChDTU) + J. Yugar, L. Vokorokos, S. Marchevskyi (TU
Košice). Key publications --- article in \emph{Journal of Electrical
Engineering} (Bratislava, 2016) {[}26{]} and in \emph{IET Signal
Processing} (2017) {[}27{]}.

\textbf{Germany} --- Hannover University, Institute of Communication
Technology, where Kunchenko was a guest in 2001--2002. Direct result of
the visit --- publication of canonical monograph in English in 2002 in
Shaker Verlag (Aachen) {[}9{]}.

\subsubsection{5.4. Institutional artifacts and memorial
infrastructure}\label{institutional-artifacts-and-memorial-infrastructure}

After the death of Kunchenko in 2006, a memorial infrastructure of the
school was created in Cherkasy: the Museum of the History of Radio
Technology named after Yu.P. Kunchenko (since 2013 at the Department of
Radio Engineering of ChDTU); Charitable Foundation ``Science School
named after Yuriy Petrovych Kunchenko'', whose president was his wife
V.I. Kunchenko-Kharchenko; a memorial plaque on the house where he
lived; exposition fund in the Cherkasy Regional Museum of Local History
(opened on May 20, 2008) {[}7{]}. In the 2010s, the International
Scientific and Practical Conference ``Signal Processing and Non-Gaussian
Processes'' (OSNP), dedicated to the memory of Professor Yu.P. Kunchenko,
was regularly held at the department. Its seventh meeting was held in
2019 {[}29{]} and its eighth on May 25--26, 2021 {[}42{]}.
After 2021, the conference was not held, which is due to the well-known
circumstances of the war, which from 2022 will significantly limit
academic life in Ukraine.

\subsection{6. Three branches of the modern school
2006--2026}\label{three-branches-of-the-modern-school-20062026}

The school's internal taxonomy, recorded by students {[}4{]},
distinguishes two conceptual branches of the apparatus of stochastic
polynomials --- statistical estimation of parameters through PMM and
statistical hypothesis testing, and the latter has two sub-branches:
decomposition of the logarithm of the likelihood ratio in a stochastic
series with coefficients optimal by moment criteria, and decomposition
in space with related element for pattern recognition. In terms of
active research groups 2006--2026, this gives three branches of the
school with different centers of gravity.

\subsubsection{6.1. Regression-evaluation
branch}\label{regression-evaluation-branch}

The most active in 2018--2026 in terms of the volume of publications ---
group S.V. Zabolotnii with the Polish partner Z.L. Warsza. The group in
Cherkasy also includes direct graduate students of Zabolotnii, whose
dissertations directly developed the regression-evaluation line:
\textbf{V.O. Selin} (2013, polynomial trends with non-Gaussian
stochastic components), \textbf{A.V. Chepynoga} (2016, polynomial
estimation of parameters of polygaussian models with moment-cumulant
description) and \textbf{O.M. Tkachenko} (PhD 2021, polynomial methods
for estimating regression parameters with non-Gaussian errors). It was
the works of Tkachenko, together with Zabolotnii and Warsza, that brought
this branch into the modern regression setting --- from estimation of
distribution parameters to linear, non-linear and time-series
regression. Topic Core: Polynomial parameter estimates under asymmetric
and platykurtic error distributions, with a focus on modern regression
models and time series.

Key publications: \emph{Polynomial parameter estimation of exponential
power distribution data} (2018) {[}30{]}, \emph{Polynomial estimation of
linear regression parameters for asymmetric PDF} (Springer Automation
2018) {[}22{]}, \emph{Estimation of linear regression parameters for
symmetric non-Gaussian errors} (Springer Automation 2019) {[}23{]},
\emph{Estimating parameters of linear regression with an exponential
power distribution of errors by using a polynomial maximization method}
(2021) {[}31{]}, \emph{Polynomial maximization method for estimation of
asymmetric non-Gaussian MA models} (Springer Automation 2023) {[}24{]},
\emph{Application of PMM for estimating nonlinear regression parameters}
(Springer Automation 2024) {[}25{]}. A separate line consists of works
on the verification of statistical hypotheses about the average in space
with a generating element --- in particular, the article by
Zabolotnii--Martynenko--Salypa in \emph{Radioelectronics and
Communications Systems} (2018) {[}10a{]}.

The most recent point of this branch --- 2025 preprint on arXiv:
\emph{Applying PMM to estimate ARIMA models with asymmetric non-Gaussian
innovations} {[}32{]}. This is the first appearance of PMM/PMM on arXiv
as a full-length research article, symbolically opening the school to
the global statistical community. The preprint runs 128,000 Monte Carlo
simulations for ARIMA(p,d,q) with gamma, lognormal, and chi-square
innovations, showing PMM2 gains 30--50\% over classical estimators under
moderate skewness.

The software infrastructure for this branch is the R package
\textbf{EstemPMM}, released to CRAN in November 2025. {[}33{]}. The
package implements PMM2 (for asymmetric distributions) and PMM3 (for
symmetric platykurtic distributions), with the automatic selection
function pmm\_dispatch that evaluates \(\gamma_{3},\gamma_{4}\) by
sampling and chooses between OLS, PMM2, and PMM3 by the \(g_{S}\)
criterion. This makes PMM reproducible for the wider scientific
community without requiring the user to manually implement the normal
equations. The engineering infrastructure of the school also includes
the declarative patent of Ukraine for the utility model ``Method of
generating random variables'' (Zabolotnii, Chepinoga, Salipa, 2010/2011)
{[}33a{]}.

\subsubsection{6.2. Signal detector
branch}\label{signal-detector-branch}

Center --- V.V. Palagin, Cherkasy, with an international coalition (TU
Košice --- Y. Yugar, L. Vokorokos, S. Marchevskyi; ChDTU --- O.V.
Ivchenko, D.A. Vedernikov, D.O. Smirnov, O. Palagina, A. Goncharov, V.
Umanets). To Palagin's Cherkasy group, in addition to O.V. Ivchenko,
belong to: \textbf{S.A. Leleko} (Ph.D.~2018, moment quality
criterion of signal detection), \textbf{D.A. Vedernikov} (PhD 2021,
constant signal parameter estimation), \textbf{D.O. Smirnov} (PhD 2025,
continuous signal detection) and \textbf{O.S. Zorin} (PhD defense 2026,
metrological application). All these dissertations consistently deploy
the apparatus of stochastic polynomials to a wider class of signal
processing problems on a correlated non-Gaussian background --- from
passive detection to joint estimation of signal and interference
parameters.

Key publications: 2014--2017 joint works with cumulant detectors on
correlated non-Gaussian data {[}34{]}; an article in the \emph{Journal
of Electrical Engineering} (2016) on joint estimation of signal
parameters in non-Gaussian noise {[}26{]}; paper in \emph{IET Signal
Processing} (2017) on estimation of correlated non-Gaussian processes
--- the school's first appearance in a Q1 level journal {[}27{]}; work
in \emph{Circuits, Systems, and Signal Processing} (2017) on signal
detection in correlated non-Gaussian noise via higher-order statistics
{[}35{]}; \emph{Mathematical modeling of signal detection in
non-Gaussian correlated noise} (Springer LNNS 2023) {[}36{]};
\emph{Cumulant Detector of Non-Gaussian Signals against Background of
Non-Gaussian Interferences} (Radioelectronics and Communications
Systems, 2024) {[}37{]}.

\subsubsection{6.3. Pattern recognition and template matching
branch}\label{pattern-recognition-and-template-matching-branch}

The smallest branch in terms of volume, but the most original in terms
of conceptual progress. The central idea is the use of a distribution in
a generating element space to construct features that minimize the
reconstruction error for a particular class of signals and maximize it
for other classes. If, according to the class of features, the
reconstruction of a given object turns out to be more accurate than
according to others, the object is classified as belonging to this
class.

An independent group led by O. Chertov at NTU ``KPI'' developed this
branch in two parallel lines. \textbf{The Chertov--Tavrov line}
(statistical data analysis, group anonymity, microdata): the pioneering
article \emph{Use Kunchenko Polynomials for the Analysis of the
Statistical Data} in \emph{Eastern-European Journal of Enterprise
Technologies} (2010) {[}38{]}. \textbf{The Chertov--Slipets line}
(template matching, biomedical signals): preprint \emph{Kunchenko's Polynomials for Template Matching}
(arXiv 2011, the first appearance of the name Kunchenko on arXiv in
general) {[}19{]}; article \emph{Epileptic seizures diagnose using
Kunchenko polynomials template matching} in Springer ECMI 2012 {[}20{]}.
Application --- diagnosis of epileptic attacks on EEG, analysis of
statistical data, in the future --- biomedical signals in general.

Internally at the school, this branch is closely related to the doctoral
work of S.V. Zabolotnii 2015 {[}4{]} on probabilistic diagnosis of
perturbation of parameters of non-Gaussian sequences: distribution in
space with a generating element is used as pre-filtering to reduce
process variance before applying non-parametric tests of distribution
change. This is a well-designed combination of the decomposition apparatus in
space with a generating element and the classical theory of change-point
detection (CUSUM, GRSh, Kolmogorov-Smirnova), which opens up the
possibility of hybrid detectors --- parametric within the class of
normal operation, non-parametric for changes outside the class.

\subsection{7. Kunchenko's school among related
theories}\label{kunchenkos-school-among-related-theories}

Kunchenko's apparatus of stochastic polynomials did not arise in a
vacuum. The beginning of the 1970s, when Yurii Petrovych defended his
PhD thesis, was a period of simultaneous development of several
independent, but conceptually close directions in the processing of
non-Gaussian random processes. Bringing these directions into a single
map allows us to see more clearly the unique position of the Kunchenko
school and answer the question: why, with a 50-year-old apparatus, the
school still remains inconspicuous in the global statistical literature?

\subsubsection{7.1. The classical Volterra--Wiener theory: a precursor, not
an
alternative}\label{the-classical-wolter-wiener-theory-a-precursor-not-an-alternative}

The most obvious predecessor of the school is the classical theory of
functional series of Volterra--Wiener, formulated in the monographs of W.J.
Rugh \emph{Nonlinear System Theory: The Volterra/Wiener Approach} (1981)
{[}43{]}, M. Schetzen \emph{The Volterra and Wiener Theories of
Nonlinear Systems} (1980) {[}44{]}, V.J. Mathews and G.L. Sicuranza
\emph{Polynomial Signal Processing} (2000) {[}45{]}. This is the
dominant Western canon of non-linear signal processing, in which
Volterra series are used primarily to \textbf{identify non-linear
systems with memory} --- transmission channels, power amplifiers,
acoustic paths.

Candidate Kunchenko of 1972/1973 proceeded from this theoretical base,
but made a conceptual turn: Volterra's series was not used for system
identification, but for \textbf{finding estimates of unknown parameters
of random processes}. This is a fundamental change in the subject area
--- from a system to a statistical problem. As shown in § 3, this turn
opened the way to a further generalization to a stochastic polynomial
from the moment characteristics, which freed the apparatus from
being bound to the physics of the channel.

The Western school did not make this turn. Volterra kernels in the works
of Schetzen, Mathews, Sicuranza remained functions of time delays
\(h_{n}\left( \tau_{1},\ldots,\tau_{n} \right)\), and statistical
problems (parameter estimation, hypothesis testing) were solved by
separate devices --- the method of least squares, the method of maximum
likelihood, M-estimation. As a result, Kunchenko's school and the
Western Volterraan tradition developed in parallel, without formal
bridges. Until 2026 --- the year of the appearance of the article by
Gariachyo/Shcherbinin {[}6{]}, which we consider in § 8, --- there is
not a single publication in the global literature where Volterra's
series and Kunchenko's stochastic polynomials would be considered as two
poles of the same conceptual space.

\subsubsection{7.2. Higher orders of statistics: a parallel
path}\label{higher-orders-of-statistics-a-parallel-path}

The closest Western parallel to Kunchenko's school is the tradition of
\textbf{higher-order statistics} (HOS, statistics of higher orders),
designed in the works of J.M. Mendel, C.L. Nikias, A.N. Petropoulou,
G.B. Giannakis, A.K. Swami in the 1980s-1990s {[}46--48{]}. This school
makes extensive use of cumulants and third- and higher-order moments for
non-Gaussian signal analysis, with applications to blind deconvolution,
system identification, signal detection in non-Gaussian noise.

The conceptual overlap with Kunchenko's school is deep. Both directions
are based on the same observation: for non-Gaussian processes, the
limitation with a linear description and second moments loses a
fundamental part of information; higher moments and cumulants carry
statistical information that cannot be replaced by other means. Both
directions use these points to build more efficient estimators and
detectors.

The difference is based on the method of utilization of moment
information. The HOS school operates mainly in the \textbf{spectral
domain} --- bispectra, trispectra, polyspectra as direct generalizations
of the energy spectrum to higher orders. The Kunchenko school operates
in the \textbf{space of polynomial functions from moments}, with the
selection of a basis (\(\xi^{i}\), \(\cos(ik\xi)\), etc.), optimization
of coefficients according to moment criteria and explicit work with the
body of a stochastic polynomial as a quadratic form. This makes the
Kunchenko apparatus closer to the classical method of moments, but with
a significant nonlinear expansion due to the polynomial structure.

There are few direct formal bridges between these two schools in the
literature. Separate appearances of the Kunchenko school in journals of
the traditional HOS-profile (\emph{Circuits, Systems, and Signal
Processing} 2017 {[}35{]}, \emph{IET Signal Processing} 2017 {[}27{]})
are sporadic rather than a systematic integration. This remains an open
gap: a systematic comparison of stochastic Kunchenko polynomial with
spectral HOS estimators in modern signal processing --- a potential
topic of a separate study.

\subsubsection{7.3. Robust statistics and Huber's
M-estimators}\label{robust-statistics-and-hubers-m-estimators}

Another parallel is the tradition of robust estimation started by P.J.
Huber in the work \emph{Robust Estimation of a Location Parameter}
(1964) and developed in the monograph \emph{Robust Statistics} (1981)
{[}49{]}. Robust methods are based on the criterion of minimizing the
sum of the ψ-functions from the residuals, with the selection of the
ψ-function so as to limit the impact of outliers and deviations from
normality.

Affinity with the Kunchenko school --- at the level of methodological
approach: both approaches are resistant to deviations from Gaussian
idealization. The difference is in the form of parameterization of this
stability. Huber's M-estimators use ψ-functions that are specified a
priori and do not depend on the specific form of the non-Gaussian data
distribution. PMM Kunchenko, on the other hand, uses
\textbf{distribution information}, but in a limited form --- through
moments and cumulants of a certain order. This makes PMM potentially
more efficient (provided the moments are estimated accurately) and the
robust methods --- potentially more robust (provided the moment
structure can be distorted by coarse outliers).

There are few systematic comparisons of the effectiveness of these two
approaches on the same data in the literature. In modern works of the
{[}31, 32{]} school, it is stated that PMM-2 does not lose to OLS with
Gaussian innovations and gives a win with asymmetric distributions,
which puts it in the same weight category as robust estimators. However,
there are no direct benchmarks of PMM vs Huber M-estimator vs Hosking's
L-moments on modern realistic datasets yet. This is another blank spot
of the school (§ 11).

\subsubsection{7.4. Method of generalized moments (GMM Hansen) and
L-moments}\label{method-of-generalized-moments-gmm-hansen-and-l-moments}

L. Hansen in the classic work of 1982 \emph{Large Sample Properties of
Generalized Method of Moments Estimators} {[}50{]} formulated the method
of generalized moments (Generalized Method of Moments, GMM) --- an
apparatus that has become a standard of modern econometrics. In GMM, the
estimate is found as a solution of a system of moment conditions with a
weight matrix optimized to minimize the asymptotic variance. This is
actually a generalization of the classical method of moments for the
case when the number of moment conditions exceeds the number of
evaluated parameters.

The ideological affinity of GMM and PMM is obvious: both exploit higher
moments, both give estimators that asymptotically tend to the efficiency
limit, both do not require full knowledge of the distribution function.
Difference in technique: GMM uses linear (or quasi-linear) functionals
from moment terms, while PMM constructs \textbf{polynomial functions
from moments} with dispersion-optimized coefficients. In a sense, PMM
can be interpreted as a specific form of GMM in which the Hansen weight
matrix is \hspace{0pt}\hspace{0pt}replaced by a matrix
\(F_{S}\left( \overrightarrow{\theta} \right)\) of centered stochastic
polynomial correlations. However, this is an interpretive move that has
not yet been formally bridged in the literature.

Close to this apparatus is the \textbf{L-moments of J. Hosking} (1990)
{[}51{]} --- linear combinations of order statistics that give an
alternative parameterization of non-Gaussian distributions. L-moments
are more robust to outliers than classical moments and are mainly used
in hydrology, meteorology, and actuarial statistics. Kunchenko's school
of L-moments has no formal bridges; the possibility of systematic
integration of these devices is an open prospect.

Close to GMM, but conceptually even closer to the Kunchenko apparatus,
is the Western tradition \textbf{SLS (Second-order Least Squares)},
started by L. Wang in 2003 and systematized in the next two decades as a
standard for nonlinear regression problems with non-Gaussian errors. The
relationship PMM2 ↔ SLS is not only conceptual, but also quantitative in
nature --- both methods give practically identical MSEs on the joint
problems {[}25{]}. This makes SLS the most direct point of the formal
bridge between the Western statistical literature and the apparatus of
the Kunchenko school, therefore, we move the consideration of this
parallel to a separate subsection § 7.5.

\subsubsection{7.5. Second-order Least Squares (SLS): parallel
trajectory of the
West}\label{second-order-least-squares-sls-parallel-trajectory-of-the-west}

The most conceptually close Western analogue of the Kunchenko apparatus,
which is not included in the four previous subsections of this
paragraph, is the method of \textbf{Second-order Least Squares} (SLS),
initiated by L. Wang in 2003 on the material of nonlinear regression
models with measurement errors of the Berksonian type {[}52{]} and
systematized by Wang and Leblanc in 2008 p.~as a universal procedure for
nonlinear estimation under asymmetric errors {[}53{]}. In the next
decade, the method received theoretical development --- analysis of
asymptotic efficiency {[}54{]}, new optimization criteria {[}55{]},
extension to the ARCH model and time series {[}56{]}, adaptation to
models with multiple curves {[}57{]}. Today, SLS is a standard tool of
modern econometrics and biostatistics for nonlinear problems with
non-Gaussian errors.

The conceptual idea of \hspace{0pt}\hspace{0pt}SLS is put into one
sentence: to the optimization criterion of the standard OLS, which
relies only on the first conditional moment
\(E\left\lbrack y|x \right\rbrack = R(\theta,x)\), the second one ---
\(E\left\lbrack y^{2}|x \right\rbrack = R^{2}(\theta,x) + \sigma^{2}\)
--- is added, and the parameters of \(\theta\) are evaluated by
minimizing the combined functional from the deviations of both moments.
In formal form:

\[
{\widehat{\theta}}_{SLS} = \arg\min_{\theta}\sum_{v = 1}^{N}\left\lbrack \left( y_{v} - R\left( \theta,x_{v} \right) \right)^{2} + \omega \cdot \left( y_{v}^{2} - R^{2}\left( \theta,x_{v} \right) - \sigma^{2} \right)^{2} \right\rbrack,
\]

with weighting factor \(\omega\) optimizing for asymptotic efficiency.

Conceptual affinity with PMM degree \(S = 2\) (PMM2) of the Kunchenko
school is deep and structurally symmetrical. Both methods proceed from
the same observation: for non-Gaussian asymmetric errors, the classical
OLS loses a fundamental part of information because it ignores the
second and third moments of deviations. Both methods compensate for this
by adding moment information --- SLS through an additional quadratic
term in the loss functional, PMM2 through the construction of a
stochastic polynomial of the 2nd degree with coefficients optimized
according to the moment-cumulant description. Both reduce to OLS in
the Gaussian case (where \(\gamma_{3} = 0\)) and give a significant gain
in the variance of estimates for nonzero asymmetry.

A specific comparison of the performance of SLS and PMM2 on joint tasks
was performed in a recent joint work of Zabolotnii's group with Polish
colleagues {[}25{]} on two nonlinear regression models with
χ²-distributed errors: the exponential
\(y = \theta_{1}\exp\left( \theta_{2}x \right) + \xi\) and the growth
model
\(y = \theta_{1}/\left\lbrack 1 + exp\left( \theta_{2} + \theta_{3}x \right) \right\rbrack + \xi\).
Monte Carlo simulations with sample sizes \(N = 30,50,100,200\) and
replicates \(m = 1000\) gave a result that can be summarized in one
sentence: \textbf{PMM2 and SLS give almost identical MSE metrics}
(difference within 1--3\% depending on parameter and sample size), and
both significantly outperform OLS --- by 30\% gain with \(N = 30\) to
50\% of the gain with \(N = 200\), which coincides with the theoretical
prediction of the coefficient of variation reduction
\(g_{2} \approx 0.56\) for χ²(3)-errors.

The authors of {[}25{]} themselves interpret this coincidence as
\textbf{a consequence of using the same information}: ``both estimators
use the same additional information about the probabilistic nature of
the random component of the regression model in the form of the moments
of the regression residuals up to the 4th order.'' This is a correct
technical explanation, but it also captures a \textbf{deeper fact}: SLS
and PMM2 are \textbf{two different mathematical paths to the same
practical result} that have been developed in parallel and independently
for twenty years. SLS --- in the Western econometric tradition, since
2003; PMM2 --- at the Cherkasy school of Kunchenko, with the canonical
monograph of 2002 and preliminary results from the 1990s. Both
traditions solve the same problem and get the same result without
knowing about each other.

This makes the pair PMM2 ↔ SLS an exceptionally attractive point for
building a \textbf{formal bridge} between Kunchenko's school and modern
world statistics. There is a conceptual parallel; an empirical
comparison in {[}25{]} has already been made; systematic theoretical
summary --- the next natural step (see § 11.2). A potentially formally
provable advantage of PMM over SLS is \textbf{scalability}: while SLS is
standardly restricted to second order (possible third-order
generalizations exist in the literature but have not been systematically
developed), PMM naturally extends to \(S = 3,4,\ldots\) through the
perforation of the cumulant description and the selection basic
functions. This gives Kunchenko's school a theoretical reserve of
effectiveness that SLS does not possess.

\subsubsection{7.6. Semiparametric position of the
school}\label{semiparametric-position-of-the-school}

Summarizing all the listed areas into a single map reveals the unique
position of the Kunchenko school among methods of statistical processing
{[}4, Fig. 1.4{]}. The classic scheme divides these methods into three
categories:

\begin{itemize}
\item
  \textbf{Parametric methods} (likelihood methods, Bayesian procedures)
  --- require full a priori information about the distribution law;
  provide optimal estimates subject to the assumptions being fulfilled;
  are critically sensitive to model misspecification.
\item
  \textbf{Non-parametric methods} (Kolmogorov-Smirnov, Mann-Whitney
  tests; rank methods; CUSUM, GRSh in the non-parametric version) --- do
  not require a priori information about the distribution; resistant to
  deviations from the model; efficiency is often far from optimal
  values.
\item
  \textbf{Semiparametric methods} (Kunchenko stochastic polynomials +
  moment-cumulant description) --- use a \textbf{limited} amount of a
  priori information in the form of a sequence of moments or cumulants;
  provide a compromise between the amount of necessary information,
  computational complexity and accuracy; support self-learning and
  adaptability.
\end{itemize}

It is the semiparametric position that makes Kunchenko's apparatus
conceptually valuable for modern problems where complete a priori
information is not available, but it would also be wasteful to
completely ignore information about higher moments. This positioning of
the school, clearly recorded in {[}4{]}, is fully consistent with the
trajectory of the robust-semiparametric revolution in modern statistics
(quasi-likelihood, GMM, M-estimators, SLS) --- however, the school
itself does not yet systematically use this connection for its
international positioning.

\subsection{8. Case 2026: reopening the original
problem}\label{case-2026-reopening-the-original-problem}

In 2026, the Ukrainian scientific and technical journal ``Measuring and
Computing Techniques in Technological Processes'' (VOTTP, No.~1,
pp.~47--60) published an article by M. Garyachy and S. Shcherbinin (Air
Force Research Center of Ivan Kozhedub Kharkiv National Air Force
University) entitled ``The method of generating stochastic signals using
Volterra series'' {[}6{]}. The article concentratedly demonstrates the
state in which the applied Ukrainian radio engineering school found
itself in relation to the conceptual toolkit of the Cherkasy Kunchenko
Scientific School 50 years after its founding candidate's thesis.
Consideration of this paper is important not as a criticism of the
authors --- its authors represent a serious defense radio engineering
center and are working on a very real problem --- but as a
\textbf{symptomatic example} of a wider trend, the implications of which
we discuss in § 11.

\subsubsection{8.1. Content of the article of
2026}\label{content-of-the-article-of-2026}

Garyachy and Shcherbinin solve the problem of formation and processing
of stochastic signals for information and control systems in the
conditions of nonlinear distortions, effects of channel memory and
limited frequency resource. Classical linear filtering methods, as the
authors note, cannot compensate for nonlinear distortions and memory
effects, so it is suggested to switch to the 2nd-order Volterra series
apparatus with adaptive kernel parameter setting.

Volterra's discrete model used in the article looks like this:

\[
y\lbrack n\rbrack = h_{0} + \sum_{i = 0}^{M_{1} - 1}h_{1}\lbrack i\rbrack\, x\lbrack n - i\rbrack + \sum_{i = 0}^{M_{2} - 1}\sum_{j = 0}^{M_{2} - 1}h_{2}\lbrack i,j\rbrack\, x\lbrack n - i\rbrack\, x\lbrack n - j\rbrack,
\]

where \(h_{1}\lbrack i\rbrack\), \(h_{2}\lbrack i,j\rbrack\) --- core
coefficients, \(M_{1},M_{2}\) --- memory depths of the corresponding
cores. Adaptation of the coefficients is performed according to the
criterion MMSE --- the minimum mean square error:

\[
MMSE = \min_{\{ h_{1},h_{2}\}}E\left\lbrack \left( y\lbrack n\rbrack - \widehat{y}\lbrack n\rbrack \right)^{2} \right\rbrack.
\]

At each adaptation step, the local error is calculated, and the kernel
weights are updated through the system of normal equations
\(C_{x} \cdot \overrightarrow{h} = {\overrightarrow{r}}_{yx}\), where
\(C_{x}\) --- input covariance matrix, \({\overrightarrow{r}}_{yx}\) ---
cross-correlation vector.

The numerical simulation performed by the authors gives an interesting
result. For structured signals (OFDM, 64 subcarriers, QAM-16, sampling
frequency 960 kHz), which have a pronounced harmonic structure, the
2nd-order Volterra method provides a \textbf{decrease in the effective
spectrum width by 1.9--8.6\%} and a corresponding increase in spectral
efficiency by 1.9--5\%. However, for wideband stochastic (noisy)
signals, the same method gives an \textbf{increase in effective spectral
width of about 15\%}, with a corresponding drop in spectral efficiency
of \textasciitilde13\% {[}6, Table 1{]}. The authors candidly note
this limitation and draw the conclusion: Volterra's method is not
universal, and for its stable operation on signals with different
statistics, ``decorrelation and reduction of spectral redundancy'' is
required.

\subsubsection{8.2. Conceptual
reconstruction}\label{conceptual-reconstruction}

Without going into technical details, the conceptual content of the
{[}6{]} 2026 article can be reconstructed into four points.

\textbf{First.} The formulation of the problem --- Volterra processing
of a stochastic signal with adaptation of kernel parameters according to
the statistical characteristics of the input --- formally echoes the
type of Kunchenko's candidate problem in 1972/1973, where Volterra
series were used to find estimates of unknown parameters of random
processes. After 50 years, Ukrainian radio engineering researchers
independently return to a close conceptual point: a nonlinear functional
from a random process is used as an evaluation or adaptation tool.

\textbf{Second.} The adaptation method ---
\(C_{x} \cdot \overrightarrow{h} = {\overrightarrow{r}}_{yx}\) normal
equations --- has a related linear-algebraic form to the equations for
the coefficients of the polynomial estimators, but optimizes a different
object. In the 2026 paper, this is MMSE-regression for the Volterra
filter kernel: the parameters are \(h_{1},h_{2}\) and the criterion is
the model output error. In the mature Kunchenko apparatus, the parameter
is a characteristic of the process or distribution, and the matrix of
centered correlations is included in the moment estimation procedure. It
is this difference that prevents VOTTP-approach from being identified
with PMM, but makes it a natural point for such a generalization.

\textbf{Third.} The limitation identified by the authors --- 13\%
spectral degradation on noisy signals --- can be interpreted as a
symptom of insufficient matching of the kernel with the moment
structure of the signal. The method optimizes kernels according to the
L2-criterion, which works through covariance projection and does not
directly use \(\gamma_{3}\) asymmetry, \(\gamma_{4}\) kurtosis, and
other higher-order characteristics. Kunchenko's apparatus suggests
another possible trajectory: to build adaptation not only by
covariances, but also by a matrix of centered correlations of basic
functions. Under appropriate moment conditions and the non-degeneracy of
this matrix, such a procedure can reduce the variance of estimates in
those classes of non-Gaussian distributions where \(g_{2}<1\).

\textbf{Fourth.} The bibliography of the article contains 20 references,
mostly classical: Rugh 1981 {[}43{]}, Stenger and Rabenstein, Favier,
Orcioni, Gibiino, Zhu, Chang, Gariachy and Shcherbinin (their previous
work 2025), Kapustii, Karhunen-Loève transform, Wheeler and Holder,
Sosulin, Kharkevich, Pugachov. \textbf{No reference to the Kunchenko
school.} No mention of the Shaker 2002 monograph, no mention of the IEEE
ISIT report 1997. Ukrainian authors in the Ukrainian radio engineering
journal cite classic literature, but not Kunchenko 2002 from Aachen ---
even though Kunchenko was present in the IEEE literature already since
1997. However, they miss not only the Kunchenko school, but also
\textbf{its Western counterpart} --- the SLS (Second-order Least
Squares) tradition, which since 2003 has been systematically developing
in Western econometrics and biostatistics precisely for problems
conceptually close to the one solved in {[}6{]} (see § 7.5). This
emphasizes the systemic nature of the infrastructural gap: between the
apparatus that is really needed to solve the given task and the current
toolkit of Ukrainian applied radio engineering researchers, there is a
gap that is not closed not only by the local (Cherkasy) school, but also
by its global Western counterpart.

\subsubsection{8.3. Why is this a symptom and not a
reproach}\label{why-is-this-a-symptom-and-not-a-reproach}

I emphasize: the situation is not the fault of the authors. Garyachy and
Shcherbinin represent the Kharkiv Radiophysics Center --- the same one
in whose specialized academic council Kunchenko defended his doctoral
dissertation in 1988. Between these two points --- Kharkiv 1988 and
Kharkiv 2026 --- there are decades during which the canon of the school
was insufficiently incorporated into mainstream Ukrainian radio
engineering education. This is a structural, not a personal problem.

The symptom is as follows. Kunchenko's school, despite a 50-year
history, six monographs, a multi-generational dissertation structure (§
5), access to international journals (§ 6) and presence in
IEEE-literature since 1997 {[}12a{]}, remains \textbf{locally invisible
in its own applied environment}. Ukrainian radio engineering
researchers, working on tasks for which the school apparatus was
literally created, do not know about its existence. This cannot be
chalked up to ``not enough publicity'' --- the school gave Shaker Verlag
2002, IET Signal Processing 2017, arXiv 2025, public R package EstemPMM
at CRAN. The problem is deeper: between the apparatus of the school and
its applied consumers in Ukraine there is an \textbf{infrastructural
gap}, the consequences of which we discuss in § 11.

The article by Gariachyo/Shcherbinin of 2026 --- the most direct
diagnostic point of this gap. And that is why it becomes a natural
motivation for our memorial article: not to point out the oversight of
the authors, but in order to \textbf{through the demonstration of a
formal bridge} between their task and the school apparatus to close this
gap at least to one point.

We build this bridge in § 9.

\subsection{9. Mathematical bridge: Volterra's model as a stochastic
polynomial and the difference of evaluation
criteria}\label{mathematical-bridge-volterras-model-as-a-stochastic-polynomial-and-the-difference-of-evaluation-criteria}

This section is the methodological core of the article. Its purpose is
not only to draw a historical parallel between Volterra's series and
Kunchenko's stochastic polynomials, but to formally show that the class
of finite Volterra models naturally nests into the class of generalized
stochastic polynomials on a vector argument. This nesting is a statement
about the basis and form of the model, not a statement about the
identity of the methods. The two criteria can then be clearly separated:
MMSE/L2 as a fixed covariance projection for kernel adaptation and PMM
as a parametrically dependent moment procedure for estimating a process
or distribution parameter.

\subsubsection{9.1. Assertion about the embedding of the Volterra
model}\label{assertion-about-the-embedding-of-the-volterra-model}

\textbf{Theorem 1 (The Volterra model as a generalized stochastic
polynomial).} Let there be a discrete process \(x[n]\) for which there
exist moments of order sufficient to construct all products to the power
\(N\). Also let the finite memory \(M\) and the vector of observations
be given

\[
\overrightarrow{x}_{n} = (x[n],x[n-1],\ldots,x[n-M+1])^{T}.
\]

Then any finite Volterra model of degree \(N\) with memory \(M\) can be
written as a generalized stochastic Kunchenko polynomial on the vector
argument \(\overrightarrow{x}_{n}\) with a basis of monomial products

\[
\{\varphi_{k}\} = \left\{x[n-i_{1}]x[n-i_{2}]\cdots x[n-i_{p}]:\; p=1,\ldots,N,\;0\leq i_{1}\leq\cdots\leq i_{p}\leq M-1\right\}.
\]

The number of basis functions is equal to

\[
S = \sum_{p=1}^{N}\binom{M+p-1}{p}.
\]

For the case \(N=2\) we have \(S=M+M(M+1)/2\) if the symmetry of the
quadratic kernel is taken into account.

The proof is a direct rewriting of the Volterra expansion as a linear
combination of basis functions from the vector
\(\overrightarrow{x}_{n}\). For the second order, the discrete model has
the form

\[
y[n] = h_{0}+\sum_{i=0}^{M_{1}-1}h_{1}[i]x[n-i]+\sum_{i=0}^{M_{2}-1}\sum_{j=0}^{M_{2}-1}h_{2}[i,j]x[n-i]x[n-j].
\]

After entering \(M=\max(M_{1},M_{2})\), it is rewritten as

\[
y[n] = h_{0}+\sum_{i=0}^{M-1}h_{1}[i]\varphi_{1}^{(i)}(\overrightarrow{x}_{n})+\sum_{i,j=0}^{M-1}h_{2}[i,j]\varphi_{2}^{(i,j)}(\overrightarrow{x}_{n}),
\]

where \(\varphi_{1}^{(i)}(\overrightarrow{x}_{n})=x[n-i]\) and
\(\varphi_{2}^{(i,j)}(\overrightarrow{x}_{n})=x[n-i]x[n-j]\). So, at the
level of Volterra's basic representation, the model is a special case of
a generalized stochastic polynomial with a monomial basis and a vector
argument.

\subsubsection{9.2. Difference between MMSE/L2 and
PMM-criterion}\label{difference-between-mmsel2-and-pmm-criterion}

\textbf{Proposition 2 (criterion difference).} The MMSE-adaptation of
the Volterra kernel and the PMM-estimator use a similar linear algebraic
form of the normal equations, but optimize different statistical
objects.

In the MMSE formulation, the kernel coefficients are from the problem

\[
\min_{h}E\{(y[n]-\widehat{y}[n])^{2}\},
\]

leading to the system

\[
\mathbf{C}_{x}\overrightarrow{h}=\overrightarrow{r}_{yx},
\]

where \(\mathbf{C}_{x}\) is the covariance matrix of the observed basis
functions, and \(\overrightarrow{r}_{yx}\) is the cross-correlation
vector. This is an L2 projection on a fixed basis.

In the PMM-formulation, the optimal coefficients are determined through
a parametrically dependent matrix of centered correlations

\[
\mathbf{F}_{S}(\theta)\overrightarrow{h}^{*}=\overrightarrow{b}(\theta),
\]

where the elements \(\mathbf{F}_{S}(\theta)\) are constructed from the
moments of the basis functions, and the vector
\(\overrightarrow{b}(\theta)\) is related to the derivatives of the
moment characteristics of the parameter being evaluated. Therefore, PMM
is not just an alternative way to solve the same least-squares problem:
it changes the criterion itself, transferring the problem from a fixed
L2-projection to a parametrically adapted moment estimation.

To estimate the mean \(\theta=\mu\) in classes of distributions for
which the standard moment conditions and PMM regularity conditions are
satisfied, the asymptotic variance of the second power estimate can be
written in terms of the variance reduction factor

\[
\sigma_{\widehat{\mu},\mathrm{PMM2}}^{2}=\frac{c_{2}}{N}g_{2},\qquad g_{2}=1-\frac{\gamma_{3}^{2}}{2+\gamma_{4}}.
\]

This formula should be read conditionally: the gain statement PMM is
correct for those non-Gaussian classes where the correlation matrix is
\hspace{0pt}\hspace{0pt}non-degenerate, the required moments exist, and
\(g_{2}<1\). In this form, the result is a statement about a class of
estimators and their asymptotic efficiency, not a universal declaration
for any unconditional non-Gaussian distribution.

\subsubsection{9.3. A corollary for the Volterra-kernel adaptation
problem}\label{a-corollary-for-the-volterra-kernel-adaptation-problem}

The task considered in the article {[}6{]} has exactly such a
methodological meaning. For structured signals of the type OFDM MMSE,
the Volterra kernel adaptation works satisfactorily because the second
order compensates for part of the cross-harmonic interactions. For noisy
broadband signals, the same L2-criterion can worsen the spectral
efficiency, because it does not use the moment-cumulant structure of
the signal.

The mathematical possibility opened by the PMM apparatus is to adapt the
Volterra kernel not only by the covariance matrix of the input, but by
the parametric matrix of centered correlations of the basis functions
estimated for the corresponding class of non-Gaussian process. The
hypothesis to be tested numerically is this: for signals with
significantly non-zero higher-order moments, the PMM Volter-kernel
adaptation can reduce the variance of the estimates compared to the
MMSE/L2 adaptation, and thus potentially eliminate the spectral
degradation observed in {[}6{]}.

This should not be presented as a complete empirical result: the current
article builds a formal bridge and a methodological hypothesis. Its
validation requires Monte Carlo comparisons on the same types of signals
used in {[}6{]}: OFDM, filtered white noise, hybrid harmonic-noise
signal, different levels of nonlinearity and channel memory.

\subsubsection{9.4. Methodological
conclusion}\label{methodological-conclusion}

A cautious but productive conclusion follows from this: the apparatus of
the Kunchenko school can be transferred to the Volterra model only after
the explicit construction of a suitable basis, a matrix of centered
correlations, moment-cumulant characteristics, non-degeneracy checks,
comparison of variances and validation on benchmark scenarios. It is
this formulation that shifts the center of gravity of the article from a
historical reminder to the formalization of the class of evaluators,
criteria and verified conditions of their effectiveness.

\subsection{10. Research program: from historical bridge to testable
models}\label{research-program-from-historical-bridge-to-testable-models}

The formal bridge built in § 9 translates the historical overview into a
work program. Its logic is simple: first, it is necessary to check
whether the replacement of the criterion gives a real gain on the same
signals, where the classic Volterra adaptation showed limitations;
further --- extend the basis to describe the transition between
structured and noise-like signals; finally --- to combine the evaluation
and detection branches of the school in the task of detecting a regime
change. Below, these steps are presented as a sequence of tested models,
not as a declaration of finished results.

\subsubsection{10.1. PMM-adaptation of the Volterra kernel as a
moment-based generalization of the
MMSE-criterion}\label{pmm-adaptation-of-the-volterra-kernel-as-a-moment-generalization-of-the-mmse-criterion}

The immediate task --- direct experimental verification of the thesis §
9.3. The working hypothesis is carefully formulated: replacing the
L2-criterion in the adaptation of the 2nd-order Volterra kernel by the
PMM-criterion with coefficients optimized for moments of higher orders
can eliminate spectral degradation on noisy signals in those classes of
processes where the PMM moment conditions are fulfilled and variational
reduction is observed.

The experimental design reproduces the conditions of the article
{[}6{]}: OFDM (64 subcarriers, QAM-16, 960 kHz), filtered white noise
(LPF with normalized cutoff frequency 0.4), hybrid signal (harmonic +
noise), with different levels of channel nonlinearity and memory
effects. Two variants of adaptation are compared: basic MMSE-normal
equations and moment generalization, in which the system of
PMM-equations
\(F_{S}\left( \overrightarrow{\theta} \right) \cdot {\overrightarrow{h}}^{*} = \overrightarrow{b}\left( \overrightarrow{\theta} \right)\)
is constructed with \(\gamma_{3},\gamma_{4}\) estimation directly from
the sample. Comparison metrics: effective spectrum width, spectral
efficiency, dispersion of kernel estimates.

Implementation --- based on the R-package EstemPMM {[}33{]} with an
extension to the vector-valued argument and the Volterra-structure of
the basis. Approximate time of preparation of experiment and article ---
2--3 months after completion of this review work. Target journal --- IET
Signal Processing or Circuits, Systems, and Signal Processing where the
school already has an entry point through {[}27, 35{]}. Alternative ---
direct publication in VOTTP as a direct dialogue with the article by
Gariachy and Shcherbinin.

\subsubsection{10.2. PATP-kernel
parameterization}\label{patp-kernel-parameterization}

The second problem arises from the limitation of the article {[}6{]},
formulated by the authors themselves: Volterra's method works
differently for structured signals (OFDM) and noise-like ones.
Therefore, the PATP-parameterization of the kernel is considered here as
a continuous control of the transition between structuredness and
noise-likeness. The current Kunchenko school proposes a discrete
distinction through the perforation of the cumulant description ---
asymmetric, excess, asymmetric-excess classes {[}4{]}. This is a
discrete choice that does not reflect the continuous nature of real
signals, which may have intermediate degrees of structure.

An alternative approach --- introduction of \textbf{parametrically
adaptive transition polynomial} (PATP, Parametrically-Adaptive
Transition Polynomial). This is a hypothetical generalization of the
basic structure of the stochastic polynomial, in which instead of a
discrete choice from several classes (``power'', ``trigonometric'',
``fractal''), a continuous parameter \(\alpha \in \lbrack 0,1\rbrack\)
is used. This parameter determines the type of basis functions smoothly:
in the extreme case \(\alpha = 0\), the basis consists of fractal
functions of the form \(|\xi|^{1/i}\), which better describe noisy
signals with heavy tails; in the case of \(\alpha = 1\) --- from
standard power \(\xi^{i}\), optimal for structured signals of type OFDM;
intermediate values \hspace{0pt}\hspace{0pt}of \(\alpha\) give transient
bases that adapt to partially structured signals. The parameter
\(\alpha\) is optimized automatically --- by cross-validation on the
calibration sample.

Conceptually, this is an extension of the school's apparatus to
continuous parameterization of the basis. It does not contradict
Kunchenko's classical formulation, but complements it: discrete classes
of perforation become partial cases of continuous PATP-space.
Technically --- requires reformulation of the centered correlation
matrix \(F_{S}\) as a function of two arguments: the estimated parameter
\(\overrightarrow{\theta}\) and the basis parameter \(\alpha\).
Implementation and numerical verification --- a separate article for
4--6 months. Target journals --- Signal Processing (Elsevier) or IEEE
Transactions on Signal Processing, where continuous basis
parameterization as an adaptive processing tool will find a natural
audience.

\subsubsection{10.3. GSA-CUSUM detector instead of heuristic instability
indicator}\label{gsa-cusum-detector-instead-of-heuristic-instability-indicator}

The third task --- replacing the heuristic indicator of instability
\(K(t)\) from the article {[}6{]} with a formal polynomial CUSUM
detector of a change in distribution. For this statement, there is
already a close internal school precedent --- semiparametric estimation
of the moment of disturbance of the parameters of non-Gaussian sequences
by the method of polynomial maximization {[}28, 28a{]}. Here we use the
conventional working abbreviation \textbf{GSA} (Generalized Stochastic
Approximation) for that branch of Kunchenko's apparatus, which is based
on the expansion of the logarithm of the likelihood ratio into a
stochastic series with optimization of coefficients according to
moment criteria of the quality of the formation of decisive rules
{[}12, 12a, 13{]}. This abbreviation is not systematically used in the
school --- in the internal terminology, this branch is described in
detail (as ``decomposition of the logarithm of the likelihood ratio in
the form of stochastic series\ldots{}'' {[}4{]}) --- however, in the
international context, the shortened shortcut is convenient, and we
accept it as the technical term of this article.

The formalism of the polynomial CUSUM-detector (CUSUM, Cumulative Sum
--- cumulative sum, classical statistics for detecting changes in the
moment of disturbance):

\[
T_{n} = \sum_{k = 1}^{n}\Lambda_{S}^{(\mathrm{poly})}\left( x_{k} \right),\quad\tau_{\mathrm{detect}} = \min\{ n:T_{n} > h\},
\]

where \(\Lambda_{S}^{(\mathrm{poly})}\) --- a polynomial approximation
of the logarithm of the likelihood ratio, constructed by a stochastic
series with coefficients optimal according to moment criteria;
threshold \(h\) is chosen from the Chebyshev or Vysochansky-Petunin
inequality to ensure a controlled FAR (False Alarm Rate, false alarm
probability). This gives a \textbf{statistically sound} distribution
change detector instead of the heuristic one \(K(t)\), with a guarantee
of \(\mathrm{FAR}\leq\varepsilon\) and a computational complexity of the
order of \(O\left( N \cdot S^{2} \right)\).

Technically --- this is an integration of two branches of the school:
the decomposition of the logarithm of the likelihood ratio into the
Kunchenko stochastic series (primary source ISIT 1997 {[}12a{]}) and the
moment-cumulant parameter estimator (doctoral apparatus {[}4{]}).
Experimental verification on real radio engineering signals with a
change of mode. Estimated term --- 6--8 months. Target journals ---
Sequential Analysis or Journal of Statistical Planning and Inference;
alternative --- IEEE Transactions on Information Theory as a journal
with a tradition of publications on the point change.

\subsubsection{10.4. Consistency and
complementarity}\label{consistency-and-complementarity}

Three problems form a natural sequence: § 10.1 --- replacing the
criterion without changing the structure of the Volterra model; § 10.2
--- expansion of the base from discrete to continuous; § 10.3 ---
integration of assessment and detection branches of the school. In this
form, the research program does not complete the memorial article, but
opens the next cycle of works from it: from historically motivated
reconstruction to numerically verified models and open benchmark
comparison.

\subsection{11. Open methodological and infrastructural
problems}\label{open-methodological-and-infrastructural-problems}

Along with the active research program § 10, the state of the school in
2026 calls for several open problems. They do not devalue what has
already been done; on the contrary, it is the maturity of the apparatus
that allows us to formulate what it lacks for wider scientific
visibility and reproducible comparison with related approaches.

\subsubsection{11.1. Infrastructural gap with applied Ukrainian radio
technology}\label{infrastructural-gap-with-applied-ukrainian-radio-technology}

The most visible and most painful gap --- that which the article {[}6{]}
made transparent. There is a gap between the apparatus of the school and
its natural applied consumers in Ukraine, which cannot be closed either
by monographs or international publications. The outputs of the school
in IEEE ISIT 1997 {[}12a{]}, IET Signal Processing 2017 {[}27{]} and
Circuits, Systems, and Signal Processing 2017 {[}35{]} did not reach the
Kharkiv radio engineering audience; the English-language canon Shaker
2002 {[}9{]} did not reach either teaching corpora or departmental
libraries. Ukrainian Defense Radio Technical Centers in 2026 cite
classic literature, but not Kunchenko 2002.

Closing this gap --- the task is not purely scientific, but
infrastructural. It requires Ukrainian review publications (including
this one), Ukrainian textbooks based on school equipment, integration of
PMM into statistical radio engineering courses in Ukrainian
universities, direct dialogue publications in journals where applied
researchers work (VOTTP, Radioelectronic and Computer Systems). Without
this step, the international visibility of the school will continue to
grow, and the local visibility will stagnate.

\subsubsection{11.2. Formal bridges to GMM, M-estimators, L-moments and
SLS}\label{formal-bridges-to-gmm-m-estimators-l-moments-and-sls}

As indicated in § 7, the conceptual affinity between Kunchenko's
apparatus and parallel Western traditions (GMM Hansen, Huber's
M-estimators, Hosking's L-moments, SLS Wang) is obvious, but formal
bridges between these schools in the literature are mostly absent. PMM
can be interpreted as a specific form of GMM with the replacement of the
Hansen weight matrix by \(F_{S}\left( \overrightarrow{\theta} \right)\),
but this relationship has not yet been formalized in a mathematically
rigorous form. Similarly with L-moments --- a conceptually close device,
but without a formal bridge.

The comparison PMM2 ↔ SLS deserves special attention. Unlike the bridges
to GMM and L-moments, which remain purely conceptual, here the
\textbf{empirical bridge is already partially built}: the work of
{[}25{]} directly compares the performance of both methods and
demonstrates their practical equivalence on nonlinear regression
problems with χ²-errors. However, a full-format theoretical compilation
of PMM2 and SLS into a single formal framework is a separate publication
that has not yet appeared in the literature. The natural formulation of
such a work is to show that SLS is a partial case of the generalized PMM
with a weight matrix of a certain structure, and the optimal
coefficients of PMM at S=2 are an alternative parameterization of the
same family of estimators. If this theoretical summation can be built,
the Kunchenko school gets a natural entry point into the standard
Western literature on nonlinear regression --- journals at the level of
\emph{Annals of the Institute of Statistical Mathematics} (where SLS is
traditionally published), \emph{Journal of Statistical Planning and
Inference}, \emph{Econometric Theory}.

Any of these formalizations --- an independent publication in a journal
at the level of \emph{Annals of Statistics}, \emph{Journal of the
American Statistical Association}, \emph{Biometrika}, where Kunchenko's
school is not yet present. This is the highest bar of international
positioning, which is real for the school only through the construction
of such bridges.

\subsubsection{11.3. Lack of a systematic benchmark
dataset}\label{lack-of-a-systematic-benchmark-dataset}

All 2006--2026 school publications use self-generated datasets---Monte
Carlo with specifically chosen innovation distributions, synthetic OFDM
with specifically chosen nonlinearities, specific ARIMA models with
specific innovations. This makes comparison of results between school
articles difficult and makes direct comparison with external evaluators
impossible. The school does not have a publicly available benchmark
dataset like the UCI ML Repository or PhysioNet against which any
outside researcher could evaluate PMM against alternatives.

Creating such a dataset --- a realistic 9--12 month task: a selection of
8--12 realistic radio-technical, metrological and financial scenarios,
with well-documented generation, open publication via Zenodo or OSF, and
a set of basic scripts to play in R/Python. This is infrastructure work,
without which the school will remain closed to its own community.

\subsubsection{11.4. PMM + Deep Learning
hybrids}\label{pmm-deep-learning-hybrids}

Since 2018, in global statistics and signal processing, the direction
combining classic semiparametric methods with neural network
architectures --- neural ODEs with statistical regularization, deep
state-space models, neural change-point detection has been rapidly
developing. Kunchenko's school is almost not present in this area yet.
Meanwhile, the apparatus of stochastic polynomials naturally fits into
this context: the coefficients PMM can be parameterized as the outputs
of small neural networks that are trained together with the main model,
and the matrix \(F_{S}\) --- can be used as an inductive bias for loss
regularization.

This is a wide and unexplored space. The first works in this direction
could appear at the junction of EstemPMM with PyTorch/JAX through the
creation of differential modules PMM.

\subsection{12. Conclusions and
dedication}\label{conclusions-and-dedication}

Yuriy Petrovych Kunchenko's school has gone from a specific
radiophysical problem to a mature apparatus of semiparametric
non-Gaussian estimation. In the candidate's thesis of 1972/1973,
Volterra's series were applied to finding estimates of unknown
parameters of random processes. In the monographs of 1987--2006, this
idea was transformed into stochastic polynomials, the method of
polynomial maximization, the space with a generating element, and moment
criteria for hypothesis testing. In the works of students and followers
2006--2026, it continued in regression estimation, signal detector
problems, metrology, biomedical signal processing and software
implementation EstemPMM.

The main conclusion of this article is that the history of the school is
not only a history of succession. It is the trajectory of the
statistical method. The 2026 case of Volterra processing of stochastic
signals shows that the original type of Kunchenko's formulation returns
to modern applied radio engineering: the nonlinear functional from a
random process is again used for an estimative or adaptive procedure.
The formal bridge constructed in § 9 explains how the finite Volterra
model fits into the class of generalized stochastic polynomials, but
does not equate MMSE-adaptation with PMM. It is the difference between
these criteria that creates a research opportunity: to replace or
supplement the fixed L2-criterion with a parametrically dependent moment
criterion and check whether it gives a gain on problems of the {[}6{]}
type.

This conclusion should not be read as a ready promise of universal gain.
A correct formulation requires checking the moment conditions, the
non-degeneracy of the matrix of centered correlations, and a numerical
comparison on benchmark scenarios. That is why the research program § 10
is a necessary continuation of the article: it transforms the memory of
the school into work with its apparatus in the modern methodological
field.

The text is dedicated to the memory of Yuriy Petrovych Kunchenko on the
87th anniversary of his birth. His scientific legacy is important not
only because it created a school, monographs and generations of
students. It left a way of seeing a mathematical object in an applied
problem, and a future applied problem in a mathematical object. It is
this dual view that makes the apparatus of stochastic polynomials alive
today.

\subsection{References}\label{references}

\begin{otherlanguage*}{ukrainian}
\forcecyrillic

{[}1{]} Кунченко Ю.П. Застосування рядів Вольтера для знаходження оцінок
невідомих параметрів випадкових процесів: дис. \ldots{} канд. фіз.-мат.
наук: 01.04.03. Томськ: Томський державний університет, 1972/1973. 164
с. Шифр зберігання РДБ: OD Дк 73-1/1227. URL:
https://search.rsl.ru/ru/record/01009787980; автореферат:
https://search.rsl.ru/ru/record/01007428279.

{[}2{]} Кунченко Ю.П., Лега Ю.Г. Оцінка параметрів випадкових величин
методом максимізації поліному. Київ: Наукова думка, 1991/1992.

{[}3{]} Кунченко Ю.П. Поліноми наближення у просторі з породжувальним
елементом. Київ: Наукова думка, 2005. \emph{(україномовна версія)}

{[}3a{]} Кунченко Ю.П. Полиномы приближения в пространстве с порождающим
элементом. К.: Наукова думка, 2003. 243 с. \emph{(оригінальне
російськомовне видання)}

{[}4{]} Заболотній С.В. Інформаційна технологія ймовірнісного
діагностування розладнання параметрів негаусових послідовностей: дис.
\ldots{} д-ра техн. наук: 05.13.06. Львів: Українська академія
друкарства, 2015.

{[}5{]} Палагін В.В. Математичні моделі, методи та засоби виявлення і
розрізнення сигналів на фоні негаусових завад: дис. \ldots{} д-ра техн.
наук: 01.05.02. Черкаси: ЧДТУ, 2013.

{[}6{]} Гарячий М., Щербінін С. Метод формування стохастичних сигналів
із використанням рядів Вольтера // Вимірювальна та обчислювальна техніка
в технологічних процесах. 2026. № 1. С. 47--60.

{[}7{]} Меморіальна стаття про Ю.П. Кунченка. Сайт кафедри радіотехніки
ЧДТУ. URL: \url{https://rtrs.chdtu.edu.ua/history-of-the-department/}

{[}8{]} Кунченко Ю.П. Нелінійна оцінка параметрів негауссівських
радіофізичних сигналів. Київ: Вища школа, 1987. 191 с.

{[}9{]} Kunchenko Y.P. \emph{Polynomial Parameter Estimations of Close
to Gaussian Random Variables}. Aachen: Shaker Verlag, 2002. 396 p.

{[}10{]} Кунченко Ю.П., Заболотній С.В. Поліноміальні оцінки параметрів
близьких до гаусових випадкових величин. Частина 1, 2. Черкаси: ЧІТІ,
2001.

{[}10a{]} Zabolotnii S.W., Martynenko S.S., Salypa S.V. Method of
verification of hypothesis about mean value on a basis of expansion in a
space with generating element // Radioelectronics and Communications
Systems. 2018. Vol. 61, No.~5. P. 222--229. DOI:
https://doi.org/10.3103/S0735272718050060.

{[}11{]} Кунченко Ю.П. Стохастические полиномы. К.: Наукова думка, 2006.
275 с.

{[}12{]} Кунченко Ю.П., Палагін В.В. Побудова моментного критерію якості
типу Неймана-Пірсона для перевірки простих статистичних гіпотез //
Вісник Інженерної академії України. 2005. № 1. С. 26--30.

{[}12a{]} Kunchenko Y. A moment performance criteria of a
decision-making for testing simple statistical hypothesis // Proc. IEEE
International Symposium on Information Theory (ISIT), Ulm, Germany, July
1997. P. 407. DOI: https://doi.org/10.1109/ISIT.1997.613344.

{[}13{]} Кунченко Ю.П., Палагін В.В. Перевірка статистичних гіпотез при
використанні поліноміальних вирішних правил, оптимальних за моментним
критерієм суми асимптотичних ймовірностей помилок // Радіоелектроніка та
інформатика. 2006. № 3 (34). С. 4--11.

{[}14{]} Палагін В.В. Алгоритми виявлення сигналів на фоні
негауссівських завад за критерієм асимптотичної нормальності: дис.
\ldots{} канд. техн. наук: 05.12.01. Черкаси: ЧДТУ, 1999.

{[}15{]} Заболотній С.В. Нелінійні алгоритми визначення параметрів
негаусових випадкових послідовностей у каналах
інформаційно-вимірювальних систем: дис. \ldots{} канд. техн. наук:
05.11.16. Черкаси: ЧДТУ, 2000.

{[}16{]} Вєдєрніков Д.А. Математичні моделі, методи та засоби оцінювання
параметра постійного сигналу на фоні негаусових корельованих завад: дис.
\ldots{} PhD. Черкаси: ЧДТУ, 2021.

{[}17{]} Смірнов Д.О. Математичні моделі, методи та засоби виявлення
постійного сигналу на фоні негаусових корельованих завад: дис. \ldots{}
PhD. Черкаси: ЧДТУ, 2025.

{[}18{]} Ткаченко О.М. Поліноміальні методи та засоби оцінювання
параметрів регресії з використанням моделей негаусових помилок: дис.
\ldots{} PhD. Черкаси: ЧДТУ, 2021.

{[}19{]} Chertov O., Slipets T. Kunchenko's Polynomials for Template
Matching. arXiv:1107.2085, 2011.

{[}20{]} Chertov O., Slipets T. Epileptic seizures diagnose using
Kunchenko polynomials template matching // Progress in Industrial
Mathematics at ECMI 2012 / Eds. M. Fontes, M. Günther, N. Marheineke.
Springer, Cham, 2014. P. 245--248. DOI:
https://doi.org/10.1007/978-3-319-05365-3\_33.

{[}21{]} Warsza Z.L., Zabolotnii S.W. A polynomial estimation of
measurand parameters for samples of non-Gaussian symmetrically
distributed data // AISC, vol.~550. Springer, Cham, 2017. P. 468--480.
DOI: https://doi.org/10.1007/978-3-319-54042-9\_45.

{[}22{]} Warsza Z.L., Zabolotnii S.W. Estimation of measurand parameters
for data from asymmetric distributions by polynomial maximization method
// AISC, vol.~743. Springer, 2018. P. 746--757. DOI:
https://doi.org/10.1007/978-3-319-77179-3\_74.

{[}23{]} Zabolotnii S., Warsza Z.L., Tkachenko O. Polynomial Estimation
of Linear Regression Parameters for the Asymmetric PDF of Errors //
AISC, vol.~743. Springer, 2018. DOI:
https://doi.org/10.1007/978-3-319-77179-3\_75.

{[}24{]} Zabolotnii S., Warsza Z.L., Tkachenko O. Estimation of Linear
Regression Parameters of Symmetric Non-Gaussian Errors by Polynomial
Maximization Method // Advances in Intelligent Systems and Computing,
vol.~920. Springer, 2019. DOI:
https://doi.org/10.1007/978-3-030-13273-6\_59.

{[}25{]} Zabolotnii S., Tkachenko O., Nowakowski W., Warsza Z.L.
Application of the Polynomial Maximization Method for Estimating
Nonlinear Regression Parameters with Non-gaussian Asymmetric Errors //
Automation 2024: Advances in Automation, Robotics and Measurement
Techniques. Springer LNNS, vol.~1219, 2025. P. 1--15. DOI:
https://doi.org/10.1007/978-3-031-78266-4\_30.

{[}26{]} Palahin V., Juhár J. Joint signal parameter estimation in
non-Gaussian noise by the method of polynomial maximization // Journal
of Electrical Engineering. 2016. Vol. 67, No.~3. P. 217--221. DOI:
https://doi.org/10.1515/jee-2016-0031.

{[}27{]} Vokorokos L., Ivchenko A., Marchevský S., Palahina E., Palahin
V. Parameters estimation of correlated non-Gaussian processes by the
method of polynomial maximisation // IET Signal Processing. 2017. Vol.
11, No.~3. P. 313--319.

{[}28{]} Zabolotnii S.W., Warsza Z.L. Semi-parametric estimation of the
change-point of parameters of non-Gaussian sequences by polynomial
maximization method // AISC, vol.~440. Springer, Heidelberg, 2016. P.
903--919. DOI: https://doi.org/10.1007/978-3-319-29357-8\_80.

{[}28a{]} Zabolotnii S.W., Warsza Z.L. Semi-parametric polynomial method
for retrospective estimation of the change-point of parameters of
non-Gaussian sequences // Advanced Mathematical and Computational Tools
in Metrology and Testing X (AMCTM X) / Eds. F. Pavese et al.~Series on
Advances in Mathematics for Applied Sciences, vol.~86. World Scientific,
Singapore, 2015. P. 400--408.

{[}29{]} Матеріали VII Міжнародної науково-практичної конференції
«Обробка сигналів і негауссівських процесів». Черкаси: ЧДТУ, 2019. 212
с.

{[}30{]} Zabolotnii S.V., Chepynoha A.V., Bondarenko Yu.Yu., Rud M.P.
Polynomial parameter estimation of exponential power distribution data
// Visnyk NTUU KPI Seriia - Radiotekhnika Radioaparatobuduvannia. 2018.
Issue 75. P. 40--47. DOI: https://doi.org/10.20535/radap.2018.75.40-47.

{[}31{]} Zabolotnii S., Khotunov V., Chepynoha A., Tkachenko O.
Estimating parameters of linear regression with an exponential power
distribution of errors by using a polynomial maximization method //
Eastern-European Journal of Enterprise Technologies. 2021. Vol. 1,
No.~4(109). DOI: https://doi.org/10.15587/1729-4061.2021.225525.

{[}32{]} Zabolotnii S. Applying PMM to estimate ARIMA models with
asymmetric non-Gaussian innovations. arXiv:2511.07059, 2025.

{[}33{]} Zabolotnii S. EstemPMM: Polynomial Maximization Method for
Non-Gaussian Errors. R package, CRAN, 2025. URL:
https://cran.r-project.org/package=EstemPMM

{[}33a{]} Заболотній С.В., Чепинога А.В., Салипа С.В. Спосіб генерації
випадкових величин: декларац. патент України на корисну модель № 57092,
МПК G06F7/58. Заявл. 16.07.2010; Опубл. 10.02.2011. Бюл. № 3.

{[}34{]} Palahin V., Palahina I., Filipov V., Leleko S., Ivchenko A.
Modeling of joint signal detection and parameter estimation on
background of Non-Gaussian noise // Journal of Applied Mathematics and
Computational Mechanics. 2015. Vol. 14, No.~3. P. 87--94. DOI:
\url{https://doi.org/10.17512/jamcm.2015.3.09}.

{[}35{]} Palahina E., Gamcová M., Gladišová I., Gamec J., Palahin V.
Signal Detection in Correlated Non-Gaussian Noise Using Higher-Order
Statistics // Circuits, Systems, and Signal Processing. 2017. Vol. 37,
No.~4. P. 1704--1723. DOI: https://doi.org/10.1007/s00034-017-0623-5.

{[}36{]} Smirnov D., Palahina E., Palahin V. Mathematical Modeling of
Signal Detection in Non-gaussian Correlated Noise // Smart Technologies
in Urban Engineering. STUE 2022. Lecture Notes in Networks and Systems,
vol.~536. Springer, 2023. P. 65--74. DOI:
https://doi.org/10.1007/978-3-031-20141-7\_7.

{[}37{]} Krasilnikov A., Beregun V. Cumulant Detector of Non-Gaussian
Signals against Background of Non-Gaussian Interferences //
Radioelectronics and Communications Systems. 2024. Vol. 67, No.~6. P.
317--330. DOI: https://doi.org/10.3103/S0735272724060037.

{[}38{]} Чертов О.Р., Тавров Д.Ю. Use Kunchenko Polynomials for the
Analysis of the Statistical Data // Eastern-European Journal of
Enterprise Technologies. 2010. Vol. 4, Issue 4(46). P. 70--75.

{[}39{]} Захищені дисертації за напрямком наукової школи професора
Кунченка Ю.П. Сайт кафедри радіотехніки ЧДТУ. URL:
https://rtrs.chdtu.edu.ua/zahysty-dysertaczijnyh-robit/

{[}40{]} Заболотній С.В., Ткаченко О.М. Поліноміальні адаптивні
процедури регресійного аналізу із використанням моделей негаусових
помилок на основі статистик вищих порядків // Тези доповідей IV
Міжнародної науково-практичної конференції «Обчислювальний інтелект
(результати, проблеми, перспективи) --- 2017» (ComInt-2017). Київ: КНУ
ім. Т. Шевченка, 16--18 травня 2017 р. С. 113--114.

{[}41{]} Заболотній С.В., Ткаченко О.М. Застосування методу максимізації
полінома для оцінювання параметрів однофакторної лінійної регресії при
негаусовому розподілі помилок. Бібліографічний запис у переліку праць
О.М. Ткаченка, ЧДТУ. URL:
https://rtrs.chdtu.edu.ua/postgraduate-students-of-the-department/

{[}42{]} Збірка тез VIII Міжнародної науково-практичної конференції
«Обробка сигналів і негауссівських процесів» (ОСНП-2021), присвяченої
пам'яті професора Ю.П. Кунченка. Черкаси: ЧДТУ, 25--26 травня 2021 р.

{[}43{]} Rugh W.J. \emph{Nonlinear System Theory: The Volterra/Wiener
Approach}. Baltimore: Johns Hopkins University Press, 1981.

{[}44{]} Schetzen M. \emph{The Volterra and Wiener Theories of Nonlinear
Systems}. New York: Wiley, 1980.

{[}45{]} Mathews V.J., Sicuranza G.L. \emph{Polynomial Signal
Processing}. New York: Wiley, 2000.

{[}46{]} Mendel J.M. Tutorial on higher-order statistics (spectra) in
signal processing and system theory: theoretical results and some
applications // Proceedings of the IEEE. 1991. Vol. 79, No.~3. P.
278--305.

{[}47{]} Nikias C.L., Petropulu A.P. \emph{Higher-Order Spectra
Analysis: A Nonlinear Signal Processing Framework}. Prentice Hall, 1993.

{[}48{]} Giannakis G.B., Tsatsanis M.K. Signal detection and
classification using matched filtering and higher order statistics //
IEEE Transactions on Acoustics, Speech, and Signal Processing. 1990.
Vol. 38, No.~7. P. 1284--1296.

{[}49{]} Huber P.J. \emph{Robust Statistics}. New York: Wiley, 1981.

{[}50{]} Hansen L.P. Large sample properties of generalized method of
moments estimators // Econometrica. 1982. Vol. 50, No.~4. P. 1029--1054.

{[}51{]} Hosking J.R.M. L-moments: analysis and estimation of
distributions using linear combinations of order statistics // Journal
of the Royal Statistical Society, Series B. 1990. Vol. 52, No.~1. P.
105--124.

{[}52{]} Wang L. Estimation of nonlinear Berkson-type measurement error
models // Statistica Sinica. 2003. Vol. 13. P. 1201--1210.

{[}53{]} Wang L., Leblanc A. Second-order nonlinear least squares
estimation // Annals of the Institute of Statistical Mathematics. 2008.
Vol. 60. P. 883--900. DOI: https://doi.org/10.1007/s10463-007-0139-z.

{[}54{]} Kim M., Ma Y. The efficiency of the second-order nonlinear
least squares estimator and its extension // Annals of the Institute of
Statistical Mathematics. 2012. Vol. 64. P. 751--764. DOI:
https://doi.org/10.1007/s10463-011-0332-y.

{[}55{]} Gao L.L., Zhou J. New optimal design criteria for regression
models with asymmetric errors // Journal of Statistical Planning and
Inference. 2014. Vol. 149. P. 140--151. DOI:
\url{https://doi.org/10.1016/j.jspi.2014.01.005}.

{[}56{]} Salamh M., Wang L. Second-order least squares estimation in
nonlinear time series models with ARCH errors // Econometrics. 2021.
Vol. 9, No.~4. Article 41. DOI:
\url{https://doi.org/10.3390/econometrics9040041}.

{[}57{]} He L., Yue R.X., Du A. Optimal designs for comparing curves in
regression models with asymmetric errors // Journal of Statistical
Planning and Inference. 2024. Vol. 228. P. 46--58. DOI:
https://doi.org/10.1016/j.jspi.2023.06.005.

\end{otherlanguage*}

\clearpage
\selectlanguage{ukrainian}
\forcecyrillic

\phantomsection\label{ux432ux456ux434-ux440ux44fux434ux456ux432-ux432ux43eux43bux44cux442ux435ux440ux430-ux434ux43e-ux441ux442ux43eux445ux430ux441ux442ux438ux447ux43dux438ux445-ux43fux43eux43bux456ux43dux43eux43cux456ux432-ux43aux443ux43dux447ux435ux43dux43aux430-ux43fux456ux432ux441ux442ux43eux43bux456ux442ux442ux44f-ux43cux435ux442ux43eux434ux43eux43bux43eux433ux456ux457-ux43dux435ux433ux430ux443ux441ux456ux432ux441ux44cux43aux43eux433ux43e-ux43eux446ux456ux43dux44eux432ux430ux43dux43dux44f}
\addcontentsline{toc}{section}{Від рядів Вольтера до стохастичних поліномів Кунченка}
\begin{center}
{\LARGE\bfseries Від рядів Вольтера до стохастичних поліномів Кунченка:
півстоліття методології негаусівського оцінювання\par}

\vspace{0.5em}
\emph{До 87-ї річниці від дня народження засновника черкаської наукової
школи}

\vspace{0.4em}
\textbf{С. В. Заболотній}

\emph{Черкаський державний фаховий бізнес-коледж}\\
\emph{ORCID: 0000-0003-0242-2234}
\end{center}

\subsection{Анотація}\label{ux430ux43dux43eux442ux430ux446ux456ux44f}

Стаття реконструює півстолітню еволюцію наукової школи Юрія Петровича
Кунченка (1939--2006) як історію формування семіпараметричної
методології негаусівського оцінювання. Вихідною точкою є кандидатська
дисертація 1972/1973 р., у якій ряди Вольтера було застосовано до задачі
знаходження оцінок невідомих параметрів випадкових процесів; подальший
розвиток простежено до сучасних робіт учнів і послідовників школи
2006--2026 рр. Апарат стохастичних поліномів подано не лише як локальну
традицію черкаської школи, а як узгоджену сім'ю моментно-кумулянтних
процедур: метод максимізації поліному для оцінювання параметрів,
поліноміальні критерії перевірки статистичних гіпотез і декомпозицію в
просторі з порідним елементом для задач розпізнавання. Окремо враховано
інституційну структуру школи: верифіковану генеалогію з 15 захищених
дисертацій, міжнародні колаборації в Польщі, Словаччині та Німеччині, а
також сучасну програмну інфраструктуру у вигляді R-пакета EstemPMM. На
прикладі статті 2026 р. про Вольтера-обробку стохастичних сигналів
показано, що вихідна постановка Кунченка повертається в сучасну
прикладну радіотехніку на рівні типу задачі: нелінійний функціонал від
випадкового процесу використовується для оцінювальної або адаптивної
процедури. Побудовано формальний міст між скінченною моделлю Вольтера і
узагальненим стохастичним поліномом Кунченка на векторному аргументі;
водночас розмежовано MMSE/L2-критерій VOTTP як коваріаційну проєкцію для
адаптації ядра і ММПл як параметрично залежну моментну процедуру
оцінювання параметра процесу або розподілу. Твердження про ефективність
ММПл сформульовано в умовній, методологічно коректній формі: виграш
очікується для тих негаусівських класів, де існують потрібні моменти,
матриця центрованих корелянтів невироджена, а коефіцієнт варіаційної
редукції менший за одиницю. У фіналі окреслено дослідницьку програму, що
переводить історичну реконструкцію в набір перевірюваних статистичних і
сигнал-обробних задач.

\textbf{Ключові слова:} стохастичні поліноми Кунченка; метод
максимізації поліному; ряди Вольтера; негаусівські випадкові процеси;
моментно-кумулянтний опис; простір з порідним елементом;
семіпараметричні методи; нелінійна обробка сигналів; черкаська наукова
школа; адаптивне оцінювання параметрів; коефіцієнт варіаційної редукції;
перфорація кумулянтного опису; second-order least squares (SLS).

\subsection{Abstract}\label{abstract-uk}

This paper reconstructs the half-century evolution of the scientific
school founded by Yuriy P. Kunchenko (1939--2006) as the development of
a semiparametric methodology for non-Gaussian estimation. Its starting
point is Kunchenko's 1972/1973 candidate dissertation, where Volterra
series were applied to estimating unknown parameters of random
processes; the trajectory is followed through the work of his students
and successors in 2006--2026. Kunchenko stochastic polynomials are
presented not merely as a local research tradition, but as a coherent
family of moment-cumulant procedures: the polynomial maximization method
for parameter estimation, polynomial criteria for statistical hypothesis
testing, and decomposition in the space with a generating element for
recognition problems. The paper also accounts for the institutional
structure of the school: a verified genealogy of 15 defended
dissertations, international collaborations in Poland, Slovakia, and
Germany, and the recent R package EstemPMM. A 2026 paper on
Volterra-based processing of stochastic signals is used as a diagnostic
case showing that Kunchenko's original problem has reappeared in applied
radio engineering at the level of problem structure: a nonlinear
functional of a random process is used to construct an estimation or
adaptation procedure. A formal bridge is built between finite Volterra
models and generalized Kunchenko stochastic polynomials on vector
arguments; at the same time, the MMSE/L2 criterion in the 2026 paper is
separated from PMM, since the former is a covariance projection for
kernel adaptation whereas the latter is a parameter-dependent
moment-based procedure for estimating parameters of a process or
distribution. Claims about PMM efficiency are stated conditionally:
gains are expected for non-Gaussian classes where the required moments
exist, the centered correlant matrix is nondegenerate, and the variance
reduction coefficient is below one. The concluding research program
turns the historical reconstruction into a set of testable statistical
and signal-processing tasks.

\textbf{Keywords:} Kunchenko stochastic polynomials; polynomial
maximization method; Volterra series; non-Gaussian random processes;
moment-cumulant description; space with generating element;
semiparametric methods; nonlinear signal processing; Cherkasy scientific
school; adaptive parameter estimation; variance reduction coefficient;
cumulant description perforation; second-order least squares (SLS).

\subsection{1. Вступ: історія як траєкторія
методу}\label{ux432ux441ux442ux443ux43f-ux456ux441ux442ux43eux440ux456ux44f-ux44fux43a-ux442ux440ux430ux454ux43aux442ux43eux440ux456ux44f-ux43cux435ux442ux43eux434ux443}

Двадцять шостого травня 2026 року минає вісімдесят сім років від дня
народження Юрія Петровича Кунченка (1939, Ростов-на-Дону --- 2006,
Черкаси) --- українського радіофізика й математика, доктора
фізико-математичних наук, заслуженого діяча науки і техніки України,
засновника черкаської наукової школи нелінійного статистичного аналізу
негаусівських сигналів. Його рання робота з рядами Вольтера важлива не
тільки як історичний факт. У ній уже було закладено спосіб мислення,
який згодом стане впізнаваним для всієї школи: параметр випадкового
процесу треба оцінювати не лише через лінійну статистику або повністю
задану функцію розподілу, а через поліноміальні функціонали, що вміють
використовувати інформацію моментів вищих порядків.

Саме тому цю статтю доцільно читати не як біографічний нарис і не як
локальну хроніку кафедральної традиції. Її предметом є шлях
статистичного методу. Від кандидатської дисертації 1972/1973 р. до
монографій 1987--2006 рр., від оцінювання параметрів випадкових величин
до перевірки статистичних гіпотез і розпізнавання образів, від
рукописних нормальних рівнянь до відтворюваного R-пакета EstemPMM ---
перед нами розгортається одна й та сама ідея: негаусівськість не є
шумом, який треба усунути, а джерелом додаткової статистичної
інформації.

Ця оглядова стаття має три завдання. Перше --- реконструювати ідейну
еволюцію школи з опертям на верифіковані бібліографічні та інституційні
дані. Друге --- показати, що три прикладні гілки школи, а саме
оцінювання параметрів, перевірка гіпотез і розпізнавання образів, є
похідними одного формального апарату стохастичних поліномів. Третє ---
пояснити, чому свіжа задача Вольтера-обробки стохастичних сигналів,
опублікована у 2026 р., фактично повертає нас до вихідного типу
постановки Кунченка і водночас відкриває сучасну дослідницьку програму.

У структурному плані текст рухається трьома пластами. Розділи 2--4
відтворюють еволюцію від вольтерівської постановки до зрілого апарату
стохастичних поліномів. Розділи 5--6 показують школу як інституційну й
дослідницьку мережу. Розділи 7--10 переводять історію в методологію:
зіставляють апарат Кунченка із суміжними статистичними традиціями,
аналізують випадок 2026 р. і будують формальний міст між
Вольтера-моделлю та стохастичним поліномом. Завершальні розділи
називають відкриті проблеми й формулюють наступні кроки.

\subsection{2. Точка нуль: ряди Вольтера в дисертації 1972/1973
р.}\label{ux442ux43eux447ux43aux430-ux43dux443ux43bux44c-ux440ux44fux434ux438-ux432ux43eux43bux44cux442ux435ux440ux430-ux432-ux434ux438ux441ux435ux440ux442ux430ux446ux456ux457-19721973-ux440.}

Кандидатська дисертація Кунченка має такі вихідні дані: рукопис датовано
1972 роком, місце захисту --- Томський державний університет, обсяг ---
164 сторінки, спеціальність --- 01.04.03 «Радіофізика, включаючи
квантову радіофізику». За меморіальною статтею ЧДТУ, захист відбувся
1973 року в спеціалізованій раді при Томському університеті {[}7{]}. Ця
двозначність дати --- типовий для радянських дисертацій випадок
розходження між роком завершення рукопису й роком процедурного захисту
--- не несе жодного концептуального наслідку, і в подальшому ми будемо
посилатися на дисертацію як на роботу 1972/1973 р.

Що становить інтелектуальний контекст цієї роботи? Початок 1970-х в
радянській радіофізичній школі --- період зрілості теорії випадкових
процесів, оформленої в монографіях Пугачова, Тихонова, Сосуліна.
Одночасно західна школа (Wiener, Volterra, пізніше Schetzen, Rugh)
активно розвивала функціональні ряди як інструмент опису нелінійних
систем з пам'яттю. Точка перетину цих двох ліній --- а саме використання
рядів Вольтера для розв'язання статистичних задач, не
системно-ідентифікаційних, а параметрично-оцінювальних --- і складає
предметне поле дисертації Кунченка.

Кандидатська постановка формулюється так. Нехай спостерігається
випадковий процес \(\xi(t)\), статистичні характеристики якого залежать
від невідомого параметра \(\theta\) (або вектора параметрів
\(\overrightarrow{\theta}\)). Ряд Вольтера дозволяє представити
функціонал від \(\xi(t)\) у вигляді скінченновимірного нелінійного
розкладу за добутками значень процесу в різні моменти часу, з ядрами
\(h_{n}\left( \tau_{1},\ldots,\tau_{n} \right)\), що несуть інформацію
про нелінійну структуру залежності спостережуваної величини від
\(\theta\). Задача оцінювання \(\theta\) зводиться, отже, до знаходження
таких ядер \(h_{n}\), які мінімізують середньоквадратичну помилку
наближення статистики, обчисленої з вибірки, до її теоретичного значення
як функції невідомого параметра. Технічно --- це система нормальних
рівнянь типу \(C_{x} \cdot h = r_{yx}\), де \(C_{x}\) --- матриця
моментів процесу, а \(r_{yx}\) --- крос-кореляційний вектор.

У цій постановці вже присутні всі чотири елементи, які за наступні
десятиріччя стануть концептуальною підставою школи. По-перше,
\textbf{орієнтація на оцінювання, а не на ідентифікацію системи}: ядро
Вольтера у Кунченка не описує фізичний канал, а є інструментом отримання
оцінок параметра. По-друге, \textbf{поліноміальна структура наближення}:
оцінка будується не за єдиною лінійною статистикою, а за добутками
значень процесу різних порядків, що відкриває можливість залучати
інформацію вищих моментів. По-третє, \textbf{моментний характер опису}:
коефіцієнти ядер виражаються через моменти (кумулянти) процесу, а не
через його повну функцію розподілу. По-четверте, \textbf{орієнтація на
негаусівські процеси}: для гаусівських процесів вищі кумулянти тотожні
нулю, і поліноміальний розклад втрачає переваги перед лінійним; уся
ефективність методу проявляється саме за наявності негаусівських
характеристик.

Жоден із цих чотирьох елементів у 1972 р. ще не оформлений як окрема
теорія. Їхній синтез у самостійний апарат --- справа наступних 15 років,
з кульмінацією у докторській дисертації 1988 р. і монографії 1987 р. Але
важливо зафіксувати: концептуальний зародок усієї майбутньої школи вже
присутній у кандидатській. Стохастичний поліном як математичний об'єкт
ще не названий, простір з порідним елементом ще не введений,
моментно-кумулянтний опис ще не систематизований --- однак методологічна
установка вже задана: \textbf{оцінювати параметри негаусівських
випадкових процесів через нелінійні поліноміальні функціонали від
моментів спостережуваних статистик, а не через лінійну статистику й
апріорну функцію розподілу}.

Саме ця установка і є тим, що еволюціонувало в наступні десятиліття у
сім'ю взаємопов'язаних апаратів --- і саме її повторно й незалежно
реконструюють у 2026 році автори статті {[}6{]}, про яку йдеться в § 8.
Між цими двома точками лежить півстоліття внутрішньої роботи школи, яку
наступні розділи послідовно розгортають.

\subsection{3. Перехід 1973--1987: від ядер Вольтера до нелінійного
оцінювання негаусівських
сигналів}\label{ux43fux435ux440ux435ux445ux456ux434-19731987-ux432ux456ux434-ux44fux434ux435ux440-ux432ux43eux43bux44cux442ux435ux440ux430-ux434ux43e-ux43dux435ux43bux456ux43dux456ux439ux43dux43eux433ux43e-ux43eux446ux456ux43dux44eux432ux430ux43dux43dux44f-ux43dux435ux433ux430ux443ux441ux456ux432ux441ux44cux43aux438ux445-ux441ux438ux433ux43dux430ux43bux456ux432}

П'ятнадцять років, що відокремлюють кандидатську дисертацію 1972/1973 р.
від першої монографії 1987 р. «Нелінійна оцінка параметрів
негауссівських радіофізичних сигналів» (Київ: Вища школа, 191 с.)
{[}8{]}, складають найменш видимий, але концептуально центральний період
еволюції школи. Цей період охоплює завершення томського етапу біографії
Кунченка (1962--1978, кафедра радіофізики ТДУ та Сибірський
фізико-технічний інститут) і його кіровоградський етап (1979--1990,
доцент, згодом завідувач кафедри вищої математики Кіровоградського
інституту сільськогосподарського машинобудування). Завершується період
захистом докторської дисертації 1988 р. у спеціалізованій раді при
Харківському державному університеті за тією самою спеціальністю
«радіофізика, включаючи квантову радіофізику», що й кандидатська
{[}7{]}.

Документальний слід цього періоду в публічно доступних джерелах
фрагментарний. Відкриті бібліотечні каталоги дають точну назву й шифр
кандидатської (§ 2), і дозволяють локалізувати монографію 1987 р., але
не зберігають повного списку статей і доповідей між цими двома датами.
Це не виняткова ситуація --- для радянських авторів радіофізичного
профілю 1970-х--1980-х основний публікаційний канал ішов через журнали,
а також через закриті відомчі звіти НДР і збірники конференцій,
оцифровані лише вибірково.

Концептуально, однак, спрямованість руху чітко простежується за
порівнянням двох опорних точок --- кандидатської 1972/1973 і монографії
1987.

У монографії 1987 р. вже присутні всі базові елементи, які складуть
зрілий апарат школи. Поняття стохастичного поліному як випадкової
величини, що є лінійною комбінацією функцій (степеневих,
тригонометричних, експоненційних) від спостережуваної випадкової
величини, з матрицею центрованих корелянтів і об'ємом тіла поліному як
ключовими інваріантами {[}4{]}. Постановка задачі оцінювання параметрів
через максимізацію стохастичного поліному, в якій оцінки знаходяться як
точки глобального максимуму математичного сподівання поліному, що
розглядається як функція невідомого параметра. Доведення асимптотичної
ефективності таких оцінок зі зростанням степеню поліному. Перенесення
цього апарату з оцінювання параметрів випадкових величин на ширший клас
задач --- оцінювання параметрів сигналів на негаусівському тлі.

Ключовий концептуальний зсув від кандидатської 1972/1973 до монографії
1987 --- \textbf{перехід від функціонального опису ``вхід-вихід'' до
моментно-кумулянтного опису розподілу}. У вихідній вольтерівській
постановці ядра \(h_{n}\) описували фізичний канал передачі сигналу,
навіть якщо метою було оцінювання параметра, а не ідентифікація системи.
У зрілому апараті 1987 р. поліноміальна структура переноситься з
простору сигналів у простір моментів: коефіцієнти \(h_{i}\) перестають
бути функціями часових затримок \(\tau_{1},\ldots,\tau_{n}\) і стають
коефіцієнтами розкладу за моментними або кумулянтними характеристиками
випадкової величини. Цей зсув звільняє теорію від прив'язки до
конкретної фізики каналу і робить її універсальною. Один і той самий
апарат тепер працює і для оцінювання параметрів вибірки, і для
оцінювання параметрів процесу, і для побудови вирішних правил перевірки
гіпотез --- бо всі ці задачі зводяться до операцій з моментами.

Другий концептуальний зсув цього періоду --- \textbf{обмеження класу
досліджуваних розподілів через ``близькість до гаусового''}. У
монографії 1987 р. та працях наступних років послідовно розвивається
ідея перфорації кумулянтного опису {[}4{]}: для багатьох практично
важливих задач кумулянтні коефіцієнти вище деякого порядку настільки
малі, що ними можна знехтувати, обнуляючи їх у моделі. Шляхом комбінації
різних типів перфорації виділяються три класи негаусівських випадкових
величин --- асиметричні, ексцесні й асиметрично-ексцесні, які отримують
назву «близьких до гаусових». Ця класифікація стане основою для
канонічної англомовної монографії 2002 р. \emph{Polynomial Parameter
Estimations of Close to Gaussian Random Variables} (Aachen: Shaker
Verlag, 396 с.) {[}9{]}, назва якої прямо відображає цей теоретичний
хід.

Третій зсув --- \textbf{поява моментного критерію якості формування
вирішних правил}. Вже в монографії 1987 р. і в статтях кінця 1980-х ---
початку 1990-х оформлюється ідея, що поліноміальний апарат застосовний
не тільки до оцінювання, а й до перевірки гіпотез: розклад логарифму
відношення правдоподібності у стохастичний ряд з коефіцієнтами,
оптимальними за моментними критеріями, дає альтернативу класичному
критерію Неймана-Пірсона, не вимагаючи знання повної функції розподілу
{[}4{]}. Ця гілка вперше отримує систематичне оформлення у доповіді Ю.П.
Кунченка на IEEE International Symposium on Information Theory в Ульмі
1997 р. {[}12a{]} --- цей факт варто підкреслити окремо: апарат школи
був присутній у міжнародній IEEE-літературі вже за чверть століття до
сьогоднішнього періоду її продовження. Ця гілка згодом викристалізується
у працях Кунченка з В.В. Палагіним та С.С. Мартиненком 1998--2006 рр.
{[}12, 13{]} і складе фундамент сигнально-детекторної гілки школи (§
6.2).

Таким чином, період 1973--1987 рр. --- це період, у який ідея, зародок
якої присутній у кандидатській, формалізується в самостійний апарат.
Стохастичний поліном Кунченка як математичний об'єкт, ММПл як метод
оцінювання, моментно-кумулянтний опис як форма інформації, перфорація як
прийом класифікації, близькі до гаусових випадкові величини як цільовий
клас. Усе це --- наслідок відносно тихого внутрішнього ходу 15-річної
роботи, документально зафіксованого тільки на двох опорних точках, але
концептуально безперервного.

\subsection{4. Зрілість школи 1991--2006: оформлення апарату
стохастичних
поліномів}\label{ux437ux440ux456ux43bux456ux441ux442ux44c-ux448ux43aux43eux43bux438-19912006-ux43eux444ux43eux440ux43cux43bux435ux43dux43dux44f-ux430ux43fux430ux440ux430ux442ux443-ux441ux442ux43eux445ux430ux441ux442ux438ux447ux43dux438ux445-ux43fux43eux43bux456ux43dux43eux43cux456ux432}

Період між монографією 1987 р. і смертю Юрія Петровича 6 серпня 2006 р.
--- це сімнадцять років, упродовж яких школа з тематичного напряму
одного автора перетворюється на повноцінну інституційну формацію з
власною кафедрою (з 1990 р. --- кафедра радіотехніки Черкаського
інженерно-технологічного інституту, нині ЧДТУ), власною термінологією,
власним канонічним корпусом монографій та першим поколінням учнів. Цей
період відкривається переїздом Кунченка з Кіровограда до Черкас на
початку 1990 р. і закривається посмертною монографією 2006 р., яка стала
підсумковим синтезом теорії.

\subsubsection{4.1. Канонічний корпус
монографій}\label{ux43aux430ux43dux43eux43dux456ux447ux43dux438ux439-ux43aux43eux440ux43fux443ux441-ux43cux43eux43dux43eux433ux440ux430ux444ux456ux439}

Книжковий канон школи, оформлений у цей період, складається з шести
монографій, що утворюють внутрішньо узгоджену послідовність. Перша ---
вже згадана «Нелінійна оцінка параметрів негауссівських радіофізичних
сигналів» 1987 р. {[}8{]} --- слугує мостом між кандидатським періодом
та кафедральною роботою в Черкасах. Друга --- «Оцінка параметрів
випадкових величин методом максимізації поліному» (Київ: Наукова думка,
1991/1992; у співавторстві з Ю.Г. Легою) {[}2{]} --- стає першим
систематичним викладом ММПл як методу. Третя --- «Полиномиальные оценки
параметров близких к гауссовским случайных величин. Часть 1» (Черкаси,
2001) разом з «Часть 2», написаною у співавторстві з С.В. Заболотнім ---
фіксує черкаський етап і виводить класифікацію близьких до гаусових
величин (асиметричних, ексцесних, асиметрично-ексцесних) у форму,
безпосередньо орієнтовану на прикладні задачі {[}10{]}.

Четверта монографія --- \emph{Polynomial Parameter Estimations of Close
to Gaussian Random Variables} (Aachen: Shaker Verlag, 2002, 396 с.)
{[}9{]} --- складає головний англомовний канон школи. Її поява пов'язана
з гостьовим перебуванням Кунченка в Інституті техніки зв'язку
Ганноверського університету в 2001--2002 рр., де він читав курс лекцій
про метод максимізації поліному. Ця монографія залишається донині єдиним
повним викладом апарату ММПл англійською мовою, і саме її цитують усі
без винятку міжнародні публікації школи 2010--2026 рр. як головну точку
входу.

П'ята і шоста монографії 2003--2006 рр. --- «Полиномы приближения в
пространстве с порождающим элементом» (К.: Наукова думка, 2003, 243 с.)
{[}3a{]}, її україномовна версія «Поліноми наближення у просторі з
породжувальним елементом» (К.: Наукова думка, 2005) {[}3{]} та
підсумкова «Стохастические полиномы» (К.: Наукова думка, 2006, 275 с.)
{[}11{]} --- оформлюють геометричну інтерпретацію апарату. Тут вводиться
простір з порідним елементом (простір Кунченка) як абстрактне формальне
середовище, в якому стохастичний поліном існує як об'єкт незалежно від
конкретного класу базисних функцій. Саме ця геометризація відкриває
третю гілку школи --- розклад в просторі із породжувальним елементом для
розпізнавання образів, до якої повернемося в § 6.3.

\subsubsection{4.2. Формальне ядро
апарату}\label{ux444ux43eux440ux43cux430ux43bux44cux43dux435-ux44fux434ux440ux43e-ux430ux43fux430ux440ux430ux442ux443}

Зріла форма апарату стохастичних поліномів, оформлена в монографіях
2002--2006 рр., у компактному викладі учнів школи {[}4{]} виглядає так.
Нехай \(\xi\) --- спостережувана випадкова величина з невідомим
параметром \(\overrightarrow{\theta}\). Розглядається множина функцій
\(\{\varphi_{i}(\xi)\}_{i = 1}^{S}\), лінійно незалежних на області
допустимих значень параметра. \textbf{Узагальнений стохастичний поліном}
степеня \(S\) визначається як випадкова величина

\[
\eta_{S} = h_{0} + \sum_{i = 1}^{S}h_{i}\varphi_{i}(\xi),\quad\left| h_{i} \right| < \infty,
\]

де коефіцієнти \(h_{i}\) --- невипадкові константи. Якщо
\(\varphi_{i}(\xi) = \xi^{i}\), поліном називається степеневим і
повністю описується через початкові моменти
\(\alpha_{i}\left( \overrightarrow{\theta} \right) = E\{\xi^{i}\}\).
Якщо \(\varphi_{i}(\xi) = \cos(ik\xi)\) або
\(\varphi_{i}(\xi) = \sin(ik\xi)\) --- тригонометричним і описується
через характеристичну функцію.

Ключовою характеристикою поліному є \textbf{матриця центрованих
корелянтів}

\[
F_{i,j}\left( \overrightarrow{\theta} \right) = \Psi_{i,j}\left( \overrightarrow{\theta} \right) - \Psi_{i}\left( \overrightarrow{\theta} \right)\Psi_{j}\left( \overrightarrow{\theta} \right),
\]

де
\(\Psi_{i,j}\left( \overrightarrow{\theta} \right) = E\{\varphi_{i}(\xi)\varphi_{j}(\xi)\}\),
\(\Psi_{i}\left( \overrightarrow{\theta} \right) = E\{\varphi_{i}(\xi)\}\).
Дисперсія поліному виражається через матрицю
\(F_{S} = \parallel F_{i,j} \parallel_{i,j = 1}^{S}\) як квадратична
форма

\[
\sigma_{\eta}^{2} = \sum_{i = 1}^{S}\sum_{j = 1}^{S}h_{i}h_{j}F_{i,j}\left( \overrightarrow{\theta} \right).
\]

Матрицю \(F_{S}\) Кунченко називає \textbf{тілом стохастичного
поліному}, її детермінант
\(\Delta_{S}\left( \overrightarrow{\theta} \right) = \left| F_{S} \right|\)
--- \textbf{об'ємом тіла}. Умова невиродженості
\(\Delta_{S}\left( \overrightarrow{\theta} \right) > 0\), що випливає з
властивостей детермінанта Грама за лінійної незалежності
\(\{\varphi_{i}\}\), забезпечує існування і єдиність розв'язку задачі
оцінювання параметра. Для степеневого поліному корелянти виражаються
прямо через моменти: \(\Psi_{i,j} = \alpha_{i + j}\),
\(F_{i,j} = \alpha_{i + j} - \alpha_{i}\alpha_{j}\).

\subsubsection{4.3. Метод максимізації
поліному}\label{ux43cux435ux442ux43eux434-ux43cux430ux43aux441ux438ux43cux456ux437ux430ux446ux456ux457-ux43fux43eux43bux456ux43dux43eux43cux443}

ММПл будується на доведеній Кунченком властивості: за певних умов на
коефіцієнти \(h_{i}\) математичне сподівання стохастичного поліному як
функція параметра \(\theta\) має глобальний максимум в околі дійсного
значення цього параметра. Формально, оцінка
\({\widehat{\theta}}_{\text{ММПл}}\) знаходиться як точка максимуму
функції

\[
L_{S}\left( \theta;x_{1},\ldots,x_{N} \right) = \sum_{i = 1}^{S}h_{i}^{*}(\theta) \cdot \frac{1}{N}\sum_{n = 1}^{N}\varphi_{i}\left( x_{n} \right),
\]

де \(h_{i}^{*}(\theta)\) --- оптимальні коефіцієнти, що мінімізують
дисперсію оцінки за заданого \(\theta\). Розв'язок системи лінійних
рівнянь
\(F_{S}(\theta) \cdot \overrightarrow{h} = \overrightarrow{b}(\theta)\),
де \(\overrightarrow{b}(\theta)\) --- вектор, що залежить від похідних
\(\Psi_{i}\) за параметром, дає явну форму оптимальних коефіцієнтів.

Центральним результатом є асимптотична поведінка дисперсії оцінки. Для
оцінювання середнього \(\theta = \mu\) асиметричного негаусівського
розподілу з кумулянтами \(c_{2},c_{3},c_{4},\ldots\), дисперсія оцінки,
отриманої ММПл степеня \(S = 2\), виражається у вигляді

\[
\sigma_{\widehat{\mu},\text{ММПл-2}}^{2} = \frac{c_{2}}{N} \cdot g_{2},\quad g_{2}\left( \overrightarrow{\theta} \right) = 1 - \frac{c_{3}^{2}}{c_{2}\left( 2c_{2} + c_{4}/c_{2} \right)}.
\]

У стандартній безрозмірній формі через коефіцієнти асиметрії
\(\gamma_{3} = c_{3}/c_{2}^{3/2}\) та ексцесу
\(\gamma_{4} = c_{4}/c_{2}^{2}\):

\[
g_{2} = 1 - \frac{\gamma_{3}^{2}}{2 + \gamma_{4}}.
\]

Ця формула --- \textbf{коефіцієнт варіаційної редукції} ММПл другого
степеня --- займає одне із центральних місць в усьому апараті школи. Її
значення легко інтерпретувати. Для гаусівського розподілу
\(\gamma_{3} = 0\), \(\gamma_{4} = 0\), отже \(g_{2} = 1\) --- ММПл
збігається з класичним вибірковим середнім, виграшу немає. Для
розподілів, у яких виконуються потрібні моментні умови і \(g_{2} < 1\),
оцінка ММПл-2 має меншу асимптотичну дисперсію, ніж вибіркове середнє;
величина редукції визначається характеристиками розподілу, а не лише
обсягом вибірки \(N\).

Узагальнення на \(S \geq 3\) у регулярних класах може зменшувати
\(g_{S}\), наближаючи оцінку до межі ефективності за Крамером-Рао. У
межі \(S \rightarrow \infty\) оцінка ММПл асимптотично еквівалентна
оцінці максимальної правдоподібності, при цьому \textbf{не вимагаючи
знання повної функції розподілу}: достатньо знання моментів до \(2S\)-го
порядку. Саме ця властивість і робить ММПл семіпараметричним методом ---
компромісом між параметричним підходом (який потребує апріорі повного
розподілу) і непараметричним (який ігнорує інформацію вищих моментів)
{[}4{]}.

\subsubsection{4.4. Перфорація кумулянтного опису та класифікація
близьких до гаусових
величин}\label{ux43fux435ux440ux444ux43eux440ux430ux446ux456ux44f-ux43aux443ux43cux443ux43bux44fux43dux442ux43dux43eux433ux43e-ux43eux43fux438ux441ux443-ux442ux430-ux43aux43bux430ux441ux438ux444ux456ux43aux430ux446ux456ux44f-ux431ux43bux438ux437ux44cux43aux438ux445-ux434ux43e-ux433ux430ux443ux441ux43eux432ux438ux445-ux432ux435ux43bux438ux447ux438ux43d}

Практичне застосування формули (6) і її узагальнень на \(S \geq 3\)
упирається в проблему, яку Кунченко формулює як \textbf{проблему
незамкненості}: для оцінювання \(S\)-го коефіцієнта потрібні моменти до
\(2S\)-го порядку, що породжує нескінченний ланцюжок рівнянь моментів
{[}4{]}. Стандартний прийом обнуління моментів вище деякого порядку (так
звана статистична лінеаризація) часто виявляється некоректним і
призводить до значних похибок.

Альтернатива, запропонована Кунченком --- \textbf{перфорація
кумулянтного опису}. Ідея полягає у тому, що в багатьох практично
важливих задачах кумулянтні коефіцієнти вище деякого порядку малі
настільки, що їх можна обнулити, тоді як певні кумулянти вищих порядків
залишаються інформативними і враховуються при синтезі алгоритму, а решта
приймає довільні значення. Шляхом комбінації різних типів перфорації
виділяються три класи близьких до гаусових випадкових величин:

\begin{itemize}
\item
  \textbf{асиметричні} (інформативні \(\gamma_{3}\), обнулені
  \(\gamma_{4} = \gamma_{5} = \ldots = 0\));
\item
  \textbf{ексцесні} (інформативні \(\gamma_{4}\), обнулені
  \(\gamma_{3} = \gamma_{5} = \ldots = 0\));
\item
  \textbf{асиметрично-ексцесні} (інформативні обидва, \(\gamma_{3}\) і
  \(\gamma_{4}\)).
\end{itemize}

Ця класифікація --- не лише математична зручність, а конкретний
інструмент моделювання радіотехнічних сигналів і завад, для яких
емпірично відомо, що відхилення від гаусівського розподілу зосереджені
переважно у третьому й четвертому моментах. Для кожного з трьох класів
зрілий апарат школи дає замкнені формули оцінок, дисперсій, і умов
застосовності.

\subsubsection{4.5. Завершення черкаського
періоду}\label{ux437ux430ux432ux435ux440ux448ux435ux43dux43dux44f-ux447ux435ux440ux43aux430ux441ux44cux43aux43eux433ux43e-ux43fux435ux440ux456ux43eux434ux443}

Останні роки життя Юрія Петровича (2003--2006) проходять у напруженому
темпі: міжнародні відрядження, написання монографій, низка спільних
статей з учнями по моментними критеріями якості типу Неймана-Пірсона для
перевірки статистичних гіпотез {[}12, 13{]} (продовжуючи лінію,
започатковану в IEEE ISIT 1997 {[}12a{]}), робота на посаді проректора з
науково-дослідної роботи та міжнародних зв'язків ЧДТУ. У 2006 р.
безпосередньо перед смертю Кунченко довів до захисту дві дисертації:
А.В. Гончарова --- «Оцінка параметра постійного сигналу при близьких до
гаусівських адитивних завадах» (спец. 01.05.02) та Т.В. Воробкало ---
«Методи та алгоритми оцінювання кута надходження гармонічного сигналу на
антенну решітку при негаусівських завадах» (спец. 01.05.02) {[}7{]}. 20
січня 2006 р. Указом Президента України Юрію Петровичу присвоєно почесне
звання «Заслужений діяч науки і техніки України».

До моменту смерті школа існує як зріла структура з власною інституційною
основою. Шістнадцять років роботи в Черкасах. Шість кандидатських
дисертацій захищені під керівництвом Кунченка: В.В. Палагін 1999, С.В.
Заболотній 2000, О.С. Гавриш 2001, С.С. Мартиненко 2003, А.В. Гончаров
та Т.В. Воробкало 2006 {[}7{]}. Шість монографій і понад 150 наукових
праць. Виокремлений теоретичний апарат, термінологія, класифікація
задач. Перший склад кафедри радіотехніки ЧДТУ, який стане інституційною
основою школи на наступні два десятиліття. Подальшу долю школи --- як
вона пережила втрату засновника, як розвинулася в три сучасні гілки, як
вийшла на міжнародний рівень --- розглядає наступний розділ.

\subsection{5. Школа як інституційний феномен: генеалогія і
географія}\label{ux448ux43aux43eux43bux430-ux44fux43a-ux456ux43dux441ux442ux438ux442ux443ux446ux456ux439ux43dux438ux439-ux444ux435ux43dux43eux43cux435ux43d-ux433ux435ux43dux435ux430ux43bux43eux433ux456ux44f-ux456-ux433ux435ux43eux433ux440ux430ux444ux456ux44f}

Переглядаючи історію школи Кунченка з точки 2026 р., бачимо структуру,
що виходить за межі звичайного «наставник --- учні». Це багатопоколінна
мережа, побудована на трьох рівнях: безпосередні учні Кунченка, які
захищалися під його керівництвом у 1999--2006 рр.; друге покоління,
очільниками якого стали В.В. Палагін та С.В. Заболотній з захистами
докторських робіт у 2013 та 2015 рр.; і незалежні застосування апарату,
що розгортаються поза прямою генеалогічною лінією, але опертям мають
канонічні монографії школи. Офіційний реєстр захищених дисертацій
підтримується на сайті кафедри і станом на 2026 р. налічує п'ятнадцять
дисертацій за тематикою наукової школи професора Ю.П. Кунченка {[}39{]}.

Окреме зауваження. На кафедрі радіотехніки ЧДТУ протягом черкаського
періоду Кунченка були захищені також інші дисертаційні роботи, у яких
Юрій Петрович не виступав науковим керівником: дві докторські (В.М.
Шарапов, 1998 р.; Ю.Г. Лега, 2001 р.) і одна кандидатська (О.О. Ситник,
1997 р.) {[}7{]}. У переліку учнів школи в наступних підрозділах ці три
дисертації не наводяться, оскільки їхні автори не були учнями Кунченка в
формальному сенсі, хоча належать до спільного інституційного контексту
кафедри.

\subsubsection{5.1. Перше покоління: безпосередні учні
Кунченка}\label{ux43fux435ux440ux448ux435-ux43fux43eux43aux43eux43bux456ux43dux43dux44f-ux431ux435ux437ux43fux43eux441ux435ux440ux435ux434ux43dux456-ux443ux447ux43dux456-ux43aux443ux43dux447ux435ux43dux43aux430}

\textbf{Володимир Васильович Палагін}. Кандидатська 1999 р. «Алгоритми
виявлення сигналів на фоні негауссівських завад за критерієм
асимптотичної нормальності» (спец. 05.12.01 «Теоретична радіотехніка»;
науковий керівник --- Ю.П. Кунченко) {[}14, 39{]}. Палагін працював на
кафедрі радіотехніки ЧДТУ з 1992 р. (тобто був з Кунченком практично з
моменту створення кафедри). У 2013 р. захистив докторську дисертацію
«Математичні моделі, методи та засоби виявлення і розрізнення сигналів
на фоні негаусових завад» за спеціальністю 01.05.02 «Математичне
моделювання та обчислювальні методи»; науковим консультантом виступив
д.т.н., професор А.Ф. Верлань {[}5, 39{]}. З 2014 р. --- завідувач
кафедри робототехнічних і телекомунікаційних систем та кібербезпеки
ЧДТУ. Палагін є інституційним продовжувачем наукової школи в ЧДТУ.

\textbf{Сергій Васильович Заболотній} прийшов до школи Кунченка ще
студентом 3-го курсу. Дипломна робота 1995 р. була присвячена розкладу
випадкових величин і процесів у просторі з порідним елементом ---
задачі, яка згодом стала першою постановкою його кандидатської
дисертації. Тему пізніше було переформульовано Кунченком на пов'язану з
методом максимізації поліному, що відобразило один із типових моментів
роботи школи: внутрішню рухливість учнів між концептуальними гілками
апарату. Кандидатська 2000 р. «Нелінійні алгоритми визначення параметрів
негаусових випадкових послідовностей у каналах
інформаційно-вимірювальних систем» захищена за спеціальністю 05.11.16
«Інформаційно-вимірювальні системи» під науковим керівництвом Ю.П.
Кунченка {[}15, 39{]}. Докторська 2015 р. «Інформаційна технологія
ймовірнісного діагностування розладнання параметрів негаусових
послідовностей» захищена в Українській академії друкарства за
спеціальністю 05.13.06 «Інформаційні технології»; науковим консультантом
був Л.С. Сікора {[}4, 39{]}. Хоча за формальною спеціальністю докторська
вийшла за межі радіотехнічного профілю школи, вона зберегла центральну
роль апарату стохастичних поліномів і моментно-кумулянтного опису,
перенісши його у сферу інформаційних технологій діагностування.

\textbf{Юрій Григорович Лега} формально не належить до учнів Кунченка в
генеалогічному сенсі. Він був у першому складі кафедри, заснованої
Кунченком 1990 р., і виступив співавтором книги «Оцінка параметрів
випадкових величин методом максимізації поліному» 1991/1992 р. {[}2{]}.
У 2001 р. захистив докторську дисертацію за тематикою наближеною до
школи Кунченко; у 2006--2014 рр. --- ректор ЧДТУ. Лега --- інституційний
союзник і співтворець черкаського етапу школи, який після смерті
Кунченка частково переніс адміністративну вагу й у 2013 р. виступив
керівником кандидатської дисертації одного з представників другого
покоління школи --- В.В. Коваля (к.т.н. 2013 р., спец. 01.05.02, тема
«Моделі та методи поліноміального оцінювання параметрів сигналів на фоні
мультиплікативних та адитивно-мультиплікативних завад»).

Решта прямих учнів Кунченка з канонічного переліку {[}39{]}:

\begin{itemize}
\item
  \textbf{О.С. Гавриш} --- к.т.н. 2001 р., спец. «Радіофізика», тема
  «Алгоритми виміру параметрів гармонічного і полігармонічного сигналів
  при негаусівських завадах»;
\item
  \textbf{С.С. Мартиненко} --- к.т.н. 2003 р., спец. «Радіофізика», тема
  «Виявлення сигналів на фоні негаусівських завад поліноміальними
  алгоритмами»;
\item
  \textbf{А.В. Гончаров} --- к.т.н. 2006 р., спец. 01.05.02, тема
  «Оцінка параметра постійного сигналу при близьких до гаусівських
  адитивних завадах»;
\item
  \textbf{Т.В. Воробкало} --- к.т.н. 2006 р., спец. 01.05.02, тема
  «Методи та алгоритми оцінювання кута надходження гармонічного сигналу
  на антенну решітку при негаусівських завадах».
\end{itemize}

У сучасній структурі школи А.В. Гончаров відіграв радше роль
продовжувача незавершеного керівництва Кунченка: він фактично довів до
захисту В.В. Філіпова, який починав як аспірант Кунченка. Крім того,
Гончаров був формальним керівником О.М. Ткаченка, тоді як концептуальне
керівництво цією роботою належить С.В. Заболотньому.

\subsubsection{5.2. Друге покоління: дві основні
лінії}\label{ux434ux440ux443ux433ux435-ux43fux43eux43aux43eux43bux456ux43dux43dux44f-ux434ux432ux456-ux43eux441ux43dux43eux432ux43dux456-ux43bux456ux43dux456ux457}

Через два десятиліття після смерті Кунченка школа налічує щонайменше
вісім захищених дисертацій другого покоління. Її основна генеалогічна
структура розгортається за двома вертикальними лініями --- Палагіна і
Заболотнього; поруч із ними існують окремі споріднені захисти, пов'язані
з іншими представниками першого покоління та інституційним оточенням
кафедри. Ця структура відображена в офіційному реєстрі ЧДТУ {[}39{]}.

\textbf{Лінія Палагіна} --- найповніше представлена сигнально-детекторна
гілка:

\begin{itemize}
\item
  \textbf{О.В. Івченко} --- к.т.н. 2015 р., спец. 01.05.02, тема
  «Математичні моделі, методи та засоби оцінювання параметрів негаусових
  корельованих випадкових процесів»; науковий керівник --- В.В. Палагін;
\item
  \textbf{С.А. Лелеко} --- к.т.н. 2018 р., спец. 01.05.02, тема
  «Математичні моделі та методи виявлення сигналів на фоні негаусівських
  завад за моментним критерієм якості»; науковий керівник --- В.В.
  Палагін;
\item
  \textbf{Д.А. Вєдєрніков} --- PhD 2021 р., спец. 122 «Комп'ютерні
  науки», тема «Математичні моделі, методи та засоби оцінювання
  параметра постійного сигналу на фоні негаусових корельованих завад»;
  науковий керівник --- В.В. Палагін {[}16{]};
\item
  \textbf{Д.О. Смірнов} --- PhD 2025 р., спец. 122 «Комп'ютерні науки»,
  тема «Математичні моделі, методи та засоби виявлення постійного
  сигналу на фоні негаусових корельованих завад»; науковий керівник ---
  В.В. Палагін {[}17{]};
\item
  \textbf{О.С. Зорін} --- PhD-захист запланований на 2026 р., спец. 152
  «Метрологія та інформаційно-вимірювальна техніка», тема «Моделі та
  методи оцінювання та виявлення сигналів на фоні негаусових завад в
  інформаційно-вимірювальних системах»; науковий керівник --- В.В.
  Палагін {[}39{]}.
\end{itemize}

\textbf{Лінія Заболотнього} --- регресійно-оцінювальна гілка:

\begin{itemize}
\item
  \textbf{В.О. Селін} --- к.т.н. 2013 р., спец. 01.05.02, тема «Моделі
  та методи нелінійного оцінювання параметрів поліноміальних трендів при
  негаусовій стохастичній компоненті»; науковий керівник --- С.В.
  Заболотній;
\item
  \textbf{А.В. Чепинога} --- к.т.н. 2016 р., спец. 01.05.02, тема
  «Методи поліноміального оцінювання параметрів полігаусових моделей при
  моментно-кумулянтному описі»; науковий керівник --- С.В. Заболотній;
\item
  \textbf{О.М. Ткаченко} --- PhD 2021 р., спец. 122 «Комп'ютерні науки»,
  тема «Поліноміальні методи та засоби оцінювання параметрів регресії з
  використанням моделей негаусових помилок» {[}18{]}. Концептуальне
  керівництво належить С.В. Заболотньому, з яким О.М. Ткаченко спільно
  опублікував усі ключові праці своєї дисертаційної роботи 2017--2021
  рр. {[}40, 41{]}.
\end{itemize}

Окремо в межах другого покоління стоїть дисертація \textbf{В.В.
Філіпова} --- к.т.н. 2016 р., спец. 01.05.02, тема «Методи спільного
оцінювання параметрів постійного сигналу та негаусівських завад з
використанням усічених стохастичних поліномів». Формальним науковим
керівником виступив \textbf{А.В. Гончаров}, але генеалогічно цей захист
слід читати як доведення до завершення роботи аспіранта Кунченка після
смерті засновника школи. Тому він важливий не як окрема вертикальна
лінія, а як приклад спадкоємного завершення теми, започаткованої ще в
кунченківському науковому контексті.

\subsubsection{5.3. Географія міжнародних
колаборацій}\label{ux433ux435ux43eux433ux440ux430ux444ux456ux44f-ux43cux456ux436ux43dux430ux440ux43eux434ux43dux438ux445-ux43aux43eux43bux430ux431ux43eux440ux430ux446ux456ux439}

Міжнародний слід школи реальний, але не масовий. Це радше мережа
цільових двосторонніх колаборацій, ніж широке західне підхоплення
апарату. Опорні точки на карті:

\textbf{Польща} --- найпродуктивніший напрямок. Тандем С.В. Заболотній
(Черкаси) + З.Л. Варша (Промисловий інститут автоматики і вимірювань,
Варшава) дав від 2015 р. близько п'ятнадцяти спільних публікацій,
зокрема, у щорічній серії Springer AISC/LNNS \emph{Automation} (Варшава)
{[}21--25{]}. Тематика --- поліноміальне оцінювання параметрів
вимірювань з асиметричними розподілами, регресія, ARIMA-моделі.

\textbf{Словаччина} --- Технічний університет Кошице. Колаборація В.В.
Палагін (ЧДТУ) + Й. Югар, Л. Вокорокос, С. Марчевський (TU Košice).
Ключові публікації --- стаття в \emph{Journal of Electrical Engineering}
(Bratislava, 2016) {[}26{]} та в \emph{IET Signal Processing} (2017)
{[}27{]}.

\textbf{Німеччина} --- Ганноверський університет, Інститут техніки
зв'язку, де Кунченко був гостем у 2001--2002 рр. Прямий результат візиту
--- публікація канонічної англомовної монографії 2002 р. у Shaker Verlag
(Aachen) {[}9{]}.

\subsubsection{5.4. Інституційні артефакти і меморіальна
інфраструктура}\label{ux456ux43dux441ux442ux438ux442ux443ux446ux456ux439ux43dux456-ux430ux440ux442ux435ux444ux430ux43aux442ux438-ux456-ux43cux435ux43cux43eux440ux456ux430ux43bux44cux43dux430-ux456ux43dux444ux440ux430ux441ux442ux440ux443ux43aux442ux443ux440ux430}

Після смерті Кунченка в 2006 р. в Черкасах створено меморіальну
інфраструктуру школи: Музей історії радіотехніки імені Ю.П. Кунченка (з
2013 р. при кафедрі радіотехніки ЧДТУ); Благодійний фонд «Наукова школа
ім. Юрія Петровича Кунченка», президентом якого була його дружина В.І.
Кунченко-Харченко; меморіальна дошка на будинку, де він жив;
експозиційний фонд у Черкаському обласному краєзнавчому музеї (відкритий
20 травня 2008 р.) {[}7{]}. У 2010-х рр. при кафедрі регулярно
проводилася Міжнародна науково-практична конференція «Обробка сигналів і
негауссівських процесів» (ОСНП), присвячена пам'яті професора Ю.П.
Кунченка. Активно були проведені VII конференції в 2019 р. {[}29{]} та
VIII --- 25--26 травня 2021 р. {[}42{]}. Після 2021 р. конференцію не
проводилось, що зумовлено відомими обставинами війни, які з 2022 р.
суттєво обмежують академічне життя в Україні.

\subsection{6. Три гілки сучасної школи
2006--2026}\label{ux442ux440ux438-ux433ux456ux43bux43aux438-ux441ux443ux447ux430ux441ux43dux43eux457-ux448ux43aux43eux43bux438-20062026}

Внутрішня таксономія школи, зафіксована учнями {[}4{]}, розрізняє дві
концептуальні гілки апарату стохастичних поліномів --- статистичне
оцінювання параметрів через ММПл та перевірку статистичних гіпотез,
причому остання має дві підгілки: розклад логарифма відношення
правдоподібності у стохастичному ряді з оптимальними за моментними
критеріями коефіцієнтами, і розклад у просторі з порідним елементом для
розпізнавання образів. У термінах активних дослідницьких груп 2006--2026
рр. це дає три гілки школи з різними центрами тяжіння.

\subsubsection{6.1. Регресійно-оцінювальна
гілка}\label{ux440ux435ux433ux440ux435ux441ux456ux439ux43dux43e-ux43eux446ux456ux43dux44eux432ux430ux43bux44cux43dux430-ux433ux456ux43bux43aux430}

Найбільш активна у 2018--2026 рр. за обсягом публікацій --- група С.В.
Заболотнього з польським партнерством З.Л. Варші. До групи в Черкасах
належать також прямі аспіранти Заболотнього, чиї дисертації прямо
розгорнули регресійно-оцінювальну лінію: \textbf{В.О. Селін} (2013,
поліноміальні тренди при негаусівських стохастичних компонентах),
\textbf{А.В. Чепинога} (2016, поліноміальне оцінювання параметрів
полігаусових моделей при моментно-кумулянтному описі) та \textbf{О.М.
Ткаченко} (PhD 2021, поліноміальні методи оцінювання параметрів регресії
з негаусовими помилками). Саме роботи Ткаченка разом із Заболотнім і
Варшею вивели цю гілку в сучасну регресійну постановку --- від
оцінювання параметрів розподілів до лінійної, нелінійної та
часово-рядної регресії. Тематичне ядро: поліноміальні оцінки параметрів
за асиметричних і платикуртичних розподілах помилок, з прицілом на
сучасні регресійні моделі та часові ряди.

Ключові публікації: \emph{Polynomial parameter estimation of exponential
power distribution data} (2018) {[}30{]}, \emph{Polynomial estimation of
linear regression parameters for asymmetric PDF} (Springer Automation
2018) {[}22{]}, \emph{Estimation of linear regression parameters for
symmetric non-Gaussian errors} (Springer Automation 2019) {[}23{]},
\emph{Estimating parameters of linear regression with an exponential
power distribution of errors by using a polynomial maximization method}
(2021) {[}31{]}, \emph{Polynomial maximization method for estimation of
asymmetric non-Gaussian MA models} (Springer Automation 2023) {[}24{]},
\emph{Application of PMM for estimating nonlinear regression parameters}
(Springer Automation 2024) {[}25{]}. Окрему лінію складають роботи з
перевірки статистичних гіпотез про середнє в просторі з порідним
елементом --- зокрема, стаття Заболотній--Мартиненко--Салипа в
\emph{Radioelectronics and Communications Systems} (2018) {[}10a{]}.

Найновіша точка цієї гілки --- препринт 2025 р. на arXiv: \emph{Applying
PMM to estimate ARIMA models with asymmetric non-Gaussian innovations}
{[}32{]}. Це перша поява ММПл/PMM на arXiv у форматі повноформатної
дослідницької статті, що символічно відкриває школу для глобальної
статистичної спільноти. У препринті проводиться 128 тисяч симуляцій
Монте-Карло для ARIMA(p,d,q) з гамма-, логнормальними та хі-квадрат
інноваціями, демонструючи виграш PMM2 над класичними оцінювачами на
30--50\% за помірної асиметрії.

Програмною інфраструктурою цієї гілки є R-пакет \textbf{EstemPMM},
опублікований у CRAN у листопаді 2025 р. {[}33{]}. Пакет реалізує PMM2
(для асиметричних розподілів) і PMM3 (для симетричних платикуртичних), з
функцією автоматичного вибору pmm\_dispatch, що оцінює
\(\gamma_{3},\gamma_{4}\) за вибіркою і обирає між OLS, PMM2 і PMM3 за
критерієм \(g_{S}\). Це робить ММПл відтворюваним для широкої наукової
спільноти, не вимагаючи від користувача вручну реалізовувати нормальні
рівняння. До інженерної інфраструктури школи належить також
декларативний патент України на корисну модель «Спосіб генерації
випадкових величин» (Заболотній, Чепинога, Салипа, 2010/2011) {[}33a{]}.

\subsubsection{6.2. Сигнально-детекторна
гілка}\label{ux441ux438ux433ux43dux430ux43bux44cux43dux43e-ux434ux435ux442ux435ux43aux442ux43eux440ux43dux430-ux433ux456ux43bux43aux430}

Центр --- В.В. Палагін, Черкаси, з міжнародною коаліцією (TU Košice ---
Й. Югар, Л. Вокорокос, С. Марчевський; ЧДТУ --- О.В. Івченко, Д.А.
Вєдєрніков, Д.О. Смірнов, О. Палагіна, А. Гончаров, В. Уманець). До
черкаської групи Палагіна, окрім О.В. Івченка, належать: \textbf{С.А.
Лелеко} (к.т.н. 2018, моментний критерій якості виявлення сигналів),
\textbf{Д.А. Вєдєрніков} (PhD 2021, оцінювання параметра постійного
сигналу), \textbf{Д.О. Смірнов} (PhD 2025, виявлення постійного сигналу)
і \textbf{О.С. Зорін} (PhD-захист 2026, метрологічне застосування). Усі
ці дисертації послідовно розгортають апарат стохастичних поліномів на
ширший клас задач сигнальної обробки на корельованому негаусівському тлі
--- від пасивного виявлення до спільного оцінювання параметрів сигналу і
завад.

Ключові публікації: спільні роботи 2014--2017 рр. з кумулянтними
детекторами на корельованих негаусівських даних {[}34{]}; стаття в
\emph{Journal of Electrical Engineering} (2016) про сумісне оцінювання
параметрів сигналів у негаусівському шумі {[}26{]}; стаття в \emph{IET
Signal Processing} (2017) про оцінювання корельованих негаусівських
процесів --- перша поява школи в журналі рівня Q1 {[}27{]}; робота в
\emph{Circuits, Systems, and Signal Processing} (2017) про виявлення
сигналів у корельованому негаусівському шумі через статистики вищих
порядків {[}35{]}; \emph{Mathematical modeling of signal detection in
non-Gaussian correlated noise} (Springer LNNS 2023) {[}36{]};
\emph{Cumulant Detector of Non-Gaussian Signals against Background of
Non-Gaussian Interferences} (Radioelectronics and Communications
Systems, 2024) {[}37{]}.

\subsubsection{6.3. Гілка розпізнавання образів і template
matching}\label{ux433ux456ux43bux43aux430-ux440ux43eux437ux43fux456ux437ux43dux430ux432ux430ux43dux43dux44f-ux43eux431ux440ux430ux437ux456ux432-ux456-template-matching}

Найменша за обсягом, але найоригінальніша за концептуальним ходом гілка.
Центральна ідея --- використання розкладу в просторі з порідним
елементом для побудови ознак, що мінімізують реконструкційну похибку для
конкретного класу сигналів і максимізують для інших класів. Якщо за
класом ознак реконструкція даного об'єкту виявляється точнішою, ніж за
іншими, об'єкт класифікується як належний цьому класу.

Незалежна школа Чертова в НТУ «КПІ» розгорнула цю гілку у двох
паралельних лініях. \textbf{Лінія Чертова-Таврова} (статистичний аналіз
даних, груповий аналіз, мікрофайли): піонерна стаття \emph{Use Kunchenko
Polynomials for the Analysis of the Statistical Data} в
\emph{Eastern-European Journal of Enterprise Technologies} (2010)
{[}38{]}. \textbf{Лінія Чертова-Слипець} (template matching, біомедичні
сигнали): препринт \emph{Kunchenko's Polynomials for Template Matching}
(arXiv 2011, перша поява імені Кунченка на arXiv узагалі) {[}19{]};
стаття \emph{Epileptic seizures diagnose using Kunchenko polynomials
template matching} у Springer ECMI 2012 {[}20{]}. Застосування ---
діагностика епілептичних нападів на ЕЕГ, аналіз статистичних даних, в
перспективі --- біомедичні сигнали загалом.

Внутрішньо в школі ця гілка тісно пов'язана з докторською роботою С.В.
Заболотнього 2015 р. {[}4{]} про ймовірнісне діагностування розладнання
параметрів негаусівських послідовностей: розклад у просторі з порідним
елементом використовується як префільтрація для зменшення дисперсії
процесу перед застосуванням непараметричних тестів зміни розподілу. Це
елегантне поєднання апарату декомпозиції в просторі з порідним елементом
з класичною теорією виявлення розладнань (CUSUM, GRSh,
Колмогорова-Смірнова), яке відкриває можливість гібридних детекторів ---
параметричних усередині класу нормальної роботи, непараметричних за змін
поза класом.

\subsection{7. Школа Кунченка серед суміжних
теорій}\label{ux448ux43aux43eux43bux430-ux43aux443ux43dux447ux435ux43dux43aux430-ux441ux435ux440ux435ux434-ux441ux443ux43cux456ux436ux43dux438ux445-ux442ux435ux43eux440ux456ux439}

Апарат стохастичних поліномів Кунченка не виник у вакуумі. Початок
1970-х рр., коли Юрій Петрович захищав кандидатську дисертацію, був
періодом одночасного розгортання кількох незалежних, але концептуально
близьких напрямків в обробці негаусівських випадкових процесів. Зведення
цих напрямків в єдину карту дозволяє чіткіше побачити унікальну позицію
школи Кунченка та відповісти на запитання: чому, маючи 50-річний апарат,
школа все ще лишається малопомітною в глобальній статистичній
літературі?

\subsubsection{7.1. Класична теорія Вольтера-Вінера: попередник, а не
альтернатива}\label{ux43aux43bux430ux441ux438ux447ux43dux430-ux442ux435ux43eux440ux456ux44f-ux432ux43eux43bux44cux442ux435ux440ux430-ux432ux456ux43dux435ux440ux430-ux43fux43eux43fux435ux440ux435ux434ux43dux438ux43a-ux430-ux43dux435-ux430ux43bux44cux442ux435ux440ux43dux430ux442ux438ux432ux430}

Найочевидніший попередник школи --- класична теорія функціональних рядів
Вольтера-Вінера, оформлена в монографіях W.J. Rugh \emph{Nonlinear
System Theory: The Volterra/Wiener Approach} (1981) {[}43{]}, M.
Schetzen \emph{The Volterra and Wiener Theories of Nonlinear Systems}
(1980) {[}44{]}, V.J. Mathews і G.L. Sicuranza \emph{Polynomial Signal
Processing} (2000) {[}45{]}. Це домінантний західний канон нелінійної
обробки сигналів, у якому ряди Вольтера використовуються переважно для
\textbf{ідентифікації нелінійних систем з пам'яттю} --- каналів
передачі, підсилювачів потужності, акустичних трактів.

Кандидатська Кунченка 1972/1973 рр. виходила саме з цієї теоретичної
бази, але робила концептуальний поворот: ряд Вольтера використовувався
не для ідентифікації системи, а для \textbf{знаходження оцінок невідомих
параметрів випадкових процесів}. Це принципова зміна предметної області
--- від системи до статистичної задачі. Як показано в § 3, цей поворот
відкрив шлях до подальшого узагальнення на стохастичний поліном від
моментних характеристик, що звільняв апарат від прив'язки до фізики
каналу.

Західна школа цей поворот не зробила. Вольтерівські ядра в роботах
Schetzen, Mathews, Sicuranza залишились функціями часових затримок
\(h_{n}\left( \tau_{1},\ldots,\tau_{n} \right)\), а статистичні задачі
(оцінювання параметрів, перевірка гіпотез) розв'язувались окремими
апаратами --- методом найменших квадратів, методом максимальної
правдоподібності, M-оцінюванням. У результаті школа Кунченка і західна
вольтерівська традиція розвивались паралельно, без формальних мостів. До
2026 р. --- року появи статті Гарячого/Щербініна {[}6{]}, яку
розглядаємо в § 8, --- у глобальній літературі немає жодної публікації,
де ряди Вольтера і стохастичні поліноми Кунченка були б розглянуті як
два полюси одного концептуального простору.

\subsubsection{7.2. Вищі порядки статистики: паралельний
шлях}\label{ux432ux438ux449ux456-ux43fux43eux440ux44fux434ux43aux438-ux441ux442ux430ux442ux438ux441ux442ux438ux43aux438-ux43fux430ux440ux430ux43bux435ux43bux44cux43dux438ux439-ux448ux43bux44fux445}

Найближча західна паралель школи Кунченка --- традиція
\textbf{higher-order statistics} (HOS, статистика вищих порядків),
оформлена в роботах J.M. Mendel, C.L. Nikias, A.N. Petropulu, G.B.
Giannakis, А.К. Swami в 1980-х--1990-х рр. {[}46--48{]}. Ця школа
активно використовує кумулянти й моменти третього й вищого порядків для
аналізу негаусівських сигналів, із застосуваннями до сліпої
деконволюції, ідентифікації систем, виявлення сигналів у негаусівському
шумі.

Концептуальний перетин зі школою Кунченка глибокий. Обидва напрямки
виходять з того самого спостереження: для негаусівських процесів
обмеження лінійним описом і другими моментами втрачає принципову частину
інформації; вищі моменти й кумулянти несуть статистичну інформацію, яку
не можна замінити іншими засобами. Обидва напрямки використовують ці
моменти для побудови ефективніших оцінювачів і детекторів.

Розрізнення проходить по способу утилізації моментної інформації. Школа
HOS оперує переважно у \textbf{спектральному домені} --- біспектри,
триспектри, поліспектри як прямі узагальнення енергетичного спектра на
вищі порядки. Школа Кунченка оперує у \textbf{просторі поліноміальних
функцій від моментів}, із вибором базису (\(\xi^{i}\), \(\cos(ik\xi)\)
тощо), оптимізацією коефіцієнтів за моментними критеріями і явною
роботою з тілом стохастичного поліному як квадратичною формою. Це робить
апарат Кунченка ближчим до класичного методу моментів, але з істотним
нелінійним розширенням через поліноміальну структуру.

Прямих формальних мостів між цими двома школами в літературі мало.
Окремі виходи школи Кунченка в журнали традиційного HOS-профілю
(\emph{Circuits, Systems, and Signal Processing} 2017 {[}35{]},
\emph{IET Signal Processing} 2017 {[}27{]}) є радше точковими, ніж
систематичною інтеграцією. Це залишається відкритою прогалиною:
систематичне порівняння стохастичного поліному Кунченка зі спектральними
HOS-оцінювачами у сучасній обробці сигналів --- потенційна тема окремого
дослідження.

\subsubsection{7.3. Робастна статистика і M-оцінювачі
Хубера}\label{ux440ux43eux431ux430ux441ux442ux43dux430-ux441ux442ux430ux442ux438ux441ux442ux438ux43aux430-ux456-m-ux43eux446ux456ux43dux44eux432ux430ux447ux456-ux445ux443ux431ux435ux440ux430}

Ще одна паралель --- традиція робастного оцінювання, започаткована P.J.
Huber у роботі \emph{Robust Estimation of a Location Parameter} (1964) і
розвинута в монографії \emph{Robust Statistics} (1981) {[}49{]}.
Робастні методи будуються на критерії мінімізації суми ψ-функцій від
залишків, з вибором ψ-функції так, щоб обмежити вплив викидів і
відхилень від нормальності.

Спорідненість зі школою Кунченка --- на рівні методологічної установки:
обидва підходи стійкі до відхилень від гаусівської ідеалізації. Різниця
ж --- у формі параметризації цієї стійкості. M-оцінювачі Хубера
використовують ψ-функції, що задаються апріорно і не залежать від
конкретної форми негаусівського розподілу даних. ММПл Кунченка, навпаки,
використовує \textbf{інформацію про розподіл}, але в обмеженій формі ---
через моменти і кумулянти певного порядку. Це робить ММПл потенційно
ефективнішим (за умови, що моменти оцінено точно), а робастні методи ---
потенційно стійкішими (за умови, що моментна структура може бути
спотворена грубими викидами).

Систематичних порівнянь ефективності цих двох підходів на однакових
даних у літературі небагато. У сучасних роботах школи {[}31, 32{]}
заявляється, що ММПл-2 не програє МНК при гаусівських інноваціях і дає
виграш при асиметричних розподілах, що ставить його в одну вагову
категорію з робастними оцінювачами. Однак прямих бенчмарків ММПл vs
Huber M-estimator vs L-моменти Хосейна на сучасних реалістичних
датасетах поки немає. Це ще одна біла пляма школи (§ 11).

\subsubsection{7.4. Метод узагальнених моментів (GMM Hansen) і
L-моменти}\label{ux43cux435ux442ux43eux434-ux443ux437ux430ux433ux430ux43bux44cux43dux435ux43dux438ux445-ux43cux43eux43cux435ux43dux442ux456ux432-gmm-hansen-ux456-l-ux43cux43eux43cux435ux43dux442ux438}

L. Hansen у класичній роботі 1982 р. \emph{Large Sample Properties of
Generalized Method of Moments Estimators} {[}50{]} оформив метод
узагальнених моментів (Generalized Method of Moments, GMM) --- апарат,
що став стандартом сучасної економетрики. У GMM оцінка знаходиться як
розв'язок системи моментних умов з ваговою матрицею, оптимізованою для
мінімізації асимптотичної дисперсії. Це фактично узагальнення класичного
методу моментів на випадок, коли число моментних умов перевищує число
параметрів, що оцінюються.

Ідейна спорідненість GMM і ММПл очевидна: обидва експлуатують вищі
моменти, обидва дають оцінки, які асимптотично прагнуть до межі
ефективності, обидва не вимагають повного знання функції розподілу.
Розрізнення в техніці: GMM використовує лінійні (або квазілінійні)
функціонали від моментних умов, тоді як ММПл будує \textbf{поліноміальні
функції від моментів} з оптимізованими за дисперсійним критерієм
коефіцієнтами. У певному сенсі ММПл можна інтерпретувати як специфічну
форму GMM, у якій вагова матриця Hansen замінена на матрицю
\(F_{S}\left( \overrightarrow{\theta} \right)\) центрованих корелянтів
стохастичного поліному. Однак це інтерпретаційний хід, який у літературі
поки не оформлено формальним містком.

Близьким до цього апарату є \textbf{L-моменти J. Hosking} (1990)
{[}51{]} --- лінійні комбінації порядкових статистик, що дають
альтернативну параметризацію негаусівських розподілів. L-моменти
стійкіші до викидів, ніж класичні моменти, і використовуються переважно
в гідрології, метеорології, актуарній статистиці. Школа Кунченка з
L-моментами формальних мостів не має; ймовірність систематичної
інтеграції цих апаратів --- відкрита перспектива.

Близькою до GMM, але концептуально ще ближчою до апарату Кунченка є
західна традиція \textbf{SLS (Second-order Least Squares)},
започаткована L. Wang у 2003 р. і систематизована в наступні два
десятиліття як стандарт для нелінійних регресійних задач з
негаусівськими помилками. Зв'язок PMM2 ↔ SLS має не тільки
концептуальну, а й кількісну природу --- обидва методи дають практично
ідентичні MSE на спільних задачах {[}25{]}. Це робить SLS найбільш
безпосередньою точкою формального мосту між західною статистичною
літературою і апаратом школи Кунченка, тому розгляд цієї паралелі ми
виносимо в окремий підрозділ § 7.5.

\subsubsection{7.5. Second-order Least Squares (SLS): паралельна
траєкторія
Заходу}\label{second-order-least-squares-sls-ux43fux430ux440ux430ux43bux435ux43bux44cux43dux430-ux442ux440ux430ux454ux43aux442ux43eux440ux456ux44f-ux437ux430ux445ux43eux434ux443}

Найбільш концептуально близький західний аналог апарату Кунченка, який
не входить до чотирьох попередніх підрозділів цього параграфу, --- метод
\textbf{другопорядкових найменших квадратів} (Second-order Least
Squares, SLS), започаткований L. Wang у 2003 р. на матеріалі нелінійних
регресійних моделей з помилками вимірювання Берксонівського типу
{[}52{]} і систематизований Wang і Leblanc у 2008 р. як універсальна
процедура нелінійного оцінювання за асиметричних помилок {[}53{]}. У
наступному десятилітті метод отримав теоретичне розгортання --- аналіз
асимптотичної ефективності {[}54{]}, нові оптимізаційні критерії
{[}55{]}, розширення на ARCH-моделі і часові ряди {[}56{]}, адаптацію до
моделей з множинними кривими {[}57{]}. Сьогодні SLS --- стандартний
інструмент сучасної економетрики й біостатистики для нелінійних задач з
негаусівськими помилками.

Концептуальна ідея SLS вкладається в одне речення: до критерію
оптимізації стандартного OLS, який спирається тільки на перший умовний
момент \(E\left\lbrack y|x \right\rbrack = R(\theta,x)\), додається
другий ---
\(E\left\lbrack y^{2}|x \right\rbrack = R^{2}(\theta,x) + \sigma^{2}\)
--- і параметри \(\theta\) оцінюються через мінімізацію комбінованого
функціоналу від відхилень обох моментів. У формальному вигляді:

\[
{\widehat{\theta}}_{SLS} = \arg\min_{\theta}\sum_{v = 1}^{N}\left\lbrack \left( y_{v} - R\left( \theta,x_{v} \right) \right)^{2} + \omega \cdot \left( y_{v}^{2} - R^{2}\left( \theta,x_{v} \right) - \sigma^{2} \right)^{2} \right\rbrack,
\]

з ваговим коефіцієнтом \(\omega\), що оптимізується для асимптотичної
ефективності.

Концептуальна спорідненість з ММПл степеня \(S = 2\) (PMM2) школи
Кунченка глибока і структурно симетрична. Обидва методи виходять з того
самого спостереження: для негаусівських асиметричних помилок класичний
OLS втрачає принципову частину інформації, бо ігнорує другий і третій
моменти відхилень. Обидва методи компенсують це додаванням моментної
інформації --- SLS через додатковий квадратичний доданок у функціоналі
втрат, PMM2 через побудову стохастичного поліному 2-го степеня з
оптимізованими за моментно-кумулянтним описом коефіцієнтами. Обидва
редукуються до OLS у гаусівському випадку (де \(\gamma_{3} = 0\)) і
дають значний виграш у дисперсії оцінок при ненульовій асиметрії.

Конкретне порівняння ефективності SLS і PMM2 на спільних задачах
виконано у нещодавній спільній роботі групи Заболотнього з польськими
колегами {[}25{]} на двох нелінійних регресійних моделях з
χ²-розподіленими помилками: експоненціальній
\(y = \theta_{1}\exp\left( \theta_{2}x \right) + \xi\) і моделі росту
\(y = \theta_{1}/\left\lbrack 1 + exp\left( \theta_{2} + \theta_{3}x \right) \right\rbrack + \xi\).
Monte Carlo моделювання з обсягами вибірок \(N = 30,50,100,200\) і
повторами \(m = 1000\) дало результат, який можна підсумувати в одному
реченні: \textbf{PMM2 і SLS дають практично ідентичні MSE-метрики}
(різниця в межах 1--3\% залежно від параметра і розміру вибірки), і
обидва значно перевершують OLS --- від 30\% виграшу при \(N = 30\) до
50\% виграшу при \(N = 200\), що збігається з теоретичним прогнозом
коефіцієнта варіаційної редукції \(g_{2} \approx 0.56\) для
χ²(3)-помилок.

Самі автори {[}25{]} інтерпретують цей збіг як \textbf{наслідок
використання однакової інформації}: «обидва оцінювачі використовують ту
саму додаткову інформацію про ймовірнісну природу випадкової складової
регресійної моделі у формі моментів регресійних залишків до 4-го
порядку». Це коректне технічне пояснення, але воно одночасно фіксує і
\textbf{глибший факт}: SLS і PMM2 --- це \textbf{дві різні математичні
дороги до того самого практичного результату}, які розвивались
паралельно і незалежно протягом двадцяти років. SLS --- у західній
економетричній традиції, починаючи з 2003 р.; PMM2 --- у черкаській
школі Кунченка, з канонічною монографією 2002 р. і попередніми
результатами ще з 1990-х. Обидві традиції розв'язують ту саму задачу і
отримують той самий результат, не знаючи одна про одну.

Це робить пару PMM2 ↔ SLS винятково привабливою точкою для побудови
\textbf{формального мосту} між школою Кунченка і сучасною світовою
статистикою. Концептуальна паралель є; емпіричне порівняння в {[}25{]}
вже зроблено; систематичне теоретичне зведення --- наступний природний
крок (див. § 11.2). Перевага PMM перед SLS, яку потенційно можна
формально довести, --- у \textbf{здатності до масштабування}: тоді як
SLS стандартно обмежений другим порядком (можливі узагальнення на третій
порядок існують у літературі, але не отримали систематичного розвитку),
PMM природно поширюється на \(S = 3,4,\ldots\) через перфорацію
кумулянтного опису і вибір базисних функцій. Це дає школі Кунченка
теоретичний резерв ефективності, яким SLS не володіє.

\subsubsection{7.6. Семіпараметрична позиція
школи}\label{ux441ux435ux43cux456ux43fux430ux440ux430ux43cux435ux442ux440ux438ux447ux43dux430-ux43fux43eux437ux438ux446ux456ux44f-ux448ux43aux43eux43bux438}

Зведення всіх перерахованих напрямків в єдину карту виявляє унікальну
позицію школи Кунченка серед методів статистичної обробки {[}4, рис.
1.4{]}. Класична схема ділить ці методи на три категорії:

\begin{itemize}
\item
  \textbf{Параметричні методи} (методи правдоподібності, баєсові
  процедури) --- потребують повної апріорної інформації про закон
  розподілу; забезпечують оптимальні оцінки за умови виконання
  припущень; критично чутливі до неправильної специфікації моделі.
\item
  \textbf{Непараметричні методи} (тести Колмогорова-Смірнова,
  Манна-Уітні; ранг-методи; CUSUM, GRSh у непараметричній версії) --- не
  потребують апріорної інформації про розподіл; стійкі до відхилень від
  моделі; ефективність часто далека від оптимальних значень.
\item
  \textbf{Семіпараметричні методи} (стохастичні поліноми Кунченка +
  моментно-кумулянтний опис) --- використовують \textbf{обмежений} обсяг
  апріорної інформації у вигляді послідовності моментів або кумулянтів;
  забезпечують компроміс між обсягом необхідної інформації,
  обчислювальною складністю і точністю; підтримують самонавчання й
  адаптивність.
\end{itemize}

Саме семіпараметрична позиція робить апарат Кунченка концептуально
цінним для сучасних задач, де повної апріорної інформації немає, але
повністю ігнорувати відомості про вищі моменти теж було б марнотратно.
Це позиціонування школи, чітко зафіксоване в {[}4{]}, повністю
узгоджується з траєкторією робастно-семіпараметричної революції в
сучасній статистиці (квазі-правдоподібність, GMM, M-оцінювачі, SLS) ---
однак сама школа цей зв'язок поки систематично не використовує для свого
міжнародного позиціонування.

\subsection{8. Випадок 2026: повторне відкриття вихідної
задачі}\label{ux432ux438ux43fux430ux434ux43eux43a-2026-ux43fux43eux432ux442ux43eux440ux43dux435-ux432ux456ux434ux43aux440ux438ux442ux442ux44f-ux432ux438ux445ux456ux434ux43dux43eux457-ux437ux430ux434ux430ux447ux456}

У 2026 р. в українському науково-технічному журналі «Вимірювальна та
обчислювальна техніка в технологічних процесах» (ВОТТП, № 1, с. 47--60)
опубліковано статтю М. Гарячого та С. Щербініна (Науковий центр
Повітряних Сил Харківського національного університету Повітряних Сил
імені Івана Кожедуба) під назвою «Метод формування стохастичних сигналів
із використанням рядів Вольтера» {[}6{]}. Стаття концентровано
демонструє стан, у якому опинилась прикладна українська радіотехнічна
школа щодо концептуального інструментарію черкаської наукової школи
Кунченка через 50 років після її фундаторської кандидатської дисертації.
Розгляд цієї статті важливий не для критики авторів --- її автори
представляють серйозний оборонний радіотехнічний центр і працюють над
цілком реальною задачею --- а як \textbf{симптоматичний приклад} ширшої
тенденції, наслідки якої ми обговорюємо в § 11.

\subsubsection{8.1. Зміст статті 2026
р.}\label{ux437ux43cux456ux441ux442-ux441ux442ux430ux442ux442ux456-2026-ux440.}

Гарячий і Щербінін розв'язують задачу формування і обробки стохастичних
сигналів для інформаційно-керуючих систем в умовах нелінійних
спотворень, ефектів пам'яті каналу і обмеженого частотного ресурсу.
Класичні лінійні методи фільтрації, як зазначають автори, не можуть
компенсувати нелінійні спотворення і ефекти пам'яті, тому пропонується
перейти до апарату рядів Вольтера 2-го порядку з адаптивним
налаштуванням параметрів ядра.

Дискретна модель Вольтера, що використовується в статті, має вигляд:

\[
y\lbrack n\rbrack = h_{0} + \sum_{i = 0}^{M_{1} - 1}h_{1}\lbrack i\rbrack\, x\lbrack n - i\rbrack + \sum_{i = 0}^{M_{2} - 1}\sum_{j = 0}^{M_{2} - 1}h_{2}\lbrack i,j\rbrack\, x\lbrack n - i\rbrack\, x\lbrack n - j\rbrack,
\]

де \(h_{1}\lbrack i\rbrack\), \(h_{2}\lbrack i,j\rbrack\) ---
коефіцієнти ядер, \(M_{1},M_{2}\) --- глибини пам'яті відповідних ядер.
Адаптація коефіцієнтів виконується за критерієм MMSE --- мінімуму
середньоквадратичної помилки:

\[
MMSE = \min_{\{ h_{1},h_{2}\}}E\left\lbrack \left( y\lbrack n\rbrack - \widehat{y}\lbrack n\rbrack \right)^{2} \right\rbrack.
\]

На кожному кроці адаптації обчислюється локальна похибка, і ваги ядер
оновлюються через систему нормальних рівнянь
\(C_{x} \cdot \overrightarrow{h} = {\overrightarrow{r}}_{yx}\), де
\(C_{x}\) --- коваріаційна матриця входу, \({\overrightarrow{r}}_{yx}\)
--- крос-кореляційний вектор.

Чисельне моделювання, виконане авторами, дає цікавий результат. Для
структурованих сигналів (OFDM, 64 піднесучі, QAM-16, частота
дискретизації 960 кГц), які мають виражену гармонійну структуру, метод
Вольтера 2-го порядку забезпечує \textbf{зменшення ефективної ширини
спектра на 1.9--8.6\%} і відповідне зростання спектральної ефективності
на 1.9--5\%. Однак для широкосмугових стохастичних (шумоподібних)
сигналів той самий метод дає \textbf{збільшення ефективної ширини
спектра приблизно на 15\%}, з відповідним падінням спектральної
ефективності на \textasciitilde13\% {[}6, табл. 1{]}. Автори чесно
фіксують це обмеження і формулюють висновок: метод Вольтера не є
універсальним, і для його стабільної роботи на сигналах з різною
статистикою потрібно «декорелювання та зменшення спектральної
надмірності».

\subsubsection{8.2. Концептуальна
реконструкція}\label{ux43aux43eux43dux446ux435ux43fux442ux443ux430ux43bux44cux43dux430-ux440ux435ux43aux43eux43dux441ux442ux440ux443ux43aux446ux456ux44f}

Не входячи в технічні деталі, концептуальний зміст статті {[}6{]} 2026
р. можна реконструювати в чотири пункти.

\textbf{Перше.} Постановка задачі --- Вольтера-обробка стохастичного
сигналу з адаптацією параметрів ядра за статистичними характеристиками
входу --- формально перегукується з типом задачі кандидатської Кунченка
1972/1973 р., де ряди Вольтера застосовувалися для знаходження оцінок
невідомих параметрів випадкових процесів. Через 50 років українські
радіотехнічні дослідники незалежно повертаються до близької
концептуальної точки: нелінійний функціонал від випадкового процесу
використовується як інструмент оцінювання або адаптації.

\textbf{Друге.} Метод адаптації --- нормальні рівняння
\(C_{x} \cdot \overrightarrow{h} = {\overrightarrow{r}}_{yx}\) --- має
споріднену лінійно-алгебраїчну форму з рівняннями для коефіцієнтів
поліноміальних оцінювачів, але оптимізує інший об'єкт. У статті 2026 р.
це MMSE-регресія для ядра Вольтера-фільтра: параметрами є
\(h_{1},h_{2}\), а критерієм є похибка виходу моделі. У зрілому апараті
Кунченка параметром є характеристика процесу або розподілу, а матриця
центрованих корелянтів входить у моментну процедуру оцінювання. Саме ця
відмінність не дозволяє ототожнювати VOTTP-підхід із ММПл, але робить
його природною точкою для такого узагальнення.

\textbf{Третє.} Виявлене авторами обмеження --- 13\% спектральна
деградація на шумоподібних сигналах --- можна інтерпретувати як симптом
недостатнього узгодження ядра з моментною структурою сигналу. Метод
оптимізує ядра за L2-критерієм, який працює через коваріаційну проєкцію
і не використовує прямо асиметрію \(\gamma_{3}\), ексцес \(\gamma_{4}\)
та інші характеристики вищих порядків. Апарат Кунченка підказує іншу
можливу траєкторію: будувати адаптацію не лише за коваріаціями, а й за
матрицею центрованих корелянтів базисних функцій. За відповідних
моментних умов і невиродженості цієї матриці така процедура може дати
редукцію дисперсії оцінок у тих класах негаусівських розподілів, де
\(g_{2}<1\).

\textbf{Четверте.} Список літератури статті містить 20 посилань,
переважно класичних: Rugh 1981 {[}43{]}, Stenger і Rabenstein, Favier,
Orcioni, Gibiino, Zhu, Chang, Gariachy і Shcherbinin (їхня попередня
робота 2025), Kapustii, Karhunen-Loève transform, Wheeler і Holder,
Sosulin, Kharkevich, Pugachov. \textbf{Жодного посилання на школу
Кунченка немає.} Жодної згадки про монографію Shaker 2002, жодної згадки
про доповідь IEEE ISIT 1997. Українські автори в українському
радіотехнічному журналі цитують класичну літературу, але не Кунченка
2002 р. з Aachen --- попри те, що Кунченко був присутнім у
IEEE-літературі вже з 1997 р. Однак вони пропускають не тільки школу
Кунченка, але й \textbf{її західний аналог} --- традицію SLS
(Second-order Least Squares), яка з 2003 р. систематично розвивається в
західній економетриці й біостатистиці саме для задач, концептуально
близьких до тієї, що розв'язується в {[}6{]} (див. § 7.5). Це підкреслює
системний характер інфраструктурного розриву: між апаратом, що реально
потрібен для розв'язання поставленої задачі, і поточним інструментарієм
українських прикладних радіотехнічних дослідників лежить розрив, який не
закривається не тільки локальною (черкаською) школою, але й її
глобальним західним аналогом.

\subsubsection{8.3. Чому це симптом, а не
докір}\label{ux447ux43eux43cux443-ux446ux435-ux441ux438ux43cux43fux442ux43eux43c-ux430-ux43dux435-ux434ux43eux43aux456ux440}

Підкреслю: ситуація не є виною авторів. Гарячий і Щербінін представляють
харківський радіофізичний центр --- той самий, у спеціалізованій вченій
раді якого Кунченко захищав у 1988 р. свою докторську дисертацію. Між
цими двома точками --- Харків 1988 і Харків 2026 --- лежать десятиліття,
упродовж яких канон школи був недостатньо вписаний у мейнстрімну
українську радіотехнічну освіту. Це структурна, а не персональна
проблема.

Симптом полягає у такому. Школа Кунченка, попри 50-річну історію, шість
монографій, багатопоколінну дисертаційну структуру (§ 5), вихід на
міжнародні журнали (§ 6) і присутність в IEEE-літературі починаючи з
1997 р. {[}12a{]}, залишається \textbf{локально невидимою у власному
прикладному середовищі}. Українські радіотехнічні дослідники, працюючи
над задачами, для яких апарат школи буквально був створений, не знають
про його існування. Це не може бути списане на «недостатню рекламу» ---
школа дала Shaker Verlag 2002, IET Signal Processing 2017, arXiv 2025
р., публічний R-пакет EstemPMM на CRAN. Проблема глибша: між апаратом
школи і його прикладними споживачами в Україні існує
\textbf{інфраструктурний розрив}, наслідки якого ми обговорюємо в § 11.

Стаття Гарячого/Щербініна 2026 р. --- найпряміша діагностична точка
цього розриву. І саме тому вона стає природною мотивацією для нашої
меморіальної статті: не для того, щоб вказати на недогляд авторів, а для
того, щоб \textbf{через демонстрацію формального мосту} між їхньою
задачею і апаратом школи закрити цей розрив бодай на одну точку.

Цей міст ми будуємо в § 9.

\subsection{9. Математичний міст: модель Вольтера як стохастичний
поліном і різниця критеріїв
оцінювання}\label{ux43cux430ux442ux435ux43cux430ux442ux438ux447ux43dux438ux439-ux43cux456ux441ux442-ux43cux43eux434ux435ux43bux44c-ux432ux43eux43bux44cux442ux435ux440ux430-ux44fux43a-ux441ux442ux43eux445ux430ux441ux442ux438ux447ux43dux438ux439-ux43fux43eux43bux456ux43dux43eux43c-ux456-ux440ux456ux437ux43dux438ux446ux44f-ux43aux440ux438ux442ux435ux440ux456ux457ux432-ux43eux446ux456ux43dux44eux432ux430ux43dux43dux44f}

Цей розділ є методологічним ядром статті. Його мета --- не лише провести
історичну паралель між рядами Вольтера і стохастичними поліномами
Кунченка, а формально показати, що клас скінченних Вольтера-моделей
природно вкладається в клас узагальнених стохастичних поліномів на
векторному аргументі. Це вкладення є твердженням про базис і форму
моделі, а не твердженням про тотожність методів. Після цього можна чітко
розділити два критерії: MMSE/L2 як фіксовану коваріаційну проєкцію для
адаптації ядра і ММПл як параметрично залежну моментну процедуру
оцінювання параметра процесу або розподілу.

\subsubsection{9.1. Твердження про вкладення
Вольтера-моделі}\label{ux442ux432ux435ux440ux434ux436ux435ux43dux43dux44f-ux43fux440ux43e-ux432ux43aux43bux430ux434ux435ux43dux43dux44f-ux432ux43eux43bux44cux442ux435ux440ux430-ux43cux43eux434ux435ux43bux456}

\textbf{Твердження 1 (Вольтера-модель як узагальнений стохастичний
поліном).} Нехай спостерігається дискретний процес \(x[n]\), для якого
існують моменти порядку, достатнього для побудови всіх добутків до
степеня \(N\). Нехай також задано скінченну пам'ять \(M\) і вектор
спостережень

\[
\overrightarrow{x}_{n} = (x[n],x[n-1],\ldots,x[n-M+1])^{T}.
\]

Тоді будь-яка скінченна модель Вольтера степеня \(N\) з пам'яттю \(M\)
може бути записана як узагальнений стохастичний поліном Кунченка на
векторному аргументі \(\overrightarrow{x}_{n}\) з базисом мономіальних
добутків

\[
\{\varphi_{k}\} = \left\{x[n-i_{1}]x[n-i_{2}]\cdots x[n-i_{p}]:\; p=1,\ldots,N,\;0\leq i_{1}\leq\cdots\leq i_{p}\leq M-1\right\}.
\]

Кількість базисних функцій дорівнює

\[
S = \sum_{p=1}^{N}\binom{M+p-1}{p}.
\]

Для випадку \(N=2\) маємо \(S=M+M(M+1)/2\), якщо враховується симетрія
квадратичного ядра.

Доведення є прямим переписуванням Вольтера-розкладу як лінійної
комбінації базисних функцій від вектора \(\overrightarrow{x}_{n}\). Для
другого порядку дискретна модель має вигляд

\[
y[n] = h_{0}+\sum_{i=0}^{M_{1}-1}h_{1}[i]x[n-i]+\sum_{i=0}^{M_{2}-1}\sum_{j=0}^{M_{2}-1}h_{2}[i,j]x[n-i]x[n-j].
\]

Після введення \(M=\max(M_{1},M_{2})\) вона переписується як

\[
y[n] = h_{0}+\sum_{i=0}^{M-1}h_{1}[i]\varphi_{1}^{(i)}(\overrightarrow{x}_{n})+\sum_{i,j=0}^{M-1}h_{2}[i,j]\varphi_{2}^{(i,j)}(\overrightarrow{x}_{n}),
\]

де \(\varphi_{1}^{(i)}(\overrightarrow{x}_{n})=x[n-i]\), а
\(\varphi_{2}^{(i,j)}(\overrightarrow{x}_{n})=x[n-i]x[n-j]\). Отже, на
рівні базисного представлення Вольтера-модель є окремим випадком
узагальненого стохастичного поліному з мономіальним базисом і векторним
аргументом.

\subsubsection{9.2. Різниця між MMSE/L2 і
ММПл-критерієм}\label{ux440ux456ux437ux43dux438ux446ux44f-ux43cux456ux436-mmsel2-ux456-ux43cux43cux43fux43b-ux43aux440ux438ux442ux435ux440ux456ux454ux43c}

\textbf{Твердження 2 (критеріальна відмінність).} MMSE-адаптація
Вольтера-ядра і ММПл-оцінювання використовують подібну
лінійно-алгебраїчну форму нормальних рівнянь, але оптимізують різні
статистичні об'єкти.

У MMSE-формулюванні коефіцієнти ядра знаходяться з задачі

\[
\min_{h}E\{(y[n]-\widehat{y}[n])^{2}\},
\]

що приводить до системи

\[
\mathbf{C}_{x}\overrightarrow{h}=\overrightarrow{r}_{yx},
\]

де \(\mathbf{C}_{x}\) є коваріаційною матрицею спостережуваних базисних
функцій, а \(\overrightarrow{r}_{yx}\) --- крос-кореляційним вектором.
Це L2-проєкція на фіксований базис.

У ММПл-формулюванні оптимальні коефіцієнти визначаються через
параметрично залежну матрицю центрованих корелянтів

\[
\mathbf{F}_{S}(\theta)\overrightarrow{h}^{*}=\overrightarrow{b}(\theta),
\]

де елементи \(\mathbf{F}_{S}(\theta)\) побудовані з моментів базисних
функцій, а вектор \(\overrightarrow{b}(\theta)\) пов'язаний з похідними
моментних характеристик за параметром, який оцінюється. Тому ММПл не є
просто альтернативним способом розв'язати ту саму least-squares задачу:
він змінює сам критерій, переводячи задачу з фіксованої L2-проєкції у
параметрично адаптоване моментне оцінювання.

Для оцінювання середнього \(\theta=\mu\) у класах розподілів, для яких
виконуються стандартні моментні умови та умови регулярності ММПл,
асимптотична дисперсія оцінки другого степеня може бути записана через
коефіцієнт варіаційної редукції

\[
\sigma_{\widehat{\mu},\mathrm{PMM2}}^{2}=\frac{c_{2}}{N}g_{2},\qquad g_{2}=1-\frac{\gamma_{3}^{2}}{2+\gamma_{4}}.
\]

Цю формулу слід читати умовно: твердження про виграш ММПл є коректним
для тих негаусівських класів, де матриця корелянтів невироджена,
потрібні моменти існують, а \(g_{2}<1\). У такій формі результат є
твердженням про клас оцінювачів та їхню асимптотичну ефективність, а не
універсальною декларацією для будь-якого негаусівського розподілу без
умов.

\subsubsection{9.3. Наслідок для задачі адаптації
Вольтера-ядра}\label{ux43dux430ux441ux43bux456ux434ux43eux43a-ux434ux43bux44f-ux437ux430ux434ux430ux447ux456-ux430ux434ux430ux43fux442ux430ux446ux456ux457-ux432ux43eux43bux44cux442ux435ux440ux430-ux44fux434ux440ux430}

Розглянута у статті {[}6{]} задача має саме такий методологічний зміст.
Для структурованих сигналів типу OFDM MMSE-адаптація ядра Вольтера
працює задовільно, оскільки другий порядок компенсує частину перехресних
гармонічних взаємодій. Для шумоподібних широкосмугових сигналів той
самий L2-критерій може погіршувати спектральну ефективність, бо не
використовує моментно-кумулянтну структуру сигналу.

Математична можливість, яку відкриває апарат ММПл, полягає в тому, щоб
адаптувати ядро Вольтера не лише за коваріаційною матрицею входу, а за
параметричною матрицею центрованих корелянтів базисних функцій, оціненою
для відповідного класу негаусівського процесу. Гіпотеза, яку треба
перевірити чисельно, така: для сигналів з істотно ненульовими моментами
вищих порядків ММПл-адаптація Вольтера-ядра може зменшувати дисперсію
оцінок порівняно з MMSE/L2-адаптацією, а отже потенційно усувати
спектральну деградацію, зафіксовану в {[}6{]}.

Це не слід подавати як завершений емпіричний результат: у поточній
статті побудовано формальний міст і методологічну гіпотезу. Її
підтвердження вимагає Monte Carlo порівняння на тих самих типах
сигналів, що використовувались у {[}6{]}: OFDM, фільтрований білий шум,
гібридний гармонічно-шумовий сигнал, різні рівні нелінійності та пам'яті
каналу.

\subsubsection{9.4. Методологічний
висновок}\label{ux43cux435ux442ux43eux434ux43eux43bux43eux433ux456ux447ux43dux438ux439-ux432ux438ux441ux43dux43eux432ux43eux43a}

Звідси випливає обережний, але продуктивний висновок: апарат школи
Кунченка може бути перенесений на Вольтера-моделі лише після явної
побудови відповідного базису, матриці центрованих корелянтів,
моментно-кумулянтних характеристик, перевірки невиродженості, порівняння
дисперсій і валідації на benchmark-сценаріях. Саме така постановка
зміщує центр ваги статті з історичного нагадування на формалізацію класу
оцінювачів, критеріїв і перевірюваних умов їхньої ефективності.

\subsection{10. Дослідницька програма: від історичного мосту до
перевірюваних
моделей}\label{ux434ux43eux441ux43bux456ux434ux43dux438ux446ux44cux43aux430-ux43fux440ux43eux433ux440ux430ux43cux430-ux432ux456ux434-ux456ux441ux442ux43eux440ux438ux447ux43dux43eux433ux43e-ux43cux43eux441ux442ux443-ux434ux43e-ux43fux435ux440ux435ux432ux456ux440ux44eux432ux430ux43dux438ux445-ux43cux43eux434ux435ux43bux435ux439}

Формальний міст, побудований у § 9, переводить історичний огляд у робочу
програму. Її логіка проста: спочатку треба перевірити, чи дає заміна
критерію реальний виграш на тих самих сигналах, де класична
Вольтера-адаптація показала обмеження; далі --- розширити базис, щоб
описувати перехід між структурованими й шумоподібними сигналами; нарешті
--- поєднати оцінювальну і детекторну гілки школи в задачі виявлення
зміни режиму. Нижче ці кроки подано як послідовність перевірюваних
моделей, а не як декларацію готових результатів.

\subsubsection{10.1. ММПл-адаптація ядра Вольтера як моментне
узагальнення
MMSE-критерію}\label{ux43cux43cux43fux43b-ux430ux434ux430ux43fux442ux430ux446ux456ux44f-ux44fux434ux440ux430-ux432ux43eux43bux44cux442ux435ux440ux430-ux44fux43a-ux43cux43eux43cux435ux43dux442ux43dux435-ux443ux437ux430ux433ux430ux43bux44cux43dux435ux43dux43dux44f-mmse-ux43aux440ux438ux442ux435ux440ux456ux44e}

Найближча задача --- пряма експериментальна верифікація тези § 9.3.
Робоча гіпотеза формулюється обережно: заміна L2-критерію в адаптації
ядра Вольтера 2-го порядку на ММПл-критерій з оптимізованими за
моментами вищих порядків коефіцієнтами може усунути спектральну
деградацію на шумоподібних сигналах у тих класах процесів, де
виконуються моментні умови ММПл і спостерігається варіаційна редукція.

Експериментальний дизайн відтворює умови статті {[}6{]}: OFDM (64
піднесучі, QAM-16, 960 кГц), фільтрований білий шум (LPF з нормованою
частотою зрізу 0.4), гібридний сигнал (гармонічний + шум), при різних
рівнях нелінійності каналу і ефектах пам'яті. Порівнюються два варіанти
адаптації: базові MMSE-нормальні рівняння і моментне узагальнення, у
якому система ММПл-рівнянь
\(F_{S}\left( \overrightarrow{\theta} \right) \cdot {\overrightarrow{h}}^{*} = \overrightarrow{b}\left( \overrightarrow{\theta} \right)\)
будується з оцінюванням \(\gamma_{3},\gamma_{4}\) безпосередньо з
вибірки. Метрики порівняння: ефективна ширина спектра, спектральна
ефективність, дисперсія оцінок ядра.

Реалізація --- на базі R-пакета EstemPMM {[}33{]} з розширенням на
векторнозначний аргумент і Volterra-структуру базису. Орієнтовний час
підготовки експерименту і статті --- 2--3 місяці після завершення цієї
оглядової роботи. Цільовий журнал --- IET Signal Processing або
Circuits, Systems, and Signal Processing, де школа вже має вхідну точку
через {[}27, 35{]}. Альтернатива --- пряма публікація в ВОТТП як прямий
діалог зі статтею Гарячого і Щербініна.

\subsubsection{10.2. PATP-параметризація
ядра}\label{patp-ux43fux430ux440ux430ux43cux435ux442ux440ux438ux437ux430ux446ux456ux44f-ux44fux434ux440ux430}

Друга задача виходить з обмеження статті {[}6{]}, сформульованого самими
авторами: метод Вольтера працює по-різному для структурованих сигналів
(OFDM) і шумоподібних. Тому PATP-параметризація ядра розглядається тут
як неперервний контроль переходу між структурованістю і шумоподібністю.
Поточна школа Кунченка пропонує дискретне розрізнення через перфорацію
кумулянтного опису --- асиметричні, ексцесні, асиметрично-ексцесні класи
{[}4{]}. Це дискретний вибір, що не відображає неперервної природи
реальних сигналів, які можуть мати проміжну ступінь структурованості.

Альтернативний підхід --- введення \textbf{параметрично адаптивного
перехідного поліному} (PATP, Parametrically-Adaptive Transition
Polynomial). Це гіпотетичне узагальнення базисної структури
стохастичного поліному, в якому замість дискретного вибору з кількох
класів («степеневий», «тригонометричний», «фрактальний»)
використовується неперервний параметр \(\alpha \in \lbrack 0,1\rbrack\).
Цей параметр визначає тип базисних функцій плавно: у граничному випадку
\(\alpha = 0\) базис складається з фрактальних функцій вигляду
\(|\xi|^{1/i}\), які краще описують шумоподібні сигнали з важкими
хвостами; у випадку \(\alpha = 1\) --- зі стандартних степеневих
\(\xi^{i}\), оптимальних для структурованих сигналів типу OFDM; проміжні
значення \(\alpha\) дають перехідні базиси, що адаптуються до сигналів з
частковою структурою. Параметр \(\alpha\) оптимізується автоматично ---
за крос-валідацією на калібрувальній вибірці.

Концептуально це розширення апарату школи до неперервної параметризації
базису. Воно не суперечить класичному формулюванню Кунченка, а доповнює
його: дискретні класи перфорації стають частковими випадками
неперервного PATP-простору. Технічно --- потребує переформулювання
матриці центрованих корелянтів \(F_{S}\) як функції двох аргументів:
оцінюваного параметра \(\overrightarrow{\theta}\) і базисного параметра
\(\alpha\). Реалізація і чисельна верифікація --- окрема стаття на 4--6
місяців. Цільові журнали --- Signal Processing (Elsevier) або IEEE
Transactions on Signal Processing, де неперервна параметризація базису
як інструмент адаптивної обробки знайде природну авдиторію.

\subsubsection{10.3. GSA-CUSUM детектор замість евристичного індикатора
нестабільності}\label{gsa-cusum-ux434ux435ux442ux435ux43aux442ux43eux440-ux437ux430ux43cux456ux441ux442ux44c-ux435ux432ux440ux438ux441ux442ux438ux447ux43dux43eux433ux43e-ux456ux43dux434ux438ux43aux430ux442ux43eux440ux430-ux43dux435ux441ux442ux430ux431ux456ux43bux44cux43dux43eux441ux442ux456}

Третя задача --- заміна евристичного індикатора нестабільності \(K(t)\)
зі статті {[}6{]} на формальний поліноміальний CUSUM-детектор зміни
розподілу. Для цієї постановки вже існує близький внутрішній прецедент
школи --- напівпараметричне оцінювання моменту розладнання параметрів
негаусівських послідовностей методом максимізації полінома {[}28,
28a{]}. Тут ми використовуємо умовну робочу абревіатуру \textbf{GSA}
(Generalized Stochastic Approximation, узагальнена стохастична
апроксимація) для тієї гілки апарату Кунченка, що базується на розкладі
логарифма відношення правдоподібності в стохастичний ряд з оптимізацією
коефіцієнтів за моментними критеріями якості формування вирішних правил
{[}12, 12a, 13{]}. Ця абревіатура у школі систематично не вживається ---
у внутрішній термінології ця гілка описується розгорнуто (як «розклад
логарифма відношення правдоподібності у вигляді стохастичних
рядів\ldots» {[}4{]}) --- однак у міжнародному контексті стислий ярлик
зручний, і ми його приймаємо як технічний термін цієї статті.

Формалізм поліноміального CUSUM-детектора (CUSUM, Cumulative Sum ---
кумулятивна сума, класична статистика виявлення зміни моменту
розладнання):

\[
T_{n} = \sum_{k = 1}^{n}\Lambda_{S}^{(\mathrm{poly})}\left( x_{k} \right),\quad\tau_{\mathrm{detect}} = \min\{ n:T_{n} > h\},
\]

де \(\Lambda_{S}^{(\mathrm{poly})}\) --- поліноміальна апроксимація
логарифма відношення правдоподібності, побудована за стохастичним рядом
з оптимальними за моментними критеріями коефіцієнтами; поріг \(h\)
обирається з нерівності Чебишева або Височанського-Петуніна для
забезпечення контрольованої FAR (False Alarm Rate, ймовірність хибної
тривоги). Це дає \textbf{статистично обґрунтований} детектор зміни
розподілу замість евристичного \(K(t)\), з гарантією
\(\mathrm{FAR}\leq\varepsilon\) і обчислювальною складністю порядку
\(O\left( N \cdot S^{2} \right)\).

Технічно --- це інтеграція двох гілок школи: розкладу логарифма
відношення правдоподібності в стохастичний ряд Кунченко (первинне
джерело ISIT 1997 {[}12a{]}) і моментно-кумулянтного оцінювача
параметрів (докторський апарат {[}4{]}). Експериментальна верифікація на
реальних радіотехнічних сигналах з зміною режиму. Орієнтовний термін ---
6--8 місяців. Цільові журнали --- Sequential Analysis або Journal of
Statistical Planning and Inference; альтернатива --- IEEE Transactions
on Information Theory як журнал з традицією публікацій з зміни точки.

\subsubsection{10.4. Послідовність і
взаємодоповнюваність}\label{ux43fux43eux441ux43bux456ux434ux43eux432ux43dux456ux441ux442ux44c-ux456-ux432ux437ux430ux454ux43cux43eux434ux43eux43fux43eux432ux43dux44eux432ux430ux43dux456ux441ux442ux44c}

Три задачі утворюють природну послідовність: § 10.1 --- заміна критерію
без зміни структури Вольтера-моделі; § 10.2 --- розширення базису з
дискретного на неперервний; § 10.3 --- інтеграція оцінювальної та
детекторної гілок школи. У такому вигляді дослідницька програма не
завершує меморіальну статтю, а відкриває з неї наступний цикл робіт: від
історично вмотивованої реконструкції до чисельно перевірених моделей і
відкритого benchmark-порівняння.

\subsection{11. Відкриті методологічні та інфраструктурні
проблеми}\label{ux432ux456ux434ux43aux440ux438ux442ux456-ux43cux435ux442ux43eux434ux43eux43bux43eux433ux456ux447ux43dux456-ux442ux430-ux456ux43dux444ux440ux430ux441ux442ux440ux443ux43aux442ux443ux440ux43dux456-ux43fux440ux43eux431ux43bux435ux43cux438}

Поряд з активною дослідницькою програмою § 10, стан школи на 2026 р.
вимагає назвати кілька відкритих проблем. Вони не знецінюють уже
зроблене; навпаки, саме зрілість апарату дозволяє сформулювати, чого
йому бракує для ширшої наукової видимості та відтворюваного порівняння з
суміжними підходами.

\subsubsection{11.1. Інфраструктурний розрив із прикладною українською
радіотехнікою}\label{ux456ux43dux444ux440ux430ux441ux442ux440ux443ux43aux442ux443ux440ux43dux438ux439-ux440ux43eux437ux440ux438ux432-ux456ux437-ux43fux440ux438ux43aux43bux430ux434ux43dux43eux44e-ux443ux43aux440ux430ux457ux43dux441ux44cux43aux43eux44e-ux440ux430ux434ux456ux43eux442ux435ux445ux43dux456ux43aux43eux44e}

Найвидиміша і найболісніша прогалина --- те, що стаття {[}6{]} зробила
прозорим. Між апаратом школи і його природними прикладними споживачами в
Україні існує розрив, який не закривається ні монографіями, ні
міжнародними публікаціями. Виходи школи в IEEE ISIT 1997 {[}12a{]}, IET
Signal Processing 2017 {[}27{]} і Circuits, Systems, and Signal
Processing 2017 {[}35{]} не сягнули харківської радіотехнічної
аудиторії; англомовний канон Shaker 2002 {[}9{]} не сягнув ні
викладацьких корпусів, ні кафедральних бібліотек. Українські оборонні
радіотехнічні центри в 2026 р. цитують класичну літературу, але не
Кунченка 2002 р.

Зачинити цей розрив --- задача не суто наукова, а інфраструктурна. Вона
вимагає українських оглядових публікацій (включно з цією), українських
навчальних посібників на основі апарату школи, інтеграції ММПл у курси
статистичної радіотехніки в українських ВНЗ, прямих діалогових
публікацій у журналах, де працюють прикладні дослідники (ВОТТП,
Радіоелектронні і комп'ютерні системи). Без цього кроку міжнародна
видимість школи буде далі рости, а локальна --- стагнувати.

\subsubsection{11.2. Формальні мости до GMM, M-оцінювачів, L-моментів і
SLS}\label{ux444ux43eux440ux43cux430ux43bux44cux43dux456-ux43cux43eux441ux442ux438-ux434ux43e-gmm-m-ux43eux446ux456ux43dux44eux432ux430ux447ux456ux432-l-ux43cux43eux43cux435ux43dux442ux456ux432-ux456-sls}

Як зазначено в § 7, концептуальна спорідненість між апаратом Кунченка і
паралельними західними традиціями (GMM Hansen, M-оцінювачі Хубера,
L-моменти Хосейна, SLS Wang) очевидна, але формальних мостів між цими
школами в літературі здебільшого немає. ММПл можна інтерпретувати як
специфічну форму GMM з заміною Hansen-вагової матриці на
\(F_{S}\left( \overrightarrow{\theta} \right)\), але цей зв'язок поки не
оформлений у математично строгому вигляді. Аналогічно з L-моментами ---
концептуально близький апарат, але без формального містка.

Окремої уваги заслуговує співставлення PMM2 ↔ SLS. На відміну від мостів
до GMM та L-моментів, які залишаються чисто концептуальними, тут
\textbf{емпіричний міст уже частково побудований}: робота {[}25{]} прямо
порівнює ефективність обох методів і демонструє їхню практичну
еквівалентність на нелінійних регресійних задачах з χ²-помилками. Однак
повноформатне теоретичне зведення PMM2 і SLS у єдину формальну рамку ---
окрема публікація, яка в літературі поки не з'явилася. Природна
постановка такої роботи: показати, що SLS --- це частковий випадок
узагальненого PMM з ваговою матрицею певної структури, а оптимальні
коефіцієнти PMM при S=2 є альтернативною параметризацією тієї ж сім'ї
оцінювачів. Якщо це теоретичне зведення вдасться побудувати, школа
Кунченка отримує природну точку входу в стандартну західну літературу з
нелінійної регресії --- журнали рівня \emph{Annals of the Institute of
Statistical Mathematics} (де SLS публікується традиційно), \emph{Journal
of Statistical Planning and Inference}, \emph{Econometric Theory}.

Будь-яка з цих формалізацій --- самостійна публікація в журналі рівня
\emph{Annals of Statistics}, \emph{Journal of the American Statistical
Association}, \emph{Biometrika}, де школа Кунченка поки не присутня. Це
найвища планка міжнародного позиціонування, яка реальна для школи лише
через побудову таких мостів.

\subsubsection{11.3. Відсутність систематичного
benchmark-датасету}\label{ux432ux456ux434ux441ux443ux442ux43dux456ux441ux442ux44c-ux441ux438ux441ux442ux435ux43cux430ux442ux438ux447ux43dux43eux433ux43e-benchmark-ux434ux430ux442ux430ux441ux435ux442ux443}

Усі публікації школи 2006--2026 рр. використовують власні згенеровані
датасети --- Monte Carlo з конкретно обраними розподілами інновацій,
синтетичні OFDM з конкретно обраними нелінійностями, конкретні моделі
ARIMA з конкретними інноваціями. Це робить порівняння результатів між
статтями школи скрутним і унеможливлює пряме зіставлення з зовнішніми
оцінювачами. Школа не має publicly available benchmark dataset на зразок
UCI ML Repository або PhysioNet, на якому будь-який зовнішній дослідник
міг би оцінити ММПл проти альтернатив.

Створення такого датасету --- реалістична задача на 9--12 місяців:
добірка 8--12 реалістичних радіотехнічних, метрологічних і фінансових
сценаріїв, з добре документованою генерацією, відкритою публікацією
через Zenodo або OSF, і набором базових скриптів для відтворення на
R/Python. Це інфраструктурна робота, без якої школа залишатиметься
замкненою на власну спільноту.

\subsubsection{11.4. PMM + Deep Learning
гібриди}\label{pmm-deep-learning-ux433ux456ux431ux440ux438ux434ux438}

З 2018 р. в глобальній статистиці й обробці сигналів стрімко
розвивається напрямок, що поєднує класичні семіпараметричні методи з
нейромережевими архітектурами --- neural ODEs з статистичною
регуляризацією, deep state-space моделі, neural change-point detection.
Школа Кунченка в цьому напрямку поки майже не присутня. Між тим апарат
стохастичних поліномів природно вписується в цей контекст: коефіцієнти
ММПл можуть бути параметризовані як виходи невеликих нейромереж, що
навчаються спільно з основним моделем, а матриця \(F_{S}\) ---
використовуватись як inductive bias для регуляризації втрат.

Це широкий і недосліджений простір. Перші роботи в цьому напрямку могли
б з'явитись на стику EstemPMM з PyTorch/JAX через створення
диференційовних модулів ММПл.

\subsection{12. Висновки і
присвята}\label{ux432ux438ux441ux43dux43eux432ux43aux438-ux456-ux43fux440ux438ux441ux432ux44fux442ux430}

Школа Юрія Петровича Кунченка пройшла шлях від конкретної радіофізичної
задачі до зрілого апарату семіпараметричного негаусівського оцінювання.
У кандидатській дисертації 1972/1973 р. ряди Вольтера були застосовані
до знаходження оцінок невідомих параметрів випадкових процесів. У
монографіях 1987--2006 рр. ця ідея перетворилася на стохастичні
поліноми, метод максимізації поліному, простір з порідним елементом і
моментні критерії перевірки гіпотез. У роботах учнів і послідовників
2006--2026 рр. вона продовжилася в регресійному оцінюванні,
сигнально-детекторних задачах, метрології, обробці біомедичних сигналів
і програмній реалізації EstemPMM.

Головний висновок цієї статті полягає в тому, що історія школи не є лише
історією спадкоємності. Вона є траєкторією статистичного методу. Випадок
2026 р. з Вольтера-обробкою стохастичних сигналів показує, що вихідний
тип постановки Кунченка повертається в сучасну прикладну радіотехніку:
нелінійний функціонал від випадкового процесу знову використовується для
оцінювальної або адаптивної процедури. Формальний міст, побудований у §
9, пояснює, як скінченна Вольтера-модель вкладається в клас узагальнених
стохастичних поліномів, але не ототожнює MMSE-адаптацію з ММПл. Саме
розходження між цими критеріями і створює дослідницьку можливість:
замінити або доповнити фіксований L2-критерій параметрично залежним
моментним критерієм і перевірити, чи дає це виграш на задачах типу
{[}6{]}.

Цей висновок не слід читати як готову обіцянку універсального виграшу.
Коректна постановка потребує перевірки моментних умов, невиродженості
матриці центрованих корелянтів і чисельного порівняння на
benchmark-сценаріях. Саме тому дослідницька програма § 10 є необхідним
продовженням статті: вона перетворює пам'ять про школу на роботу з її
апаратом у сучасному методологічному полі.

Текст присвячено пам'яті Юрія Петровича Кунченка до 87-ї річниці його
народження. Його наукова спадщина важлива не лише тим, що створила
школу, монографії та покоління учнів. Вона залишила спосіб бачити в
прикладній задачі математичний об'єкт, а в математичному об'єкті ---
майбутню прикладну задачу. Саме цей подвійний погляд робить апарат
стохастичних поліномів живим і сьогодні.

\subsection{Список
літератури}\label{ux441ux43fux438ux441ux43eux43a-ux43bux456ux442ux435ux440ux430ux442ux443ux440ux438}

{[}1{]} Кунченко Ю.П. Застосування рядів Вольтера для знаходження оцінок
невідомих параметрів випадкових процесів: дис. \ldots{} канд. фіз.-мат.
наук: 01.04.03. Томськ: Томський державний університет, 1972/1973. 164
с. Шифр зберігання РДБ: OD Дк 73-1/1227. URL:
https://search.rsl.ru/ru/record/01009787980; автореферат:
https://search.rsl.ru/ru/record/01007428279.

{[}2{]} Кунченко Ю.П., Лега Ю.Г. Оцінка параметрів випадкових величин
методом максимізації поліному. Київ: Наукова думка, 1991/1992.

{[}3{]} Кунченко Ю.П. Поліноми наближення у просторі з породжувальним
елементом. Київ: Наукова думка, 2005. \emph{(україномовна версія)}

{[}3a{]} Кунченко Ю.П. Полиномы приближения в пространстве с порождающим
элементом. К.: Наукова думка, 2003. 243 с. \emph{(оригінальне
російськомовне видання)}

{[}4{]} Заболотній С.В. Інформаційна технологія ймовірнісного
діагностування розладнання параметрів негаусових послідовностей: дис.
\ldots{} д-ра техн. наук: 05.13.06. Львів: Українська академія
друкарства, 2015.

{[}5{]} Палагін В.В. Математичні моделі, методи та засоби виявлення і
розрізнення сигналів на фоні негаусових завад: дис. \ldots{} д-ра техн.
наук: 01.05.02. Черкаси: ЧДТУ, 2013.

{[}6{]} Гарячий М., Щербінін С. Метод формування стохастичних сигналів
із використанням рядів Вольтера // Вимірювальна та обчислювальна техніка
в технологічних процесах. 2026. № 1. С. 47--60.

{[}7{]} Меморіальна стаття про Ю.П. Кунченка. Сайт кафедри радіотехніки
ЧДТУ. URL: \url{https://rtrs.chdtu.edu.ua/history-of-the-department/}

{[}8{]} Кунченко Ю.П. Нелінійна оцінка параметрів негауссівських
радіофізичних сигналів. Київ: Вища школа, 1987. 191 с.

{[}9{]} Kunchenko Y.P. \emph{Polynomial Parameter Estimations of Close
to Gaussian Random Variables}. Aachen: Shaker Verlag, 2002. 396 p.

{[}10{]} Кунченко Ю.П., Заболотній С.В. Поліноміальні оцінки параметрів
близьких до гаусових випадкових величин. Частина 1, 2. Черкаси: ЧІТІ,
2001.

{[}10a{]} Zabolotnii S.W., Martynenko S.S., Salypa S.V. Method of
verification of hypothesis about mean value on a basis of expansion in a
space with generating element // Radioelectronics and Communications
Systems. 2018. Vol. 61, No.~5. P. 222--229. DOI:
https://doi.org/10.3103/S0735272718050060.

{[}11{]} Кунченко Ю.П. Стохастические полиномы. К.: Наукова думка, 2006.
275 с.

{[}12{]} Кунченко Ю.П., Палагін В.В. Побудова моментного критерію якості
типу Неймана-Пірсона для перевірки простих статистичних гіпотез //
Вісник Інженерної академії України. 2005. № 1. С. 26--30.

{[}12a{]} Kunchenko Y. A moment performance criteria of a
decision-making for testing simple statistical hypothesis // Proc. IEEE
International Symposium on Information Theory (ISIT), Ulm, Germany, July
1997. P. 407. DOI: https://doi.org/10.1109/ISIT.1997.613344.

{[}13{]} Кунченко Ю.П., Палагін В.В. Перевірка статистичних гіпотез при
використанні поліноміальних вирішних правил, оптимальних за моментним
критерієм суми асимптотичних ймовірностей помилок // Радіоелектроніка та
інформатика. 2006. № 3 (34). С. 4--11.

{[}14{]} Палагін В.В. Алгоритми виявлення сигналів на фоні
негауссівських завад за критерієм асимптотичної нормальності: дис.
\ldots{} канд. техн. наук: 05.12.01. Черкаси: ЧДТУ, 1999.

{[}15{]} Заболотній С.В. Нелінійні алгоритми визначення параметрів
негаусових випадкових послідовностей у каналах
інформаційно-вимірювальних систем: дис. \ldots{} канд. техн. наук:
05.11.16. Черкаси: ЧДТУ, 2000.

{[}16{]} Вєдєрніков Д.А. Математичні моделі, методи та засоби оцінювання
параметра постійного сигналу на фоні негаусових корельованих завад: дис.
\ldots{} PhD. Черкаси: ЧДТУ, 2021.

{[}17{]} Смірнов Д.О. Математичні моделі, методи та засоби виявлення
постійного сигналу на фоні негаусових корельованих завад: дис. \ldots{}
PhD. Черкаси: ЧДТУ, 2025.

{[}18{]} Ткаченко О.М. Поліноміальні методи та засоби оцінювання
параметрів регресії з використанням моделей негаусових помилок: дис.
\ldots{} PhD. Черкаси: ЧДТУ, 2021.

{[}19{]} Chertov O., Slipets T. Kunchenko's Polynomials for Template
Matching. arXiv:1107.2085, 2011.

{[}20{]} Chertov O., Slipets T. Epileptic seizures diagnose using
Kunchenko polynomials template matching // Progress in Industrial
Mathematics at ECMI 2012 / Eds. M. Fontes, M. Günther, N. Marheineke.
Springer, Cham, 2014. P. 245--248. DOI:
https://doi.org/10.1007/978-3-319-05365-3\_33.

{[}21{]} Warsza Z.L., Zabolotnii S.W. A polynomial estimation of
measurand parameters for samples of non-Gaussian symmetrically
distributed data // AISC, vol.~550. Springer, Cham, 2017. P. 468--480.
DOI: https://doi.org/10.1007/978-3-319-54042-9\_45.

{[}22{]} Warsza Z.L., Zabolotnii S.W. Estimation of measurand parameters
for data from asymmetric distributions by polynomial maximization method
// AISC, vol.~743. Springer, 2018. P. 746--757. DOI:
https://doi.org/10.1007/978-3-319-77179-3\_74.

{[}23{]} Zabolotnii S., Warsza Z.L., Tkachenko O. Polynomial Estimation
of Linear Regression Parameters for the Asymmetric PDF of Errors //
AISC, vol.~743. Springer, 2018. DOI:
https://doi.org/10.1007/978-3-319-77179-3\_75.

{[}24{]} Zabolotnii S., Warsza Z.L., Tkachenko O. Estimation of Linear
Regression Parameters of Symmetric Non-Gaussian Errors by Polynomial
Maximization Method // Advances in Intelligent Systems and Computing,
vol.~920. Springer, 2019. DOI:
https://doi.org/10.1007/978-3-030-13273-6\_59.

{[}25{]} Zabolotnii S., Tkachenko O., Nowakowski W., Warsza Z.L.
Application of the Polynomial Maximization Method for Estimating
Nonlinear Regression Parameters with Non-gaussian Asymmetric Errors //
Automation 2024: Advances in Automation, Robotics and Measurement
Techniques. Springer LNNS, vol.~1219, 2025. P. 1--15. DOI:
https://doi.org/10.1007/978-3-031-78266-4\_30.

{[}26{]} Palahin V., Juhár J. Joint signal parameter estimation in
non-Gaussian noise by the method of polynomial maximization // Journal
of Electrical Engineering. 2016. Vol. 67, No.~3. P. 217--221. DOI:
https://doi.org/10.1515/jee-2016-0031.

{[}27{]} Vokorokos L., Ivchenko A., Marchevský S., Palahina E., Palahin
V. Parameters estimation of correlated non-Gaussian processes by the
method of polynomial maximisation // IET Signal Processing. 2017. Vol.
11, No.~3. P. 313--319.

{[}28{]} Zabolotnii S.W., Warsza Z.L. Semi-parametric estimation of the
change-point of parameters of non-Gaussian sequences by polynomial
maximization method // AISC, vol.~440. Springer, Heidelberg, 2016. P.
903--919. DOI: https://doi.org/10.1007/978-3-319-29357-8\_80.

{[}28a{]} Zabolotnii S.W., Warsza Z.L. Semi-parametric polynomial method
for retrospective estimation of the change-point of parameters of
non-Gaussian sequences // Advanced Mathematical and Computational Tools
in Metrology and Testing X (AMCTM X) / Eds. F. Pavese et al.~Series on
Advances in Mathematics for Applied Sciences, vol.~86. World Scientific,
Singapore, 2015. P. 400--408.

{[}29{]} Матеріали VII Міжнародної науково-практичної конференції
«Обробка сигналів і негауссівських процесів». Черкаси: ЧДТУ, 2019. 212
с.

{[}30{]} Zabolotnii S.V., Chepynoha A.V., Bondarenko Yu.Yu., Rud M.P.
Polynomial parameter estimation of exponential power distribution data
// Visnyk NTUU KPI Seriia - Radiotekhnika Radioaparatobuduvannia. 2018.
Issue 75. P. 40--47. DOI: https://doi.org/10.20535/radap.2018.75.40-47.

{[}31{]} Zabolotnii S., Khotunov V., Chepynoha A., Tkachenko O.
Estimating parameters of linear regression with an exponential power
distribution of errors by using a polynomial maximization method //
Eastern-European Journal of Enterprise Technologies. 2021. Vol. 1,
No.~4(109). DOI: https://doi.org/10.15587/1729-4061.2021.225525.

{[}32{]} Zabolotnii S. Applying PMM to estimate ARIMA models with
asymmetric non-Gaussian innovations. arXiv:2511.07059, 2025.

{[}33{]} Zabolotnii S. EstemPMM: Polynomial Maximization Method for
Non-Gaussian Errors. R package, CRAN, 2025. URL:
https://cran.r-project.org/package=EstemPMM

{[}33a{]} Заболотній С.В., Чепинога А.В., Салипа С.В. Спосіб генерації
випадкових величин: декларац. патент України на корисну модель № 57092,
МПК G06F7/58. Заявл. 16.07.2010; Опубл. 10.02.2011. Бюл. № 3.

{[}34{]} Palahin V., Palahina I., Filipov V., Leleko S., Ivchenko A.
Modeling of joint signal detection and parameter estimation on
background of Non-Gaussian noise // Journal of Applied Mathematics and
Computational Mechanics. 2015. Vol. 14, No.~3. P. 87--94. DOI:
\url{https://doi.org/10.17512/jamcm.2015.3.09}.

{[}35{]} Palahina E., Gamcová M., Gladišová I., Gamec J., Palahin V.
Signal Detection in Correlated Non-Gaussian Noise Using Higher-Order
Statistics // Circuits, Systems, and Signal Processing. 2017. Vol. 37,
No.~4. P. 1704--1723. DOI: https://doi.org/10.1007/s00034-017-0623-5.

{[}36{]} Smirnov D., Palahina E., Palahin V. Mathematical Modeling of
Signal Detection in Non-gaussian Correlated Noise // Smart Technologies
in Urban Engineering. STUE 2022. Lecture Notes in Networks and Systems,
vol.~536. Springer, 2023. P. 65--74. DOI:
https://doi.org/10.1007/978-3-031-20141-7\_7.

{[}37{]} Krasilnikov A., Beregun V. Cumulant Detector of Non-Gaussian
Signals against Background of Non-Gaussian Interferences //
Radioelectronics and Communications Systems. 2024. Vol. 67, No.~6. P.
317--330. DOI: https://doi.org/10.3103/S0735272724060037.

{[}38{]} Чертов О.Р., Тавров Д.Ю. Use Kunchenko Polynomials for the
Analysis of the Statistical Data // Eastern-European Journal of
Enterprise Technologies. 2010. Vol. 4, Issue 4(46). P. 70--75.

{[}39{]} Захищені дисертації за напрямком наукової школи професора
Кунченка Ю.П. Сайт кафедри радіотехніки ЧДТУ. URL:
https://rtrs.chdtu.edu.ua/zahysty-dysertaczijnyh-robit/

{[}40{]} Заболотній С.В., Ткаченко О.М. Поліноміальні адаптивні
процедури регресійного аналізу із використанням моделей негаусових
помилок на основі статистик вищих порядків // Тези доповідей IV
Міжнародної науково-практичної конференції «Обчислювальний інтелект
(результати, проблеми, перспективи) --- 2017» (ComInt-2017). Київ: КНУ
ім. Т. Шевченка, 16--18 травня 2017 р. С. 113--114.

{[}41{]} Заболотній С.В., Ткаченко О.М. Застосування методу максимізації
полінома для оцінювання параметрів однофакторної лінійної регресії при
негаусовому розподілі помилок. Бібліографічний запис у переліку праць
О.М. Ткаченка, ЧДТУ. URL:
https://rtrs.chdtu.edu.ua/postgraduate-students-of-the-department/

{[}42{]} Збірка тез VIII Міжнародної науково-практичної конференції
«Обробка сигналів і негауссівських процесів» (ОСНП-2021), присвяченої
пам'яті професора Ю.П. Кунченка. Черкаси: ЧДТУ, 25--26 травня 2021 р.

{[}43{]} Rugh W.J. \emph{Nonlinear System Theory: The Volterra/Wiener
Approach}. Baltimore: Johns Hopkins University Press, 1981.

{[}44{]} Schetzen M. \emph{The Volterra and Wiener Theories of Nonlinear
Systems}. New York: Wiley, 1980.

{[}45{]} Mathews V.J., Sicuranza G.L. \emph{Polynomial Signal
Processing}. New York: Wiley, 2000.

{[}46{]} Mendel J.M. Tutorial on higher-order statistics (spectra) in
signal processing and system theory: theoretical results and some
applications // Proceedings of the IEEE. 1991. Vol. 79, No.~3. P.
278--305.

{[}47{]} Nikias C.L., Petropulu A.P. \emph{Higher-Order Spectra
Analysis: A Nonlinear Signal Processing Framework}. Prentice Hall, 1993.

{[}48{]} Giannakis G.B., Tsatsanis M.K. Signal detection and
classification using matched filtering and higher order statistics //
IEEE Transactions on Acoustics, Speech, and Signal Processing. 1990.
Vol. 38, No.~7. P. 1284--1296.

{[}49{]} Huber P.J. \emph{Robust Statistics}. New York: Wiley, 1981.

{[}50{]} Hansen L.P. Large sample properties of generalized method of
moments estimators // Econometrica. 1982. Vol. 50, No.~4. P. 1029--1054.

{[}51{]} Hosking J.R.M. L-moments: analysis and estimation of
distributions using linear combinations of order statistics // Journal
of the Royal Statistical Society, Series B. 1990. Vol. 52, No.~1. P.
105--124.

{[}52{]} Wang L. Estimation of nonlinear Berkson-type measurement error
models // Statistica Sinica. 2003. Vol. 13. P. 1201--1210.

{[}53{]} Wang L., Leblanc A. Second-order nonlinear least squares
estimation // Annals of the Institute of Statistical Mathematics. 2008.
Vol. 60. P. 883--900. DOI: https://doi.org/10.1007/s10463-007-0139-z.

{[}54{]} Kim M., Ma Y. The efficiency of the second-order nonlinear
least squares estimator and its extension // Annals of the Institute of
Statistical Mathematics. 2012. Vol. 64. P. 751--764. DOI:
https://doi.org/10.1007/s10463-011-0332-y.

{[}55{]} Gao L.L., Zhou J. New optimal design criteria for regression
models with asymmetric errors // Journal of Statistical Planning and
Inference. 2014. Vol. 149. P. 140--151. DOI:
\url{https://doi.org/10.1016/j.jspi.2014.01.005}.

{[}56{]} Salamh M., Wang L. Second-order least squares estimation in
nonlinear time series models with ARCH errors // Econometrics. 2021.
Vol. 9, No.~4. Article 41. DOI:
\url{https://doi.org/10.3390/econometrics9040041}.

{[}57{]} He L., Yue R.X., Du A. Optimal designs for comparing curves in
regression models with asymmetric errors // Journal of Statistical
Planning and Inference. 2024. Vol. 228. P. 46--58. DOI:
https://doi.org/10.1016/j.jspi.2023.06.005.

\end{document}